\documentclass[12pt,english]{iopart}
\usepackage{cite}
\usepackage{iopams}
\usepackage{graphicx}
\usepackage{bm}

\usepackage[toc,page]{appendix}
\usepackage{etoolbox}
\appto\appendix{\addtocontents{toc}{\protect\setcounter{tocdepth}{0}}}

\usepackage{fancyhdr}
\usepackage[usenames,dvipsnames,svgnames,table]{xcolor}
\usepackage[unicode=true,
            pdfusetitle,
            bookmarks=true,
            bookmarksnumbered=false,
            bookmarksopen=false,
            breaklinks=true,
            pdfborder={0 0 0},
            backref=false,
            colorlinks=true]{hyperref}
\hypersetup{linkcolor=NavyBlue,urlcolor=NavyBlue,citecolor=NavyBlue}

\newtheorem{prop}{Proposition}[section]
\newtheorem{theorem}{Theorem}[section]
\newtheorem{corollary}{Corollary}[theorem]

\usepackage{scrextend}
\deffootnote{0em}{1.6em}{\thefootnotemark.\enskip}

\providecommand{\openone}{\leavevmode\hbox{\small1\kern-4.3pt\normalsize1}}

\begin{document}

\title{Quantum Fisher information matrix and multiparameter estimation}

\author{Jing Liu$^{1}$, Haidong Yuan$^{2}$, Xiao-Ming Lu$^{3}$, Xiaoguang Wang$^{4}$}
\address{$^{1}$MOE Key Laboratory of Fundamental Physical Quantities Measurement \& Hubei
Key Laboratory of Gravitation and Quantum Physics, PGMF and School of Physics,
Huazhong University of Science and Technology, Wuhan 430074, P. R. China}
\address{$^{2}$Department of Mechanical and Automation Engineering, The Chinese
University of Hong Kong, Shatin, Hong Kong}
\address{$^{3}$Department of Physics, Hangzhou Dianzi University, Hangzhou 310018, China}
\address{$^{4}$Zhejiang Institute of Modern Physics, Department of Physics, Zhejiang University,
Hangzhou 310027, China}

\ead{liujingphys@hust.edu.cn, hdyuan@mae.cuhk.edu.hk, lxm@hdu.edu.cn, xgwang1208@zju.edu.cn}

\begin{abstract}
Quantum Fisher information matrix (QFIM) is a core concept in theoretical
quantum metrology due to the significant importance of quantum Cram\'{e}r-Rao bound in quantum
parameter estimation. However, studies in recent years have revealed wide connections
between QFIM and other aspects of quantum mechanics, including quantum thermodynamics,
quantum phase transition, entanglement witness, quantum speed limit and
non-Markovianity. These connections indicate that QFIM is more than a concept in
quantum metrology, but rather a fundamental quantity in quantum
mechanics. In this paper, we summarize the properties and existing calculation
techniques of QFIM for various cases, and review the development of QFIM in some
aspects of quantum mechanics apart from quantum metrology. On the other hand, as the main
application of QFIM, the second part of this paper reviews the quantum multiparameter
Cram\'{e}r-Rao bound, its attainability condition and the associated optimal measurements.
Moreover, recent developments in a few typical scenarios of quantum multiparameter estimation
and the quantum advantages are also thoroughly discussed in this part.
\end{abstract}

\maketitle

\tableofcontents

\section{Introduction}

After decades of rapid development, quantum mechanics has now gone deep into almost every corner
of modern science, not only as a fundamental theory, but also as a technology. The technology
originated from quantum mechanics is usually referred to as quantum technology, which is aiming
at developing brand new technologies or improving current existing technologies with the association
of quantum resources, quantum phenomena or quantum effects. Some aspects of quantum technology,
such as quantum communications, quantum computation, quantum cryptography
and quantum metrology, have shown great power in both theory and laboratory to lead
the next industrial revolution. Among these aspects, quantum metrology is the most
promising one that can step into practice in the near future.

\begin{figure}[tp]
\centering\includegraphics[width=14cm]{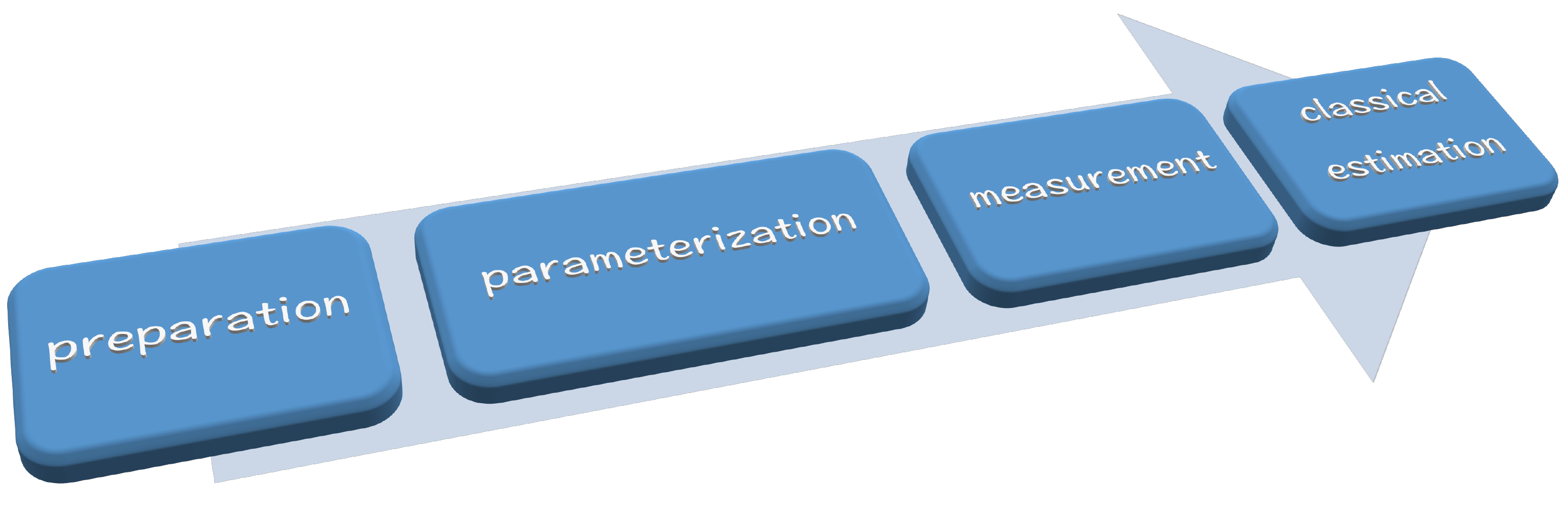}
\caption{Schematic of a complete quantum metrological process, which contains four steps:
(1) preparation of the probe state;
(2) parameterzation; (3) measurement; (4) classical estimation.}
\label{fig:metrology_flowchart}
\end{figure}

Quantum metrology focuses on making high precision measurements of given parameters
using quantum systems and quantum resources. Generally, a complete quantum metrological
process contains four steps: (1) preparation of the probe state; (2) parameterzation;
(3) measurement and (4) classical estimation, as shown in figure~\ref{fig:metrology_flowchart}.
The last step has been well studies in classical statistics, hence, the major concern of
quantum metrology is the first three steps.

Quantum parameter estimation is the theory for quantum metrology, and quantum
Cram\'{e}r-Rao bound is the most well-studied mathematical tool for quantum
parameter estimation~\cite{Helstrom,Holevo}. In quantum Cram\'{e}r-Rao bound,
the quantum Fisher information (QFI) and quantum Fisher information matrix
(QFIM) are the key quantities representing the precision limit for single
parameter and multiparameter estimations. In recent years, several outstanding
reviews on quantum metrology and quantum parameter estimation have been provided from
different perspectives and at different time, including the ones given by Giovannetti et al.
on the quantum-enhanced measurement~\cite{Giovannetti2004} and the advances
in quantum metrology~\cite{Giovannetti2011}, the ones given by Paris~\cite{Paris2009}
and Toth et al.~\cite{Toth2014} on the QFI and its applications in quantum metrology,
the one by Braun et al. on the quantum enhanced metrology without entanglement~\cite{Braunrmp},
the ones by Pezz\`{e} et al.~\cite{Pezze2018} and Huang et al.~\cite{Huang2014} on
quantum metrology with cold atoms, the one by Degen et al. on general quantum sensing~\cite{Degen2017},
the one by Pirandola et al. on the photonic quantum sensing~\cite{Pirandola2018},
the ones by Dowling on quantum optical metrology with high-N00N state~\cite{Dowling2008}
and Dowling and Seshadreesan on theoretical and experimental optical technologies
in quantum metrology, sensing and imaging~\cite{Dowling2015},
the one by Demkowicz-Dobrza\'{n}ski et al. on the quantum limits in optical
interferometry~\cite{Demkowicz2015}, the one by Sidhu and Kok on quantum parameter
estimation from a geometric perspective~\cite{Sidhu2019}, and the one by Szczykulska
et al. on simultaneous multiparameter estimation~\cite{Szczykulska2016}.
Petz et al. also wrote a thorough technical introduction on QFI~\cite{Petz2011}.

Apart from quantum metrology, the QFI also connects to other aspects of quantum physics,
such as quantum phase transition~\cite{Gu2010,Wang2014,Marzolino2017} and entanglement
witness~\cite{Hyllus2012,Toth2012}. The widespread application of QFI may be due to its
connection to the Fubini-study metric, a K\"{a}hler metric in the complex
projective Hilbert space. This relation gives the QFI a strong geometric meaning
and makes it a fundamental quantity in quantum physics. Similarly, the QFIM shares
this connection since the diagonal entries of QFIM simply gives the QFI. Moreover, the
QFIM also connects to other fundamental quantity like the quantum
geometric tensor~\cite{Venuti2007}. Thus, besides the role in multiparameter
estimation, the QFIM should also be treated as a fundamental quantity in
quantum mechanics.

In recent years, the calculation techniques of QFIM have seen a rapid development
in various scenarios and models. However, there lack papers that thoroughly
summarize these techniques in a structured manner for the community. Therefore, this
paper not only reviews the recent developments of quantum multiparameter
estimation, but also provides comprehensive techniques on the calculation
of QFIM in a variety of scenarios. For this purpose, this paper is presented in a
way similar to a textbook with many technical details given in the appendices, which
could help the readers to follow and better understand the corresponding results.

\section{Quantum Fisher information matrix}

\subsection{Definition} \label{sec:definition}

Consider a vector of parameters $\vec{x}=(x_{0},x_{1},...,x_{a},...)^{\mathrm{T}}$ with $x_{a}$
the $a$th parameter. $\vec{x}$ is encoded in the density matrix $\rho=\rho(\vec{x})$. In the
entire paper we denote the QFIM as $\mathcal{F}$, and an entry of $\mathcal{F}$ is
defined as~\cite{Helstrom,Holevo}
\begin{equation}
\mathcal{F}_{ab}:=\frac{1}{2}\mathrm{Tr}\left(\rho\{L_{a},L_{b}\}\right), \label{eq:def_QFIM}
\end{equation}
where $\{\cdot,\cdot\}$ represents the anti-commutation and $L_{a}$ ($L_{b}$)
is the symmetric logarithmic derivative (SLD) for the parameter $x_{a}$ ($x_{b}$),
which is determined by the equation~\footnote[1]{In the entire paper the notation
$\partial_{a}$ ($\partial_{t}$) is used as an abbreviation of
$\frac{\partial}{\partial x_{a}}$ $\left(\frac{\partial}{\partial t}\right)$.}
\begin{equation}
\partial_{a}\rho=\frac{1}{2}\left(\rho L_{a}+L_{a}\rho\right).
\label{eq:SLD}
\end{equation}
The SLD operator is a Hermitian operator and the expected value $\mathrm{Tr}(\rho L_{a})=0$.
Utilizing the equation above, $\mathcal{F}_{ab}$ can also be expressed by~\cite{Amari2000}
\begin{equation}
\mathcal{F}_{ab}=\mathrm{Tr}\left(L_{b}\partial_{a}\rho\right)
=-\mathrm{Tr}(\rho\partial_{a}L_{b}).
\end{equation}
Based on equation~(\ref{eq:def_QFIM}), the diagonal entry of QFIM is
\begin{equation}
\mathcal{F}_{aa}=\mathrm{Tr}\left(\rho L^{2}_{a}\right),
\end{equation}
which is exactly the QFI for parameter $x_{a}$.

The definition of Fisher information matrix originated from classical statistics.
For a probability distribution $\{p(y|\vec{x})\}$ where $p(y|\vec{x})$ is the conditional
probability for the outcome result $y$, an entry of Fisher information matrix is
defined as
\begin{equation}
\mathcal{I}_{ab}:=\int \frac{\left[\partial_{a}p(y|\vec{x})\right]
\!\left[\partial_{b}p(y|\vec{x})\right]}
{p(y|\vec{x})}~\mathrm{d}y.
\end{equation}
For discrete outcome results, it becomes $\mathcal{I}_{ab}:=\sum_{y}
\frac{[\partial_{a}p(y|\vec{x})][\partial_{b}p(y|\vec{x})]}
{p(y|\vec{x})}$. With the development of quantum metrology, the Fisher information
matrix concerning classical probability distribution is usually referred to as \emph{classical
Fisher information matrix} (CFIM), with the diagonal entry referred to
as~\emph{classical Fisher information} (CFI). In quantum mechanics, it is well known
that the choice of measurement will affect the obtained probability distribution,
and thus result in different CFIM. This fact indicates
the CFIM is actually a function of measurement. However, while
the QFI is always attained by optimizing over the measurements~\cite{Braunstein1994},
i.e., $\mathcal{F}_{aa}=\max_{\{\Pi_{y}\}}\mathcal{I}_{aa}(\rho, \{\Pi_{y}\})$,
where $\{\Pi_{y}\}$ represents a positive-operator valued measure (POVM), in general there
may not be any measurement that can attain the QFIM.

The QFIM based on SLD is not the only quantum version of CFIM. Another well-used
ones are based on the right and left logarithmic derivatives~\cite{Holevo,Yuen1973},
defined by $\partial_{a}\rho=\rho R_{a}$ and $\partial_{a}\rho=R^{\dagger}_{a}\rho$,
with the corresponding QFIM $\mathcal{F}_{ab}=\mathrm{Tr} (\rho R_{a}R^{\dagger}_{b})$.
Different with the one based on SLD, which are real symmetric, the QFIM based
on right and left logarithmic derivatives are complex and Hermitian. All versions
of QFIMs belong to a family of Riemannian monotone metrics established by Petz~\cite{Petz1996,Petz2008}
in 1996, which will be further discussed in section~\ref{sec:geometry}. All the QFIMs
can provide quantum versions of Cram\'{e}r-Rao bound, yet with different achievability.
For instance, for the D-invariant models only the one based on right logarithmic
derivative provides an achievable bound~\cite{Suzuki2016}. The quantum Cram\'{e}r-Rao
bound will be further discussed in section~\ref{sec:QCRB}. For pure states, Fujiwara
and Nagaoka~\cite{Fujiwara1995} also extended the SLD to a family via
$\partial_a\rho=\frac{1}{2} (\rho L_a +L^{\dagger}_a \rho)$, in which $L_a$ is not
necessarily to be Hermitian, and when it is, it reduces to the SLD. An
useful example here is the anti-symmetric logarithmic derivative $L^{\dagger}_a=-L_a$.
This paper focuses on the QFIM based on the SLD, thus, the QFIM in the following
only refers to the QFIM based on SLD without causing any confusion.

The properties of QFI have been well organized by G. T\'{o}th~\emph{et al.} in
reference~\cite{Toth2014}. Similarly, the QFIM also has some powerful properties that have
been widely applied in practice. Here we organize these properties as below.
\begin{prop} \label{prop:QFIM}
Properties and useful formulas of the QFIM.
\begin{itemize}
\item $\mathcal{F}$ is real symmetric, i.e., $\mathcal{F}_{ab}=\mathcal{F}_{ba}
\in \mathbb{R}$~\footnote{$\mathbb{R}$ represents the set of real numbers.}.
\item $\mathcal{F}$ is positive semi-definite, i.e., $\mathcal{F}\geq 0$. If $\mathcal{F}>0$,
then $[\mathcal{F}^{-1}]_{aa}\geq 1/\mathcal{F}_{aa}$ for any $a$.
\item $\mathcal{F}(\rho)=\mathcal{F}(U\rho U^{\dagger})$ for a $\vec{x}$-independent
unitary operation $U$.
\item If $\rho=\bigotimes_{i} \rho_{i}(\vec{x})$, then $\mathcal{F}(\rho)=\sum_{i}\mathcal{F}(\rho_{i})$.
\item If $\rho =\bigoplus_{i}\mu_{i}\rho_{i}(\vec{x})$ with $\mu_{i}$ a $\vec{x}$-independent weight,
then $\mathcal{F}(\rho )=\sum_{i}\mu_{i}\mathcal{F}(\rho_{i})$.
\item Convexity: $\mathcal{F}(p\rho_{1}+(1-p)\rho_{2})
\leq p\mathcal{F}(\rho_{1})+(1-p)\mathcal{F}(\rho_{2})$ for $p\in[0,1]$.
\item $\mathcal{F}$ is monotonic under completely positive and trace preserving map $\Phi$,
i.e., $\mathcal{F}(\Phi(\rho))\leq \mathcal{F}(\rho)$~\cite{Petz2008}.
\item If $\vec{y}$ is function of $\vec{x}$, then the QFIMs with respect to $\vec{y}$
and $\vec{x}$ satisfy $\mathcal{F}(\rho(\vec{x}))=J^{\mathrm{T}}\mathcal{F}(\rho(\vec{y}))J$,
with $J$ the Jacobian matrix, i.e., $J_{ij}=\partial y_i/\partial x_j$.
\end{itemize}
\end{prop}

\subsection{Parameterization processes}

\begin{figure}[tp]
\centering\includegraphics[width=12cm]{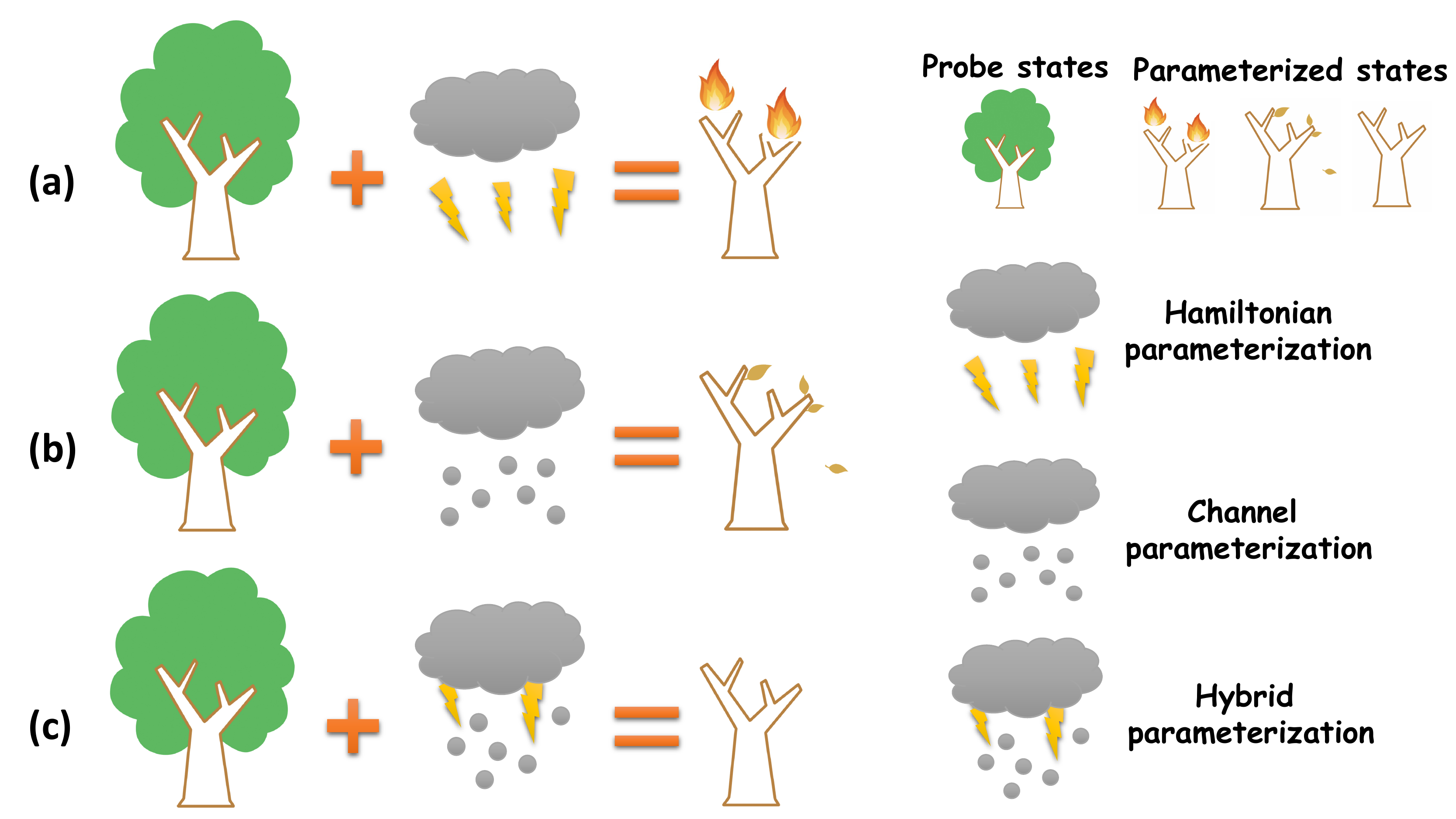}
\caption{The schematic of multiparameter parameterization processes.
(a) Hamiltonian parameterization (b) Channel parameterization (c) Hybrid parameterization.}
\label{fig:sec_intro_schematic}
\end{figure}

Generally, the parameters are encoded into the probe state via a parameter-dependent dynamics.
According to the types of dynamics, there exist three types of parameterization processes:
Hamiltonian parameterization, channel parameterization and hybrid parameterization,
as shown in figure~\ref{fig:sec_intro_schematic}. In the Hamiltonian parameterization,
$\vec{x}$ is encoded in the probe state $\rho_{0}$ through the Hamiltonian
$H_{\vec{x}}$. The dynamics is then governed by the Schr\"{o}dinger equation
\begin{equation}
\partial_{t}\rho =-i[H_{\vec{x}},\rho ],
\end{equation}
and the parameterized state can be written as
\begin{equation}
\rho =e^{-iH_{\vec{x}}t}\rho_{0}e^{iH_{\vec{x}}t}.
\end{equation}
Thus, the Hamiltonian parameterization is a unitary process. In some other scenarios
the parameters are encoded via the interaction with another system,
which means the probe system here has to be treated as an open system and the
dynamics is governed by the master equation. This is the channel parameterization.
The dynamics for the channel parameterization is
\begin{equation}
\partial_{t}\rho =-i[H,\rho ]+\mathcal{L}_{\vec{x}}(\rho ), \label{eq:channel_mastereq}
\end{equation}
where $\mathcal{L}_{\vec{x}}(\rho )$ represents the decay term dependent on $\vec{x}$.
A well-used form of $\mathcal{L}_{\vec{x}}$ is the Lindblad form
\begin{equation}
\mathcal{L} (\rho )=\sum_{j}\gamma_{j}\left(\Gamma_{j}\rho \Gamma^{\dagger}_{j}
-\frac{1}{2}\left\{\Gamma^{\dagger}_{j}\Gamma_{j},\rho \right\}\right),
\end{equation}
where $\Gamma_{j}$ is the $j$th Lindblad operator and $\gamma_{j}$ is the $j$th decay
rate. All the decay rates are unknown parameters to be estimated.
The third type is the hybrid parameterization, in which both the
Hamiltonian parameters and decay rates in equation~(\ref{eq:channel_mastereq})
are unknown and need to be estimated.

\subsection{Calculating QFIM}

In this section we review the techniques in the calculation of QFIM and some analytic
results for specific cases.

\subsubsection{General methods} \label{sec:general_method}

The traditional derivation of QFIM usually assumes the rank of the density matrix is full, i.e., all the
eigenvalues of the density matrix are positive. Specifically if we write $\rho =\sum^{\dim(\rho)-1}_{i=0}
\lambda_{i}|\lambda_{i}\rangle\langle\lambda_{i}|$, with $\lambda_{i}$ and $|\lambda_{i}\rangle$
the eigenvalue and the corresponding eigenstate, it is usually assumed that $\lambda_{i}>0$ for
all $0\leq i\leq \dim(\rho)-1$. Under this assumption the QFIM can be obtained as follows.
\begin{theorem} \label{theorem:traditional_form}
The entry of QFIM for a full-rank density matrix with the spectral decomposition
$\rho =\sum^{d-1}_{i=0}\lambda_{i}|\lambda_{i}\rangle\langle\lambda_{i}|$ can be written as
\begin{equation}
\mathcal{F}_{ab}=\sum^{d-1}_{i,j=0}\frac{2\mathrm{Re}(\langle\lambda_{i}
|\partial_{a}\rho |\lambda_{j}\rangle\langle\lambda_{j}|\partial_{b}
\rho |\lambda_{i}\rangle)}{\lambda_{i}+\lambda_{j}},
\end{equation}
where $d:=\dim(\rho)$ is the dimension of the density matrix.
\end{theorem}

One can easily see that if the density matrix is not of full rank, there can be divergent
terms in the above equation. To extend it to the general density matrices which may not have
full rank, we can manually remove the divergent terms as
\begin{equation}
\mathcal{F}_{ab}=\sum^{d-1}_{i,j=0,\lambda_{i}+\lambda_{j}\neq 0}
\frac{2\mathrm{Re}(\langle\lambda_{i}|\partial_{a}\rho |\lambda_{j}\rangle
\langle\lambda_{j}|\partial_{b}\rho |\lambda_{i}\rangle)}{\lambda_{i}+\lambda_{j}}.
\label{eq:tradition_QFIM}
\end{equation}
By substituting the spectral decomposition of $\rho$ into the equation above, it can
be rewritten as~\cite{Paris2009}
\begin{equation}
\mathcal{F}_{ab}=\!\!\!\sum^{d-1}_{i=0}\!\frac{(\partial_{a}\lambda_{i})
(\partial_{b}\lambda_{i})}{\lambda_{i}}+\!\!\!\sum_{i\neq j,\lambda_{i}+\lambda_{j}\neq 0}
\!\!\!\!\frac{2(\lambda_{i}-\lambda_{j})^{2}}{\lambda_{i}
+\!\lambda_{j}}\mathrm{Re}\!\left(\langle\lambda_{i}|\partial_{a}\lambda_{j}\rangle
\langle\partial_{b}\lambda_{j}|\lambda_{i}\rangle\right)\!.
\label{eq:QFIM_Paris}
\end{equation}
Recently, it has been rigorously proved that the QFIM for a finite dimensional density
matrix can be expressed with the support of the density matrix~\cite{LiuPhysicaA}.
The support of a density matrix, denoted by $\mathcal{S}$, is defined as
$\mathcal{S}:=\{\lambda_{i}\in\{\lambda_{i}\}|\lambda_{i}\neq 0\}$
($\{\lambda_{i}\}$ is the full set of $\rho$'s eigenvalues),
and the spectral decomposition can then be modified as
$\rho =\sum_{\lambda_{i}\in\mathcal{S}}\lambda_{i}|\lambda_{i}\rangle\langle \lambda_{i}|$.
The QFIM can then be calculated via the following theorem.
\begin{theorem}
Given the spectral decomposition of a density matrix, $\rho=\sum_{\lambda_{i}\in\mathcal{S}}
|\lambda_{i}\rangle\langle\lambda_{i}|$
where $\mathcal{S}=\{\lambda_{i}\in\{\lambda_{i}\}|\lambda_{i}\neq 0\}$ is the support,
an entry of QFIM can be calculated as~\cite{LiuPhysicaA}
\begin{eqnarray}
\mathcal{F}_{ab}&=&\sum_{\lambda_{i}\in
\mathcal{S}}\frac{(\partial_{a}\lambda_{i})(\partial_{b}\lambda_{i})}
{\lambda_{i}}+\sum_{\lambda_{i}\in\mathcal{S}}4\lambda_{i}\mathrm{Re}
\left(\langle\partial_{a}\lambda_{i}
|\partial_{b}\lambda_{i}\rangle\right) \nonumber \\
& & -\sum_{\lambda_{i},\lambda_{j}\in \mathcal{S}}\frac{8\lambda_{i}\lambda_{j}}
{\lambda_{i}+\lambda_{j}}\mathrm{Re}(\langle\partial_{a}\lambda_{i}|\lambda_{j}\rangle
\langle\lambda_{j}|\partial_{b}\lambda_{i}\rangle).
\label{eq:QFIM_nonfullrank}
\end{eqnarray}
\end{theorem}
The detailed derivation of this equation can be found in~\ref{apx:nonfull_QFIM}.
It is a general expression of QFIM for a finite-dimensional density matrix of arbitrary rank.
Due to the relation between the QFIM and QFI, one can easily obtain the following corollary.
\begin{corollary}
Given the spectral decomposition of a density matrix, $\rho=\sum_{\lambda_{i}\in\mathcal{S}}
|\lambda_{i}\rangle\langle\lambda_{i}|$, the QFI for the parameter $x_{a}$ can be
calculated as~\cite{Knysh2011,Liu2013,Zhang2013,Liu2014,Jing2014}
\begin{equation}
\mathcal{F}_{aa}=\sum_{\lambda_{i}\in \mathcal{S}}\frac{(\partial_{a}\lambda_{i})^{2}}
{\lambda_{i}}+\sum_{\lambda_{i}\in \mathcal{S}}4\lambda_{i}\langle\partial_{a}\lambda_{i}
|\partial_{a}\lambda_{i}\rangle-\!\!\sum_{\lambda_{i},\lambda_{j}
\in \mathcal{S}}\frac{8\lambda_{i}\lambda_{j}}{\lambda_{i}+\lambda_{j}}|\langle
\partial_{a}\lambda_{i}|\lambda_{j}\rangle|^{2}\!\!.
\label{eq:QFI_nonfullrank}
\end{equation}
\end{corollary}

The first term in equations~(\ref{eq:QFIM_Paris}) and (\ref{eq:QFIM_nonfullrank})
can be viewed as the counterpart of the classical Fisher information as it only contains
the derivatives of the eigenvalues which can be regarded as the counterpart of the probability
distribution. The other terms are purely quantum ~\cite{Paris2009,Liu2014}. The derivatives
of the eigenstates reflect the local structure of the eigenspace on $\vec{x}$.
The effect of this local structure on QFIM can be easily observed via
equations~(\ref{eq:QFIM_Paris}) and (\ref{eq:QFIM_nonfullrank}).

The SLD operator is important since it is not only related
to the calculation of QFIM, but also contains the information of the optimal
measurements and the attainability of the quantum Cram\'{e}r-Rao bound, which will be
further discussed in sections~\ref{sec:attainability} and~\ref{sec:opt_measurement}. In
terms of the eigen-space of $\rho $, the entries of the SLD operator can be obtained as
follows~\footnote{The derivation is in~\ref{apx:nonfull_QFIM}.}
\begin{equation}
\langle\lambda_{i}|L_{a}|\lambda_{j}\rangle=\delta_{ij}\frac{\partial_{a}\lambda_{i}}
{\lambda_{i}}+\frac{2(\lambda_{j}-\lambda_{i})}{\lambda_{i}+\lambda_{j}}\langle\lambda_{i}
|\partial_{a}\lambda_{j}\rangle;
\end{equation}
for $\lambda_{i}\in\mathcal{S}$ and $\lambda_{j}\not\in\mathcal{S}$,
$\langle\lambda_{i}|L_{a}|\lambda_{j}\rangle=-2\langle\lambda_{i}|\partial_{a}\lambda_{j}\rangle$;
and for $\lambda_{i},\lambda_{j}\not\in\mathcal{S}$, $\langle\lambda_{i}|L_{a}|\lambda_{j}\rangle$
can take arbitrary values. Fujiwara and Nagaoka~\cite{Fujiwara1995,Fujiwara1999}
first proved that this randomness does not affect the value of QFI and
all forms of SLD provide the same QFI. As a matter of fact, this conclusion
can be extended to the QFIM for any quantum state~\cite{LiuPhysicaA,Liu2016},
i.e., the entries that can take arbitrary values do not affect the value of QFIM.
Hence, if we focus on the calculation of QFIM we can just set them zeros. However,
this randomness plays a role in the search of optimal measurement, which will be
further discussed in section~\ref{sec:opt_measurement}.

In control theory, equation~(\ref{eq:SLD}) is also
referred to as the Lyapunov equation and the solution can be obtained as~\cite{Paris2009}
\begin{equation}
L_{a}=2\int^{\infty}_{0}e^{-\rho s}\left(\partial_{a}\rho \right)e^{\rho s}\mathrm{d}s,
\end{equation}
which is independent of the representation of $\rho $. This can
also be written in an expanded form~\cite{Liu2016}
\begin{equation}
L_{a}=-2\lim_{s\rightarrow\infty}\sum_{n=0}^{\infty}\frac{\left(-s\right)^{n+1}}{(n+1)!}
\mathcal{R}^{n}_{\rho}(\partial_{a}\rho),
\label{eq:SLD_expand}
\end{equation}
here  $\mathcal{R}_{\rho}(\cdot):=\{\rho,\cdot\}$ denotes the anti-commutator.
Using the fact that $\mathcal{R}^{n}_{\rho}(\partial_{a}\rho)
=\sum^{n}_{m=0}{n\choose m}\rho^{m}\left(\partial_{a}\rho \right)\rho^{n-m}$,
where ${n\choose m}=\frac{n!}{m!(n-m)!}$, equation~(\ref{eq:SLD_expand}) can be rewritten as
\begin{equation}
L_{a}=-2\lim_{s\rightarrow\infty}\sum_{n=0}^{\infty}\sum^{n}_{m=0}
\frac{\left(-s\right)^{n+1}}{(n+1)!}{n\choose m}\rho^{m}\left(\partial_{a}\rho \right)\rho^{n-m}.
\end{equation}
This form of SLD can be easy to calculate if $\rho^{m}(\partial_{a}\rho)\rho^{n-m}$ is only
non-zero for limited number of terms or has some recursive patterns.

Recently, Safr\'{a}nek~\cite{Safranek} provided another method to compute the QFIM utilizing
the density matrix in Liouville space. In Liouville space, the density matrix is a vector
containing all the entries of the density matrix in Hilbert space. Denote $\mathrm{vec}(A)$
as the \emph{column} vector of $A $ in Liouville space and $\mathrm{vec}(A)^{\dagger}$ as the
conjugate transpose of $\mathrm{vec}(A)$. The entry of $\mathrm{vec}(A)$
is $[\mathrm{vec}(A)]_{id+j}=A_{ij}$ ($i,j\in[0,d-1]$). The QFIM
can be calculated as follows.
\begin{theorem}
For a full-rank density matrix, the QFIM can be expressed by~\cite{Safranek}
\begin{equation}
\mathcal{F}_{ab}=2\mathrm{vec}(\partial_{a}\rho)^{\dagger}\left(\rho \otimes\normalfont{\openone}
+\normalfont{\openone}\otimes\rho^{*}\right)^{-1}\mathrm{vec}(\partial_{b}\rho),
\end{equation}
where $\rho^{*}$ is the conjugate of $\rho$, and the SLD operator in Liouville space,
denoted by $\mathrm{vec}(L_{a})$, reads
\begin{equation}
\mathrm{vec}(L_{a})=2\left(\rho \otimes\normalfont{\openone}
+\normalfont{\openone}\otimes\rho^{*}\right)^{-1}\mathrm{vec}(\partial_{a}\rho).
\end{equation}
\end{theorem}
This theorem can be proved by using the facts that $\mathrm{vec}(AB\openone)
=(A\otimes\openone)\mathrm{vec}(B)=(\openone\otimes B^{\mathrm{T}})\mathrm{vec}(A)$
($B^{\mathrm{T}}$ is the transpose of $B$)~\cite{Gilchrist2009,Wood2015,Havel2003} and
$\mathrm{Tr}(A^{\dagger}B)=\mathrm{vec}(A)^{\dagger}\mathrm{vec}(B)$.

Bloch representation is another well-used tool in quantum information theory.
For a $d$-dimensional density matrix, it can be expressed by
\begin{equation}
\rho=\frac{1}{d}\left(\openone+\sqrt{\frac{d(d-1)}{2}}\vec{r}\cdot\vec{\kappa}\right),
\end{equation}
where $\vec{r}=(r_1,r_2...,r_m,...)^{\mathrm{T}}$ is the Bloch vector
($|\vec{r}|^2\leq 1$) and $\vec{\kappa}$ is a $(d^{2}-1)$-dimensional vector
of $\frak{su}(d)$ generator satisfying $\mathrm{Tr}(\kappa_i)=0$. The anti-commutation
relation for them is $\{\kappa_i,\kappa_j\}=\frac{4}{d}\delta_{ij}\openone
+\sum^{d^2-1}_{m=1}\mu_{ijm}\kappa_m$, and the commutation relation is
$\left[\kappa_i,\kappa_j\right]= i\sum^{d^2-1}_{m=1}\epsilon_{ijm}\kappa_m$,
where $\mu_{ijm}$ and $\epsilon_{ijm}$ are the symmetric and antisymmetric
structure constants.
Watanabe et al. recently~\cite{Watanabe2010,Watanabe2011,Watanabe2014} provided
the formula of QFIM for a general Bloch vector by considering the Bloch vector
itself as the parameters to be estimated. Here we extend their result to a
general case as the theorem below.
\begin{theorem}
In the Bloch representation of a $d$-dimensional density matrix, the QFIM can
be expressed by
\begin{equation}
\mathcal{F}_{ab}=(\partial_{b} \vec{r})^{\mathrm{T}}
\left(\frac{d}{2(d-1)}G-\vec{r}\,\vec{r}^{\,\mathrm{T}}\right)^{-1}\partial_{a}\vec{r},
\end{equation}
where $G$ is a real symmetric matrix with the entry
\begin{equation}
G_{ij}=\frac{1}{2}\mathrm{Tr}(\rho\{\kappa_i,\kappa_j\})
=\frac{2}{d}\delta_{ij}+\sqrt{\frac{d-1}{2d}}\sum_{m}\mu_{ijm}r_m.
\end{equation}
\end{theorem}
The most well-used scenario of this theorem is single-qubit systems, in
which $\rho=(\openone+\vec{r}\cdot\vec{\sigma})/2$ with
$\vec{\sigma}=(\sigma_{x},\sigma_{y},\sigma_{z})$ the vector of Pauli matrices.
For a single-qubit system, we have the following corollary.
\begin{corollary}
For a single-qubit mixed state, the QFIM in Bloch representation can be expressed by
\begin{equation}
\mathcal{F}_{ab}= (\partial_{a}\vec{r})\cdot(\partial_{b}\vec{r})
+\frac{(\vec{r}\cdot\partial_{a}\vec{r})(\vec{r}\cdot\partial_{b}\vec{r})}
{1-|\vec{r}|^{2}},
\label{eq:bloch_singlequbit}
\end{equation}
where $|\vec{r}|$ is the norm of $\vec{r}$. For a single-qubit pure state,
$\mathcal{F}_{ab}=(\partial_{a}\vec{r})\cdot(\partial_{b}\vec{r})$.
\end{corollary}
The diagonal entry of equation~(\ref{eq:bloch_singlequbit}) is exactly the one
given by reference~\cite{Zhong2013}. The proofs of the theorem and corollary
are provided in~\ref{apx:QFIM_Bloch}.

\subsubsection{Pure states} \label{sec:pure_state}
A pure state satisfies $\rho=\rho^{2}$, i.e., the purity $\mathrm{Tr}(\rho^{2})$ equals 1.
For a pure state $|\psi\rangle$, the dimension of the support is 1, which means only one eigenvalue
is non-zero (it has to be 1 since $\mathrm{Tr\rho=1}$), with which the corresponding eigenstate
is $|\psi\rangle$. For pure states, the QFIM can be obtained as follows.
\begin{theorem}
The entries of the QFIM for a pure parameterized state $|\psi\rangle:=|\psi(\vec{x})\rangle$
can be obtained as~\cite{Helstrom,Holevo}
\begin{equation}
\mathcal{F}_{ab}=4\mathrm{Re}(\langle\partial_{a}\psi|\partial_{b}\psi\rangle-
\langle\partial_{a}\psi|\psi\rangle\langle\psi|\partial_{b}\psi\rangle).
\label{eq:purestate_general}
\end{equation}
The QFI for the parameter $x_{a}$ is just the diagonal element of the QFIM, which is given by
\begin{equation}
\mathcal{F}_{aa}=4(\langle\partial_{a}\psi|\partial_{a}\psi\rangle
-|\langle\partial_{a}\psi|\psi\rangle|^{2}),
\end{equation}
and the SLD operator corresponding to $x_a$ is $L_{a}=2\left(|\psi\rangle\langle\partial_{a}\psi|
+|\partial_{a}\psi\rangle\langle\psi|\right)$.
\end{theorem}
The SLD formula is obtained from the fact $\rho^{2}=\rho$ for a pure state, then
$\partial_a \rho=\rho \partial_a \rho+(\partial_a \rho) \rho$. Compared this equation
to the definition equation, it can be seen that $L=2\partial_a \rho$.
A simple example is $|\psi\rangle=e^{-i\sum_{j}H_{j}x_{j}t}|\psi_{0}\rangle$ with
$[H_{a},H_{b}]=0$ for any $a$ and $b$, here $|\psi_{0}\rangle$ denotes the initial
probe state. In this case, the QFIM reads
\begin{equation}
\mathcal{F}_{ab}=4t^{2}\mathrm{cov}_{|\psi_{0}\rangle}(H_{a},H_{b}),
\end{equation}
where $\mathrm{cov}_{|\varphi\rangle}(A,B)$ denotes the covariance between $A$ and $B$ on
$|\varphi\rangle$, i.e.,
\begin{equation}
\mathrm{cov}_{|\varphi\rangle}(A,B):=\frac{1}{2}\langle\varphi|\{A,B\}|\varphi\rangle
-\langle\varphi|A|\varphi\rangle\langle\varphi|B|\varphi\rangle.
\label{eq:cov_definition}
\end{equation}
A more general case where $H_{a}$ and $H_{b}$ do not commute will be discussed
in section~\ref{sec:unitary_process}.

\subsubsection{Few-qubit states}

The simplest few-qubit system is the single-qubit system. A single-qubit pure state can always
be written as $\cos\theta|0\rangle+\sin\theta e^{i\phi}|1\rangle$ ($\{|0\rangle,|1\rangle\}$
is the basis), i.e., it only has two degrees of freedom, which means only two independent parameters
($\vec{x}=(x_{0},x_{1})^{\mathrm{T}}$) can be encoded in a single-qubit pure state.
Assume $\theta$, $\phi$ are the parameters to be estimated, the QFIM can then be obtained
via equation~(\ref{eq:purestate_general}) as
\begin{equation}
\mathcal{F}_{\theta\theta}=4,~\mathcal{F}_{\phi\phi}=\sin^{2}(2\theta),
~\mathcal{F}_{\theta\phi}=0.
\end{equation}
If the unknown parameters are not $\theta,\phi$, but functions of $\theta,\phi$, the QFIM
can be obtained from formula above with the assistance of Jacobian matrix.

For a single-qubit mixed state, when the number of encoded parameters is larger
than three, the determinant of QFIM would be zero, indicating that these parameters
cannot be simultaneously estimated. This is due to the fact that there only exist
three degrees of freedom in a single-qubit mixed state, thus, only three or fewer
independent parameters can be encoded into the density matrix $\rho$. However,
more parameters may be encoded if they are not independent. Since $\rho$ here
only has two eigenvalues $\lambda_{0}$ and $\lambda_{1}$, equation~(\ref{eq:QFIM_nonfullrank})
then reduces to
\begin{equation}
\mathcal{F}_{ab}=\frac{(\partial_{a}\lambda_{0})(\partial_{b}\lambda_{0})}
{\lambda_{0}(1-\lambda_{0})}+4(1-2\lambda_{0})^{2}\mathrm{Re}
\left(\langle\partial_{a}\lambda_{0}|\lambda_{1}\rangle\langle
\lambda_{1}|\partial_{b}\lambda_{0}\rangle\right).
\label{eq:qubit_mixed}
\end{equation}

In the case of single qubit, equation~(\ref{eq:qubit_mixed}) can also be written
in a basis-independent formula~\cite{Dittmann1999} below.
\begin{theorem} \label{theorem:one_qubit_QFIM}
The basis-independent expression of QFIM for a single-qubit mixed state $\rho $
is of the following form
\begin{equation}
\mathcal{F}_{ab}=\mathrm{Tr}\left[(\partial_{a}\rho )(\partial_{b}\rho )\right]
+\frac{1}{\det(\rho)}\mathrm{Tr}\left[\rho(\partial_{a}\rho)\rho(\partial_{b}\rho)\right],
\label{eq:qubit_QFIM}
\end{equation}
where $\det(\rho)$ is the determinant of $\rho$. For a single-qubit pure state,
$\mathcal{F}_{ab}=2\mathrm{Tr}[(\partial_a\rho)(\partial_b\rho)]$.
\end{theorem}

Equation~(\ref{eq:qubit_QFIM}) is the reduced form of the one given in
reference~\cite{Dittmann1999}. The advantage of the basis-independent
formula is that the diagonalization of the density matrix is avoided.
Now we show an example for single-qubit. Consider a spin in a magnetic
field which is in the $z$-axis and suffers from dephasing noise also in
the $z$-axis. The dynamics of this spin can then be expressed by
\begin{equation}
\partial_{t}\rho=-i[B\sigma_{z},\rho]+\frac{\gamma}{2}(\sigma_{z}\rho\sigma_{z}-\rho),
\end{equation}
where $\sigma_{z}$ is a Pauli matrix. $B$ is the amplitude of the field. Take $B$ and
$\gamma$ as the parameters to be estimated. The analytical solution for $\rho(t)$ is
\begin{equation}
\rho(t)=\left(\begin{array}{cc}
\rho_{00}(0) & \rho_{01}(0)e^{-i2Bt-\gamma t}\\
\rho_{10}(0)e^{i2Bt-\gamma t} & \rho_{11}(0)
\end{array}\right).
\label{eq:qubit_dephasing}
\end{equation}
The derivatives of $\rho(t)$ on both $B$ and $\gamma$ are simple in this basis. Therefore,
the QFIM can be directly calculated from equation~(\ref{eq:qubit_QFIM}), which is a
diagonal matrix ($\mathcal{F}_{B\gamma}=0$) with the diagonal entries
\begin{eqnarray}
\mathcal{F}_{BB} &=& 16 |\rho_{01}(0)|^{2} e^{-2\gamma t}t^{2}, \\
\mathcal{F}_{\gamma\gamma} &=& \frac{4\rho_{00}(0)\rho_{11}(0)|\rho_{01}(0)|^{2}t^{2}}
{\rho_{00}(0)\rho_{11}(0)e^{2\gamma t}-|\rho_{01}(0)|^{2}}.
\end{eqnarray}

For a general two-qubit state, the calculation of QFIM requires the diagonalization
of a 4 by 4 density matrix, which is difficult to solve analytically. However, some
special two-qubit states, such as the X state, can be diagonalized analytically.
An X state has the form (in the computational basis
$\{|00\rangle, |01\rangle,|10\rangle, |11\rangle\}$) of
\begin{equation}
\rho =\left(\begin{array}{cccc}
\rho_{00} & 0 & 0 & \rho_{03}\\
0 & \rho_{11} & \rho_{12} & 0\\
0 & \rho_{21} & \rho_{22} & 0\\
\rho_{30} & 0 & 0 & \rho_{33}
\end{array}\right).
\end{equation}
By changing the basis into $\{|00\rangle, |11\rangle, |01\rangle,|10\rangle\}$, this state
can be rewritten in the block diagonal form as $\rho =\rho^{(0)}\oplus\rho^{(1)}$,
where $\oplus$ represents the direct sum and
\begin{equation}
\rho^{(0)}=\left(\begin{array}{cc}
\rho_{00} & \rho_{03} \\
\rho_{30} & \rho_{33} \\
\end{array}\right),\quad
\rho^{(1)}=\left(\begin{array}{cc}
\rho_{11} & \rho_{12} \\
\rho_{21} & \rho_{22} \\
\end{array}\right).
\end{equation}
Note that $\rho^{(0)}$ and $\rho^{(1)}$ are not  density matrices as their trace
is not  normalized. The QFIM for this block diagonal state can be written as
$\mathcal{F}_{ab}=\mathcal{F}^{(0)}_{ab}+\mathcal{F}^{(1)}_{ab}$~\cite{Liu2014},
where $\mathcal{F}^{(0)}_{ab}$ ($\mathcal{F}^{(1)}_{ab}$) is the QFIM for $\rho^{(0)}$ ($\rho^{(1)}$).
The eigenvalues of $\rho^{(i)}$ are $\lambda^{(i)}_{\pm}=\frac{1}{2}
\Big(\mathrm{Tr}\rho^{(i)}\pm\sqrt{\mathrm{Tr}^{2}\rho^{(i)}
-4\det\rho^{(i)}}\Big)$ and corresponding eigenstates are
\begin{equation}
|\lambda^{(i)}_{\pm}\rangle= \mathcal{N}^{(i)}_{\pm}\left(\frac{1}{2\mathrm{Tr}(\rho^{(i)}\sigma_{+})}
\left[\mathrm{Tr}\left(\rho^{(i)}\sigma_{z}\right)\pm
\sqrt{\mathrm{Tr}^{2}\rho^{(i)}-4\det\rho^{(i)}}\right]\!\!, 1\right)^{\mathrm{T}},
\end{equation}
for non-diagonal $\rho^{(i)}$ with $\mathcal{N}^{(i)}_{\pm}$ ($i=0,1$) the normalization coefficient.
Here the specific form of $\sigma_{z}$ and $\sigma_{+}$ are
\begin{equation}
\sigma_{z}=\left(\begin{array}{cc}
1 & 0\\
0 & -1 \\
\end{array}\right),\quad \sigma_{+}=\left(\begin{array}{cc}
0 & 1\\
0 & 0 \\
\end{array}\right).
\end{equation}
Based on above information, $\mathcal{F}^{(i)}_{ab}$ can be specifically written as
\begin{eqnarray}
\mathcal{F}^{(i)}_{ab} &=& \sum_{k=\pm}\frac{(\partial_{a}\lambda^{(i)}_{k})(\partial_{b}
\lambda^{(i)}_{k})}{\lambda^{(i)}_{k}}+\lambda^{(i)}_{k}\mathcal{F}_{ab}
\left(|\lambda^{(i)}_{k}\rangle\right) \nonumber \\
& & -\frac{16\det\rho^{(i)}}{\mathrm{Tr}\rho^{(i)}}\mathrm{Re}(\langle\partial_a \lambda^{(i)}_{+}|
\lambda^{(i)}_{-}\rangle\langle\lambda^{(i)}_{-}|\partial_b \lambda^{(i)}_{+}\rangle),
\end{eqnarray}
where $\mathcal{F}_{ab}(|\lambda^{(i)}_{k}\rangle)$ is the QFIM entry for the state
$|\lambda^{(i)}_{k}\rangle$. For diagonal $\rho^{(i)}$, $|\lambda^{(i)}_{\pm}\rangle$
is just $(0,1)^{\mathrm{T}}$ and only the classical contribution term
remains in above equation.

\subsubsection{Unitary processes} \label{sec:unitary_process}

Unitary processes are the most fundamental dynamics in quantum mechanics since it can
be naturally obtained via the Schr\"{o}dinger equation. For a $\vec{x}$-dependent
unitary process $U=U(\vec{x})$, the parameterized state $\rho $ can be written
as $\rho =U\rho_{0}U^{\dagger}$, where $\rho_{0}$ is the initial probe state
which is $\vec{x}$-independent. For such a process, the QFIM can be calculated
via the following theorem.
\begin{theorem}
For a unitary parametrization process $U$, the entry of QFIM can be
obtained as~\cite{LiuSR}
\begin{eqnarray}
\mathcal{F}_{ab} &=& \sum_{\eta_{i}\in\mathcal{S}}4\eta_{i}
\mathrm{cov}_{|\eta_{i}\rangle}(\mathcal{H}_{a},\mathcal{H}_{b}) \nonumber \\
& & -\sum_{\eta_{i},\eta_{j}\in\mathcal{S},
i\neq j}\frac{8\eta_{i}\eta_{j}}{\eta_{i}
+\eta_{j}}\mathrm{Re}\left(\langle\eta_{i}|\mathcal{H}_{a}
|\eta_{j}\rangle\langle\eta_{j}|\mathcal{H}_{b}|\eta_{i}
\rangle\right),
\label{eq:QFIM_unitary}
\end{eqnarray}
where $\eta_{i}$ and $|\eta_{i}\rangle$ are $i$th eigenvalue and
eigenstate of the initial probe state $\rho_{0}$. $\mathrm{cov}_{|\eta_{i}\rangle}
(\mathcal{H}_{a},\mathcal{H}_{b})$ is defined in equation~(\ref{eq:cov_definition}).
The operator $\mathcal{H}_{a}$ is defined as~\cite{Taddei2013,Boixo2007}
\begin{equation}
\mathcal{H}_{a}:=i\left(\partial_{a}U^{\dagger}\right)U
=-iU^{\dagger}\left(\partial_{a}U\right).
\label{eq:unitary_H}
\end{equation}
\end{theorem}
$\mathcal{H}_{a}$ is a Hermitian operator for any parameter $x_{a}$
due to above definition.

For the unitary processes, the parameterized state will remain pure for a pure probe state.
The QFIM for this case is given as follows.
\begin{corollary} \label{corollary_unitary}
For a unitary process $U$ with a pure probe state $|\psi_{0}\rangle$,
the entry of QFIM is in the form
\begin{equation}
\mathcal{F}_{ab}=4\mathrm{cov}_{|\psi_{0}\rangle}(\mathcal{H}_{a},\mathcal{H}_{b}),
\end{equation}
where $\mathrm{cov}_{|\psi_{0}\rangle}(\mathcal{H}_{a},\mathcal{H}_{b})$ is defined
by equation~(\ref{eq:cov_definition}) and the QFI for $x_{a}$ can then be obtained as
$\mathcal{F}_{aa}=4\mathrm{var}_{|\psi_{0}\rangle}(\mathcal{H}_{a})$. Here
$\mathrm{var}_{|\psi_{0}\rangle}(\mathcal{H}_{a}):=\mathrm{cov}_{|\psi_{0}\rangle}
(\mathcal{H}_{a},\mathcal{H}_{a})$ is the variance of $\mathcal{H}_{a}$
on $|\psi_{0}\rangle$.
\end{corollary}
For a single-qubit mixed state $\rho_{0}$ under a unitary process, the QFIM can be written as
\begin{equation}
\mathcal{F}_{ab}=4\left[2\mathrm{Tr}(\rho^{2}_{0})-1\right]
\mathrm{cov}_{|\eta_{0}\rangle}(\mathcal{H}_{a},\mathcal{H}_{b})
\end{equation}
with $|\eta_{0}\rangle$ an eigenstate of $\rho_{0}$. This equation is equivalent to
\begin{equation}
\mathcal{F}_{ab}=4\left[2\mathrm{Tr}(\rho^{2}_{0})-1\right]\mathrm{Re}
\left(\langle \eta_{0}|\mathcal{H}_{a}|\eta_{1}\rangle
\langle\eta_{1}|\mathcal{H}_{b}|\eta_{0}\rangle\right).
\end{equation}
The diagonal entry reads $\mathcal{F}_{aa}=4\left[2\mathrm{Tr}
(\rho^{2}_{0})-1\right]|\langle\eta_{0}|\mathcal{H}_{a}|\eta_{1}\rangle|^{2}.$
Recall that Theorem~\ref{theorem:one_qubit_QFIM} provides the basis-independent formula
for single-qubit mixed state, which leads to the next corollary.
\begin{corollary}
For a single-qubit mixed state $\rho_{0}$ under a unitary process, the basis-independent
formula of QFIM is
\begin{eqnarray}
\mathcal{F}_{ab}&=&\mathrm{Tr}(\rho^{2}_{0}\{\mathcal{H}_{a},\mathcal{H}_{b}\})
-2\mathrm{Tr}(\rho_{0}\mathcal{H}_{a}\rho_{0}\mathcal{H}_{b}) \nonumber \\
& & +\frac{1}{\det\rho_{0}}\left[\mathrm{Tr}\!\left(\rho_{0}\mathcal{H}_{a}\rho_{0}
\left\{\rho_{0}^{2},\mathcal{H}_{b}\right\}\right)-2\mathrm{Tr}
(\rho^{2}_{0}\mathcal{H}_{a}\rho^{2}_{0}\mathcal{H}_{b})\right].
\end{eqnarray}
The diagonal entry reads
\begin{eqnarray}
\mathcal{F}_{aa}&=& 2\mathrm{Tr}(\rho^{2}_{0}\mathcal{H}^{2}_{a})
-2\mathrm{Tr}[(\rho_{0}\mathcal{H}_{a})^{2}] \nonumber \\
& & +\frac{2}{\det\rho_{0}}\left[\mathrm{Tr}\!\left(\rho_{0}\mathcal{H}_{a}\rho^{3}_{0}
\mathcal{H}_{a}\right)-\mathrm{Tr}[(\rho^{2}_{0}\mathcal{H}_{a})^{2}]\right].
\end{eqnarray}
\end{corollary}

Under the unitary process, the QFIM for pure probe states, as given in equation~(\ref{eq:def_QFIM}),
can be rewritten as
\begin{equation}
\mathcal{F}_{ab}=\frac{1}{2}\langle\psi_{0}|\left\{ L_{a,\mathrm{eff}}, L_{b,\mathrm{eff}}
\right\}|\psi_{0}\rangle,
\end{equation}
where $L_{a,\mathrm{eff}}:=U^{\dagger} L_{a}U$ can be treated as an
effective SLD operator, which leads to the following theorem.
\begin{theorem}
Given a unitary process, $U$, with a pure probe state, $|\psi_{0}\rangle$, the effective
SLD operator $L_{a,\mathrm{eff}}$ can be obtained as
\begin{equation}
L_{a,\mathrm{eff}}=i2\left[\mathcal{H}_{a},|\psi_{0}\rangle\langle\psi_{0}|\right].
\end{equation}
\end{theorem}

From equation~(\ref{eq:QFIM_unitary}), all the information of the parameters is involved in
the operator set $\{\mathcal{H}_{a}\}$, which might benefit the analytical optimization
of the probe state in some scenarios.  Generally, the unitary operator can be written
as $\exp(-itH)$ where $H=H(\vec{x})$ is the Hamiltonian for the parametrization.
$\mathcal{H}_{a}$ can then be calculated as
\begin{equation}
\mathcal{H}_{a}=-\int^{t}_{0}e^{isH}\left(\partial_{a}H\right)
e^{-isH} \mathrm{d}s,
\end{equation}
where the technique $\partial_{x}e^{A}=\int^{1}_{0}e^{sA}\partial_{x}Ae^{(1-s)A}ds$
($A$ is an operator) is applied. Denote $H^{\times}(\cdot):=[H,\cdot]$, the
expression above can be rewritten in an expanded form~\cite{LiuSR}
\begin{equation}
\mathcal{H}_{a}=-\sum^{\infty}_{n=0}\frac{t^{n+1}}{(n+1)!}\left(i H^{\times}\right)^{n}
\partial_{a}H. \label{eq:unitary_Hx}
\end{equation}
In some scenarios, the recursive commutations in expression above display certain patterns,
which can lead to analytic expressions for the $\mathcal{H}$ operator. The simplest example
is $H=\sum_{a}x_{a}H_{a}$, with all $H_{a}$ commute with each other. In this case
$\mathcal{H}_{a}=-t H_{a}$ since only the zeroth order term in equation~(\ref{eq:unitary_Hx})
is nonzero. Another example is the interaction of a collective spin system with a
magnetic field with the Hamiltonian $H=-B J_{\vec{n}_{0}}$, where $B$ is the amplitude
of the external magnetic field, $J_{\vec{n}_{0}}=\vec{n}_{0}\cdot\vec{J}$ with
$\vec{n}_{0}=(\cos\theta,0,\sin\theta)$ and $\vec{J}=(J_{x},J_{y},J_{z})$. $\theta$
is the angle between the field and the collective spin. $J_{i}=\sum_{k}\sigma^{(k)}_{i}/2$
for $i=x,y,z$ is the collective spin operator. $\sigma^{(k)}_{i}$ is the Pauli matrix
for $k$th spin. In this case, the $\mathcal{H}$ operator for $\theta$ can be
analytically calculated via equation (\ref{eq:unitary_Hx}), which is~\cite{LiuSR}
\begin{equation}
\mathcal{H}_{\theta}=-2\sin\left(\frac{1}{2}Bt\right)J_{\vec{n}_{1}},
\end{equation}
where $J_{n_{1}}=\vec{n}_{1}\cdot\vec{J}$ with
$\vec{n}_{1}=(\cos(Bt/2)\sin\theta,\sin(Bt/2),-\cos(Bt/2)\cos\theta)$.

Recently, Sidhu and Kok~\cite{Sidhu2017,Sidhu2018} use this $\mathcal{H}$-representation
to study the spatial deformations, epecially the grid deformations of classical and
quantum light emitters. By calculating and analyzing the QFIM, they showed that the
higher average mode occupancies of the classical states performs better in estimating
the deformation when compared with single photon emitters.

An alternative operator that can be used to characterize the precision limit of
unitary process is~\cite{Boixo2007,Pang2014,Pang2016}
\begin{equation}
\mathcal{K}_{a}:=i\left(\partial_{a}U\right)U^{\dagger}
=-iU\left(\partial_{a}U^{\dagger}\right).
\end{equation}
As a matter of fact, this operator is the infinitesimal generator of $U$ of parameter $x_{a}$.
Assume $\vec{x}$ is shifted by $d x_{a}$ along the direction of $x_{a}$ and other
parameters are kept unchanged. Then $U(\vec{x}+d x_{a})$ can be expanded as
$U(\vec{x}+ d x_{a})\partial_{a}U(\vec{x})$. The density matrix
$\rho_{\vec{x}+d x_{a}}$ can then be approximately calculated as $\rho_{\vec{x}+d
x_{a}}=e^{-i\mathcal{K}_{a}dx_{a}}\rho e^{i\mathcal{K}_{a}dx_{a}}$~\cite{Pang2014},
which indicates that $\mathcal{K}_{a}$ is the generator of $U$ along parameter $x_{a}$.
The relation between $\mathcal{H}_{a}$ and $\mathcal{K}_{a}$ can be easily
obtained as
\begin{equation}
\mathcal{K}_{a}=-U\mathcal{H}_{a}U^{\dagger}.
\end{equation}
With this relation, the QFIM can be easily rewritten with $\mathcal{K}_{a}$ as
\begin{eqnarray}
\mathcal{F}_{ab}&=& \sum_{\lambda_{i}\in\mathcal{S}}4\lambda_{i}
\mathrm{cov}_{|\lambda_{i}\rangle}(\mathcal{K}_{a},\mathcal{K}_{b}) \nonumber \\
& & -\sum_{\lambda_{i},\lambda_{j}\in\mathcal{S}, i\neq j}\frac{8\lambda_{i}\lambda_{j}}{\lambda_{i}
+\lambda_{j}}\mathrm{Re}\left(\langle\lambda_{i}|\mathcal{K}_{a}|\lambda_{j}\rangle
\langle\lambda_{j}|\mathcal{K}_{b}|\lambda_{i}\rangle\right),
\end{eqnarray}
where $|\lambda_{i}\rangle=U|\eta_{i}\rangle$ is the $i$th eigenstate
of the parameterized state $\rho$. And $\mathrm{cov}_{|\lambda_{i}\rangle}
(\mathcal{K}_{a},\mathcal{K}_{b})$ is defined by equation~(\ref{eq:cov_definition}).
The difference between the calculation of QFIM with $\{\mathcal{K}_{a}\}$
and $\{\mathcal{H}_{a}\}$ is that the expectation is taken with the eigenstate
of the probe state $\rho_{0}$ for the use of $\{\mathcal{H}_{a}\}$ but with
the parameterized state $\rho$ for  $\{\mathcal{K}_{a}\}$.
For a pure probe state $|\psi_{0}\rangle$, the expression above reduces to
$\mathcal{F}_{ab}=4\mathrm{cov}_{|\psi\rangle}(\mathcal{K}_{a},\mathcal{K}_{b})$
with $|\psi\rangle=U|\psi_{0}\rangle$. Similarly, for a mixed state of single qubit,
the QFIM reads $\mathcal{F}_{ab}=4\left(2\mathrm{Tr}\rho^{2}_{0}-1\right)
\mathrm{cov}_{|\lambda_{0}\rangle}(\mathcal{K}_{a},\mathcal{K}_{b})$.

\subsubsection{Gaussian states} \label{subsec:gaussian_state}

Gaussian state is a widely-used quantum state in quantum physics, particularly in quantum optics,
quantum metrology and continuous variable quantum information processes. Consider a $m$-mode bosonic
system with $a_{i}$ ($a^{\dagger}_{i}$) as the annihilation (creation) operator
for the $i$th mode. The quadrature operators are~\cite{Scullybook,Braunstein2005}
$\hat{q}_{i}:=\frac{1}{\sqrt{2}}(a_{i}+a^{\dagger}_{i})$ and $\hat{p}_{i}:=\frac{1}
{i\sqrt{2}}(a_{i}-a^{\dagger}_{i})$, which satisfy the commutation relation
$\left[\hat{q}_{i}, \hat{p}_{j}\right]=i\delta_{ij}$ ($\hbar=1$).
A vector of quadrature operators, $\vec{R}=(\hat{q}_{1},\hat{p}_{1},...,
\hat{q}_{m},\hat{p}_{m})^{\mathrm{T}}$ satisfies
\begin{equation}
\left[R_{i}, R_{j}\right]=i\Omega_{ij}
\end{equation}
for any $i$ and $j$ where $\Omega$ is the symplectic matrix defined as $\Omega:=i\sigma^{\oplus m}_{y}$
with $\oplus$ denote the direct sum. Now we introduce the
covariance matrix $C(\vec{R})$ with the entries defined as
$C_{ij}:=\mathrm{cov}_{\rho }(R_{i},R_{j})=\frac{1}{2}\mathrm{Tr}(\rho \{R_{i},R_{j}\})
-\mathrm{Tr}(\rho R_{i})\mathrm{Tr}(\rho R_{j})$.
$C$ satisfies the uncertainty relation $C+\frac{i}{2}\Omega \geq 0$~\cite{Simon1994,Weedbrook2012}.
According to the Williamson’s theorem, the covariance matrix can be diagonalized utilizing
a symplectic matrix $S$~\cite{Weedbrook2012,Williamson1936}, i.e.,
\begin{equation}
C=SC_{\mathrm{d}}S^{\mathrm{T}},
\end{equation}
where $C_{\mathrm{d}}=\bigoplus^{m}_{k=1}c_{k}\openone_{2}$ with $c_{k}$ the $k$th symplectic eigenvalue.
$S$ is a $2m$-dimensional real matrix which satisfies $S\Omega S^{\mathrm{T}}=\Omega$.

A very useful quantity for Gaussian states is the characteristic function
\begin{equation}
\chi(\vec{s})=\mathrm{Tr}\!\left(\rho ~e^{i\vec{R}^{\,\mathrm{T}}\Omega \vec{s}}\right),
\end{equation}
where $\vec{s}$ is a $2m$-dimensional real vector. Another powerful function is the Wigner function,
which can be obtained by taking the Fourier transform of the characteristic function
\begin{equation}
W(\vec{R})=\frac{1}{(2\pi)^{2m}}\int_{\mathbb{R}^{2m}}\!e^{-i \vec{R}^{\,\mathrm{T}}
\Omega \vec{s}}~\chi(\vec{s})~\mathrm{d}^{2m}\vec{s}.
\end{equation}
Considering the scenario with first and second moments, a state is a Gaussian state if
$\chi(\vec{s})$ and $W(\vec{R})$ are Gaussian, i.e.,
~\cite{Weedbrook2012,Scullybook,Braunstein2005,Adesso2014,Wang2007}
\begin{eqnarray}
\chi(\vec{s}) &=& e^{-\frac{1}{2}\vec{s}^{\,\mathrm{T}}\Omega C\Omega^{\mathrm{T}}\vec{s}
-i(\Omega\langle\vec{R}\rangle)^{\mathrm{T}}\vec{s}},  \\
W(\vec{R}) &=& \frac{1}{(2\pi)^{m}\sqrt{\det C}}e^{-\frac{1}{2}\left(\vec{R}-\langle
\vec{R}\rangle\right)^{\!\!\mathrm{T}}C^{-1}\left(\vec{R}-\langle \vec{R}\rangle\right)},
\end{eqnarray}
where $\langle\vec{R}\rangle_{j}=\mathrm{Tr}(R_{j}\rho )$ is the first moment. A pure state
is Gaussian if and only if its Wigner function is non-negative~\cite{Weedbrook2012}.

The study of QFIM for Gaussian states started from the research of QFI. The expression of
QFI was first given in 2013 by Monras for the multi-mode case\cite{Monras2013} and Pinel
et al. for the single-mode case~\cite{Pinel2013}. In 2018, Nichols et al.~\cite{Nichols2018}
and \v{S}afr\'{a}nek~\cite{Safranek2018} provided the expression of QFIM for multi-mode
Gaussian states independently, which was obtained based on the calculation of SLD~\cite{Braunrmp}.
The SLD operator for Gaussian states has been given in references~\cite{Monras2013,Nichols2018,Serafini2017},
and we organize the corresponding results in the following theorem.
\begin{theorem} \label{SLD_Gaussian}
For a continuous variable bosonic $m$-mode Gaussian state with the displacement vector (first  moment)
$\langle \vec{R}\rangle$ and the covariance matrix (second moment) $C$, the SLD
operator is~\cite{Monras2013,Nichols2018,Serafini2017,Safranek2018,Gao2014,Safranek2015}
\begin{equation}
L_{a}=L^{(0)}_{a}\openone_{2m}+\vec{L}^{(1),\mathrm{T}}_{a}\vec{R}
+\vec{R}^{\,\mathrm{T}}G_{a}\vec{R}, \label{eq:SLD_Gaussian}
\end{equation}
where $\openone_{2m}$ is the $2m$-dimensional identity matrix and the coefficients read
\begin{eqnarray}
G_{a} &=& \sum^{m}_{j,k=1}\sum^{3}_{l=0}\frac{g^{(jk)}_{l}}
{4 c_{j}c_{k}+(-1)^{l+1}}\left(S^{\mathrm{T}}\right)^{-1}A^{(jk)}_{l}S^{-1}, \\
\vec{L}^{(1)}_{a} &=& C^{-1}(\partial_{a}{\langle\vec{R}\rangle})-2G_{a}
\langle\vec{R}\rangle, \\
L^{(0)}_{a} &=& \langle\vec{R}\rangle^{\mathrm{T}}G_{a}\langle\vec{R}\rangle
-(\partial_{a}\langle\vec{R}\rangle)^{\mathrm{T}}C^{-1}\langle\vec{R}\rangle
-\mathrm{Tr}(G_{a}C).
\end{eqnarray}
Here
\begin{equation}
A^{(jk)}_{l}=\frac{1}{\sqrt{2}} i\sigma^{(jk)}_{y},~\frac{1}{\sqrt{2}}
\sigma^{(jk)}_{z},~\frac{1}{\sqrt{2}}\openone^{(jk)}_{2},~\frac{1}{\sqrt{2}} \sigma^{(jk)}_{x}
\end{equation}
for $l=0,1,2,3$ and $g^{(jk)}_{l}=\mathrm{Tr}[S^{-1}(\partial_{a}C)(S^{\mathrm{T}})^{-1}
A^{(jk)}_{l}]$. $\sigma^{(jk)}_{i}$ is a $2m$-dimensional matrix with all the entries zero
expect a $2\times 2$ block, shown as below
\begin{equation}
\sigma^{(jk)}_{i} = \left(\begin{array}{ccccc}
 & 1\mathrm{st} & \cdots & k\mathrm{th} & \cdots\\
1\mathrm{st} & 0_{2\times2} & 0_{2\times2} & 0_{2\times2} & 0_{2\times2}\\
\vdots & 0_{2\times2} & \vdots & \vdots & \vdots\\
j\mathrm{th} & 0_{2\times2} & \cdots & \sigma_{i} & \cdots\\
\vdots & \vdots & \vdots & \vdots & \vdots
\end{array}\right),
\end{equation}
where $0_{2\times 2}$ represents a 2 by 2 block with zero entries. $\openone^{(jk)}_{2}$ is
similar to $\sigma^{(jk)}_{i}$ but replace the block $\sigma_{i}$ with
$\openone_{2}$~\footnote{The derivation of this theorem is in~\ref{apx:SLD_Gaussian}.}.
\end{theorem}

Being aware of the expression of SLD operator given in Theorem~\ref{SLD_Gaussian}, the QFIM can be
calculated via equation~(\ref{eq:def_QFIM}). Here we show the result explicitly in following theorem.
\begin{theorem} \label{QFIM_Gaussian}
For a continuous variable bosonic $m$-mode Gaussian state with the displacement vector (first  moment)
$\langle \vec{R}\rangle$ and the covariance matrix (second moment) $C$, the entry of QFIM can be
expressed by~\cite{Gao2014,Nichols2018,Safranek2018}
\begin{equation}
\mathcal{F}_{ab}=\mathrm{Tr}\left(G_{a}\partial_{b}C\right)
+(\partial_{a}\langle\vec{R}\rangle^{\mathrm{T}})
C^{-1}\partial_{b}\langle\vec{R}\rangle, \label{eq:QFIM_Gaussian}
\end{equation}
and the QFI for an $m$-mode Gaussian state with respect to $x_{a}$ can be immediately
obtained as~\cite{Monras2013}
\begin{equation}
\mathcal{F}_{aa}=\mathrm{Tr}\left(G_{a}\partial_{a}C\right)+(\partial_{a}
\langle\vec{R}\rangle^{\mathrm{T}})C^{-1}\partial_{a}\langle\vec{R}\rangle.
\end{equation}
\end{theorem}

The expression of right logarithmic derivative for a general Gaussian state and
the corresponding QFIM was provided by Gao and Lee~\cite{Gao2014} in 2014, which
is an appropriate tool for the estimation of complex numbers~\cite{Yuen1973},
such as the number $\alpha$ of a coherent state $|\alpha\rangle$.
The simplest case is a single-mode Gaussian state. For such a state, $G_{a}$
can be calculated as following.
\begin{corollary}
For a single-mode Gaussian state $G_{a}$ can be expressed as~\footnote{The derivation
of this corollary is in~\ref{apx:SLD_smGuassian}.}
\begin{equation}
G_{a}=\frac{4c^{2}-1}{4c^{2}+1}\Omega(\partial_{a}J)\Omega,
\end{equation}
where $c=\sqrt{\det C}$ is the symplectic eigenvalue of $C$ and
\begin{equation}
J=\frac{1}{4c^{2}-1}C.
\end{equation}
For pure states, $\det C$ is a constant, $G_{a}$ then reduces to
\begin{equation}
G_{a}=\frac{1}{4c^{2}+1}\Omega(\partial_{a}C)\Omega.
\end{equation}
\end{corollary}
From this $G_{a}$, $\vec{L}^{(1)}_{a}$ and $L^{(0)}_{a}$ can be further obtained,
which can be used to obtain the SLD operator via equation~(\ref{eq:SLD_Gaussian})
and the QFIM via equation~(\ref{eq:QFIM_Gaussian}).

Another widely used method to obtain the QFI for Gaussian states is through the fidelity
(see section~\ref{sec:fidelity_Bures} for the relation between fidelity and QFIM).
The QFI for pure Gaussian states is studied in reference~\cite{Pinel2012}.
The QFI for single-mode Gaussian states has been obtained through the fidelity by Pinel et al.
in 2013~\cite{Pinel2013}, and for two-mode Gaussian states by \v{S}afr\'{a}nek et al. in
2015~\cite{Safranek2015} and Marian et al.~\cite{Marian2016} in 2016, based
on the expressions of the fidelity given by Scutaru~\cite{Scutaru1998} and Marian et al.
in 2012~\cite{Marian2012}. The expressions of the QFI and the fidelity for multi-mode
Gaussian states are given by Monras~\cite{Monras2013}, Safranek et al.~\cite{Safranek2015}
and Banchi et al.~\cite{Banchi2015}, and reproduced by Oh et al.~\cite{Oh2019} with
a Hermitian operator related to the optimal measurement of the fidelity.

There are other approaches, such as the exponential
state~\cite{Jiang2014}, Husimi Q function~\cite{Gagatsos2016}, that can obtain
the QFI and the QFIM for some specific types of Gaussian states. Besides, a general
method to find the optimal probe states to optimize the QFIM of Gaussian unitary
channels is also provided by \v{S}afr\'{a}nek and Fuentes in 2016~\cite{Safranek2016},
and Matsubara et al.~\cite{Matsubara2019} in 2019. Matsubara et al. performed
the optimization of the QFI for Gaussian states in a passive linear optical circuit.
For a fixed total photon number, the optimal Gaussian state is proved to be a
single-mode squeezed vacuum state and the optimal measurement is a homodyne measurement.

\subsection{QFIM and geometry of quantum mechanics} \label{sec:geometry}

\subsubsection{Fubini-Study metric}

In quantum mechanics, the pure states is a normalized vector because of the
basic axiom that the norm square of its amplitude represents
the probability. The pure states thus can be represented as rays in the projective Hilbert space,
on which Fubini-Study metric is a K\"{a}hler metric. The squared infinitesimal
distance here is usually expressed as~\cite{Facchi2010}
\begin{equation}
\mathrm{d}s^{2}=\frac{\langle\mathrm{d}\psi|\mathrm{d}\psi\rangle}
{\langle\psi|\psi\rangle}-\frac{\langle\mathrm{d}\psi|\psi\rangle
\langle\psi|\mathrm{d}\psi\rangle}{\langle\psi|\psi\rangle^{2}}.
\end{equation}
As $\langle\psi|\psi\rangle=1$ and $|\mathrm{d}\psi\rangle
=\sum_{\mu}|\partial_{x_{\mu}}\psi\rangle \mathrm{d}x_{\mu}$, $\mathrm{d}s^{2}$
can be expressed as
\begin{equation}
\mathrm{d}s^{2}=\sum_{\mu\nu}\frac{1}{4}\mathcal{F}_{\mu\nu}\mathrm{d}x_{\mu}
\mathrm{d}x_{\nu},
\end{equation}
here $\mathcal{F}_{\mu\nu}$ is the $\mu\nu$ element of the QFIM. This means the Fubini-Study metric is
a quarter of the QFIM for pure states.
This is the intrinsic reason why the QFIM can depict the precision limit.
Intuitively, the precision limit is just a matter of distinguishability.
The best precision means the maximum distinguishability, which is naturally
related to the distance between the states. The counterpart of Fubini-study
metric for mixed states is the Bures metric, a well-known metric in quantum
information and closely related to the quantum fidelity, which will be discussed below.

\subsubsection{Fidelity and Bures metric} \label{sec:fidelity_Bures}
\label{sec:Bures_metric}

In quantum information, the fidelity $f(\rho_{1},\rho_{2})$ quantifies the
similarity between two quantum states $\rho_{1}$ and $\rho_{2}$, which is
defined as~\cite{Nielsen2000}
\begin{equation}
f(\rho_{1},\rho_{2}):=\mathrm{Tr}\sqrt{\sqrt{\rho_{1}}\rho_{2}\sqrt{\rho_{1}}}.
\label{eq:fidelity_def}
\end{equation}
Here $f\in[0,1]$ and $f=1$ only when $\rho_{1}=\rho_{2}$. Although the fidelity itself is not
a distance measure, it can be used to construct the Bures distance, denoted as
$D_{\mathrm{B}}$, as~\cite{Nielsen2000}
\begin{equation}
D^{2}_{\mathrm{B}}(\rho_{1},\rho_{2})=2-2 f(\rho_{1},\rho_{2}).
\end{equation}
The relationship between the fidelity and the QFIM has been well studied in the
literature~\cite{Braunstein1994,Twamley1996,Zanardi2007,Zanardi2008,LiuPhysicaA
,Yuan2016}. In the case that the rank of $\rho(\vec{x})$ is unchanged with
the varying of $\vec{x}$, the QFIM is related to the infinitestmal Bures
distance in the same way as the QFIM related to the Fubini-study
metric~\footnote{The derivation is given in~\ref{apx:fidelity_QFIM}.}
\begin{equation}
D^{2}_{\mathrm{B}}(\rho(\vec{x}),\rho({\vec{x}+\mathrm{d}\vec{x}}))=\frac{1}{4}\sum_{\mu\nu}
\mathcal{F}_{\mu\nu} dx_{\mu}dx_{\nu}.
\end{equation}

In recent years it has been found that the fidelity susceptibility, the leading order
(the second order) of the fidelity, can be used as an indicator of the quantum phase
transitions~\cite{Gu2010}. Because of this deep connection between the Bures metric
and the QFIM, it is not surprising that the QFIM can be used in a similar way.
On the other hand, the enhancement of QFIM at the critical point indicates that
the precision limit of the parameter can be improved near the phase transition,
as shown in references~\cite{Frerot2018,Rams2018}.

In the case that the rank of $\rho(\vec{x})$ does not equal to that of
$\rho(\vec{x}+\mathrm{d}\vec{x})$, \v{S}afr\'{a}nek recently showed~\cite{Safranek2017}
that the QFIM does not exactly equal to the fidelity susceptibility. Later,
Seveso et al. further suggested~\cite{Seveso2019} that the quantum Cram\'{e}r-Rao bound
may also fail at those points.

Besides the Fubini-Study metric and the Bures metric, the QFIM is also closely
connected to the Riemannian metric due to the fact that the state space of a
quantum system is actually a Riemannian manifold. In more concrete terms,
the QFIM belongs to a family of contractive Riemannian
metric~\cite{Petz1996,Amari2000,Morozova1991,Hayashi2005},
associated with which the infinitesimal distance in state space is
$\mathrm{d}s^{2}=\sum_{\mu}g_{\mu\nu}\mathrm{d}x_{\mu}\mathrm{d}x_{\nu}$ with
$g_{\mu\nu}$ as the contractive Riemannian metric. In the eigenbasis of the
density matrix $\rho$, $g_{\mu\nu}$ takes the form as~\cite{Bengtsson2006,Sommers2003,Pires2016}
\begin{equation}
g_{\mu\nu}=\frac{1}{4}\sum_{i}\frac{\langle\lambda_{i}|\mathrm{d}\rho|\lambda_{i}\rangle^{2}}
{\lambda_{i}}+\frac{1}{2}\sum_{i<j}\frac{|\langle\lambda_{i}|\mathrm{d}\rho|\lambda_{j}\rangle|^{2}}
{\lambda_{j}h(\lambda_{i}/\lambda_{j})},
\label{eq:Bures_Riemannian}
\end{equation}
where $h(\cdot)$ is the Morozova-\v{C}encov function, which is an operator monotone
(for any positive semi-definite operators), self inverse ($xh(1/x)=1/h(x)$) and
normalized ($h(1)=1$) real function. When $h(x)=(1+x)/2$, the metric above reduces
to the QFIM (based on the SLD). The QFIMs based on right and left
logarithmic derivatives can also be obtained by taking $h(x)=x$ and $h(x)=1$.
The Wigner-Yanase information metric can be obtained from it by taking
$h(x)=\frac{1}{4}(\sqrt{x}+1)^{2}$.

\subsubsection{Quantum geometric tensor}
\label{sec:geometric_tensor}

The quantum geometric tensor originates from a complex metric in the projective Hilbert
space, and is a powerful tool in quantum information science that unifies the
QFIM and the Berry connection. For a pure state $|\psi\rangle=|\psi(\vec{x})\rangle$,
the quantum geometric tensor $Q$ is defined as~\cite{Provost1980,Venuti2007}
\begin{equation}
Q_{\mu\nu}=\langle\partial_{\mu}\psi|\partial_{\nu}\psi\rangle
-\langle\partial_{\mu}\psi|\psi\rangle\langle\psi|\partial_{\nu}\psi\rangle.
\end{equation}
Recall the expression of QFIM for pure states, given in equation~(\ref{eq:purestate_general}),
the real part of $Q_{\mu\nu}$ is actually the QFIM up to a constant factor, i.e.,
\begin{equation}
\mathrm{Re}(Q_{\mu\nu})=\frac{1}{4}\mathcal{F}_{\mu\nu}.
\end{equation}
In the mean time, due to the fact that
\begin{equation}
(\langle\partial_{\mu}\psi|\psi\rangle\langle\psi|\partial_{\nu}\psi\rangle)^{*}
=\langle\psi|\partial_{\mu}\psi\rangle\langle\partial_{\nu}\psi|\psi\rangle
=\langle\partial_{\mu}\psi|\psi\rangle\langle\psi|\partial_{\nu}\psi\rangle,
\end{equation}
i.e., $\langle\partial_{\mu}\psi|\psi\rangle\langle\psi|\partial_{\nu}\psi\rangle$
is real, the imaginary part of $Q_{\mu\nu}$ then reads
\begin{equation}
\mathrm{Im}(Q_{\mu\nu})=\mathrm{Im}(\langle\partial_{\mu}\psi|\partial_{\nu}\psi\rangle)
=-\frac{1}{2}\left(\partial_{\mu}\mathcal{A}_{\nu}-\partial_{\nu}\mathcal{A}_{\mu}\right),
\end{equation}
where $\mathcal{A}_{\mu}:=i\langle\psi|\partial_{\mu}\psi\rangle$ is the
Berry connection~\cite{Shapere1989} and $\Upsilon_{\mu\nu}:=\partial_{\mu}
\mathcal{A}_{\nu}-\partial_{\nu}\mathcal{A}_{\mu}$ is the Berry curvature. The geometric
phase can then be obtained as~\cite{Berry1984}
\begin{equation}
\phi =\ointop\mathcal{A}_{\mu} \mathrm{d}x_{\mu},
\end{equation}
where the integral is taken over a closed trajectory in the parameter space.

Recently, Guo et al.~\cite{Guo2016} connected the QFIM and the Berry curvature via the Robertson
uncertainty relation. Specifically, for a unitary process with two parameters,
$\Upsilon_{\mu\nu}=i\langle\psi_{0}| [\mathcal{H}_{\mu},\mathcal{H}_{\nu}]|\psi_{0}\rangle$
with $\mathcal{H}_{\mu}$ defined in equation~(\ref{eq:unitary_H}) and $|\psi_{0}\rangle$
the probe state, the determinants of the QFIM and the Berry curvature should satisfy
\begin{equation}
\det\mathcal{F}+4\det\Upsilon \geq 0.
\end{equation}

\subsection{QFIM and thermodynamics}

The density matrix of a quantum thermal state is
\begin{equation}
\rho=\frac{1}{Z}e^{-\beta H},
\end{equation}
where $Z=\mathrm{Tr}(e^{-\beta H})$ is the partition function and $\beta=1/(k_{\mathrm{B}}T)$.
$k_{\mathrm{B}}$ is the Boltzmann constant and $T$ is the temperature.
For such state we have
$\partial_{T}\rho=\frac{1}{T^{2}}\left(\langle H\rangle-H \right)\rho$,
where we have set $k_{\mathrm{B}}=1$. If we take the temperature as the unknown parameter,
the SLD, which is the solution to $\partial_{T}\rho=\frac{1}{2}(\rho L_T+L_T\rho)$, can then
be obtained as
\begin{equation}
L_{T}=\frac{1}{T^{2}}\left(\langle H\rangle-H\right),
\end{equation}
which commutes with $\rho$. The QFI for the temperature hence reads
\begin{equation}
\mathcal{F}_{TT}=\frac{1}{T^{4}}(\langle H^{2}\rangle-\langle H\rangle^{2}),
\end{equation}
i.e., $\mathcal{F}_{TT}$ is proportional to the fluctuation of the Hamiltonian.
Compared to the specific heat $C_{v}=\frac{\partial_{T}\langle H\rangle}{\partial T}
=\frac{1}{T^{2}}(\langle H^{2}\rangle-\langle H\rangle^{2})$,
we have~\cite{Gu2010,Paris2016,Hofer2017,Zanardi2007,Zanardi2007a}
\begin{equation}
\mathcal{F}_{TT}=\frac{1}{T^{2}} C_{v},
\end{equation}
i.e., for a quantum thermal state, the QFI for the temperature is proportional to the
specific heat of this system. For a system of which the Hamiltonian has no
interaction terms, the relation above still holds for its subsystems~\cite{Pasquale2016}.

The correlation function is an important concept in quantum physics and condensed
matter physics due to the wide applications of the linear response theory. It
is well known that the static susceptibility between two observables $A$ and $B$,
which represents the influence of $\langle A\rangle$'s perturbation on
$\langle B\rangle$ under the thermal equilibrium, is proportional to the canonical
correlation $\frac{1}{\beta}\int^{\beta}_{0}\frac{1}{Z}\mathrm{Tr}(e^{-\beta H}
e^{s H}Ae^{-s H} B)\mathrm{d}s$~\cite{Kubo1991}, which can be further written into
$\int^{1}_{0}\mathrm{Tr}(\rho^{s}A\rho^{1-s}B)\mathrm{d}s$. Denote
$\int^{1}_{0}\rho^{s}\tilde{L}_{a}\rho^{1-s}\mathrm{d}s=\partial_{a} \rho$
and replace $A$, $B$ with $\tilde{L}_{a}$ and $\tilde{L}_{b}$, the canonical
correlation reduces to the so-called Bogoliubov–Kubo–Mori Fisher information matrix
$\int^{1}_{0}\rho^{s}\tilde{L}_{a}\rho^{1-s}\tilde{L}_{b}\mathrm{d}s$~\cite{Petz1993,
Petz1994,Petz2008,Hayashi2002,Balian1986}. However, this relation does not suggest how to
connect the linear response function with the QFIM based on SLD. In 2007, You
et al.~\cite{Gu2010,You2007} first studied the connection between the fidelity
susceptibility and the correlation function, and then in 2016, Hauke et
al.~\cite{Hauke2016} extended this connection between the QFI and the symmetric
and asymmetric correlation functions to the thermal states. Here we use their
methods to establish the relation between the QFIM and the cross-correlation
functions.

Consider a thermal state corresponding to the Hamiltonian $H=\sum_{a}x_{a}O_{a}$,
where $O_{a}$ is a Hermitian generator for $x_{a}$ and $[O_{a}, O_{b}]=0$
for any $a$ and $b$, the QFIM can be expressed as~\footnote{The details
of the derivation can be found in~\ref{apx:dynamic_sus}.}
\begin{equation}
\mathcal{F}_{ab} = \frac{4}{\pi}\int^{\infty}_{-\infty}
\tanh^{2}\left(\frac{\omega}{2T}\right)\mathrm{Re}(S_{ab}(\omega))\mathrm{d}\omega,
\label{eq:correlation_func0}
\end{equation}
or equivalently,
\begin{equation}
\mathcal{F}_{ab}=\frac{4}{\pi}\int^{\infty}_{-\infty}\tanh\left(\frac{\omega}{2T}\right)
\mathrm{Im}(\chi_{ab}(\omega))\mathrm{d}\omega.
\label{eq:correlation_func1}
\end{equation}
Here $S_{ab}(\omega)$ is the symmetric cross-correlation spectrum
defined as
\begin{equation}
S_{ab}(\omega)=\int^{\infty}_{-\infty}\frac{1}{2}
\langle \{Q_{a}(t), O_{b}\}\rangle e^{i\omega t}\mathrm{d}t,
\end{equation}
where $\langle\cdot\rangle=\mathrm{Tr}(\rho\cdot)$ and $O_{a(b)}(t)
=e^{iHt}O_{a(b)}e^{-iHt}$. Its real part can also be written as
\begin{equation}
\mathrm{Re}(S_{ab}(\omega))=\int^{\infty}_{-\infty}\frac{1}{2}
\langle Q_{a}(t)O_{b}+O_{b}(t)O_{a}\rangle e^{i\omega t}\mathrm{d}t.
\end{equation}
$\chi_{ab}$ is the asymmetric cross-correlation spectrum defined as
\begin{equation}
\chi_{ab}(\omega)= \int^{\infty}_{-\infty}\frac{i}{2}
\langle [O_{a}(t), O_{b}] \rangle e^{i\omega t}\mathrm{d}t.
\end{equation}
Because of equations~(\ref{eq:correlation_func0})
and~(\ref{eq:correlation_func1}), and the fact that
$S_{ab}(\omega)$ and $\chi_{ab}(\omega)$ can be directly measured in the
experiments~\cite{Shirane2002,Stoferle2004,Roy2015,Roy2015a,Sinitsyn2016},
$\mathcal{F}_{ab}$ becomes measurable in this case, which breaks
the previous understanding that QFI is not observable since
the fidelity is not observable. Furthermore, due to the fact that
the QFI is a witness for multipartite entanglement~\cite{Toth2012},
and a large QFI can imply Bell correlations~\cite{Frowis2019},
equations~(\ref{eq:correlation_func0}) and~(\ref{eq:correlation_func1})
provide an experimentally-friendly way to witness the quantum correlations
in the thermal systems. As a matter of fact, Shitara and Ueda~\cite{Shitara2016}
showed that the relations in equations~(\ref{eq:correlation_func0})
and~(\ref{eq:correlation_func1}) can be further extended to the family of metric
described in equation~(\ref{eq:Bures_Riemannian}) by utilizing the generalized
fluctuation-dissipation theorem.

\subsection{QFIM in quantum dynamics}

Quantum dynamics is not only a fundamental topic in quantum mechanics, but also
widely connected to various topics in quantum information and quantum technology.
Due to some excellent mathematical properties, the QFIM becomes a good candidate
for the characterization of certain behaviors and phenomena in quantum dynamics.
In the following we show the roles of QFIM in quantum speed limit and the
characterization of non-Markovianity.

\subsubsection{Quantum speed limit}

Quantum speed limit aims at obtaining the smallest evolution time for quantum
processes~\cite{Taddei2013,Toth2014,Deffner2017,Gessner2018,
Campo2013,Campaioli2018,Deffner2013,Shanahan2018,Pires2016}.
It is closely related to the geometry of quantum states since the dynamical
trajectory with the minimum evolution time is actually the geodesic in the state
space, which indicates that the QFIM should be capable to quantify the speed limit.
As a matter of fact, the QFI and the QFIM have been used to bound the quantum
speed limit in recent studies~\cite{Taddei2013,Toth2014,Deffner2017,Gessner2018}.
For a unitary evolution, $U=\exp(-iHt)$, to steer a state away from the initial
position with a Bures angle $D_{\mathrm{B}}=\arccos(f)$ ($f$ is the fidelity
defined in equation~(\ref{eq:fidelity_def})), the evolution time $t$ needs to
satisfy~\cite{Toth2014,Taddei2013,Frowis2012}
\begin{equation}
t \geq \frac{2D_{\mathrm{B}}}{\sqrt{\mathcal{F}_{tt}}},
\end{equation}
where $\mathcal{F}_{tt}$ is the QFI for the time $t$. For a more general case,
that the Hamiltonian is time-dependent, Taddei et al.~\cite{Taddei2013} provided
an implicit bound based on the QFI,
\begin{equation}
\sqrt{\frac{1}{2}\frac{\mathrm{d}^2\mathcal{D}(f)}{\mathrm{d}f^2}
\left(\frac{\mathrm{d}\mathcal{D}}{\mathrm{d}f}\right)^{-3}}\Bigg|_{f\rightarrow 1}
\mathcal{D}(f(\rho(0),\rho(t)))\leq \int^{t}_0\sqrt{\frac{\mathcal{F}_{\tau\tau}}{4}}\mathrm{d}\tau,
\end{equation}
where $\mathcal{D}$ is any metric on the space of quantum states via the
fidelity $f$. In 2017, Beau and del Campo~\cite{Beau2017} discussed the
nonlinear metrology of many-body open systems and established the relation
between the QFI for coupling constants and the quantum speed limit, which
indicates that the quantum speed limit directly determines the amplitude
of the estimation error in such cases.

Recently, Pires et al.~\cite{Pires2016} established an infinite family of quantum
speed limits based on a contractive Riemannian metric discussed in
section~\ref{sec:Bures_metric}. In the case that $\vec{x}$ is
time-dependent, i.e., $\vec{x}=\vec{x}(t)$, the geodesic distance
$D(\rho_{0},\rho_{t})$ gives a lower bound of general trajectory,
\begin{equation}
D(\rho_{0},\rho_{t})\leq\int^{t}_{0}\left(\frac{\mathrm{d}s}{\mathrm{d}t}\right)\mathrm{d}t
=\int^{t}_{0}\mathrm{d}t\sqrt{\sum_{\mu\nu}g_{\mu\nu}\frac{\mathrm{d}x_{\mu}}{dt}
\frac{\mathrm{d}x_{\nu}}{dt}},
\end{equation}
where $g_{\mu\nu}$ is defined in equation~(\ref{eq:Bures_Riemannian}).
Taking the maximum Morozova-\v{C}encov function, the above inequality leads
to the quantum speed limit with time-dependent parameters given in
reference~\cite{Taddei2013}.

\subsubsection{Non-Markovianity}

Non-Markovianity is an emerging concept in open quantum systems. Many different
quantification of the non-Markovianity based on monotonic quantities under the
completely positive and trace-preserving maps have been
proposed~\cite{Breuer2016,Rivas2014,Vega2017}. The QFI can also be used to
characterize the non-Markovianity since it also satisfies the monotonicity~\cite{Toth2014}.
For the master equation
\begin{equation}
\partial_{t}\rho=-i[H_{\vec{x}},\rho]+\sum_{j}\gamma_{j}(t)
\left(\Gamma_{j}\rho\Gamma^{\dagger}_{j}-\frac{1}{2}
\left\{\Gamma^{\dagger}_{j}\Gamma_{j},\rho\right\}\right),
\end{equation}
the quantum Fisher information flow
\begin{equation}
\partial_{t}\mathcal{F}_{aa}=-\sum_{j}\gamma_{j}(t)
\mathrm{Tr}\left(\rho[L_{a},\Gamma_{j}]^{\dagger}[L_{a},\Gamma_{j}]\right)
\end{equation}
given by Lu et al.~\cite{Lu2010} in 2010 is a valid witness for non-Markovianity.
Later in 2015, Song et al.~\cite{Song2015} utilized the maximum eigenvalue of average
QFIM flow to construct a quantitative measure of non-Markovianity.
The average QFIM flow is the time derivative of average QFIM $\bar{\mathcal{F}}
=\int\mathcal{F} \mathrm{d}\vec{x}$. Denote $\lambda_{\max}(t)$ as the maximum eigenvalue
of $\partial_{t}\bar{\mathcal{F}}$ at time $t$, then the non-Markovianity can be
alternatively defined as
\begin{equation}
\mathcal{N}:=\int_{\lambda_{\max}>0} \lambda_{\max}\mathrm{d}t.
\end{equation}
One may notice that this is not the only way to define non-Makovianity with the
QFIM, similar constructions would also be
qualified measures for non-Markovianity.

\section{Quantum multiparameter estimation} \label{sec:QCRB}

\subsection{Quantum multiparameter Cram\'{e}r-Rao bound}

\subsubsection{Introduction}

The main application of the QFIM is in the quantum multiparameter estimation,
which has shown very different properties and behaviors compared to its
single-parameter counterpart~\cite{Szczykulska2016}. The quantum multiparameter
Cram\'{e}r-Rao bound, also known as Helstrom bound, is one of the most widely
used asymptotic bound in quantum metrology~\cite{Holevo,Helstrom}.

\begin{theorem}
For a density matrix $\rho $ in which a vector of unknown parameters
$\vec{x}=(x_{0},x_{1},...,x_{m},...)^{\mathrm{T}}$ is encoded, the covariance
matrix $\mathrm{cov}(\hat{\vec{x}},\{\Pi_{y}\})$ of an unbiased estimator
$\hat{\vec{x}}$ under a set of POVM, $\{\Pi_{y}\}$, satisfies the following
inequality~\footnote{The derivation of this theorem is given in~\ref{apx:multiCR}.}
\begin{equation}
\mathrm{cov}(\hat{\vec{x}},\{\Pi_{y}\}) \geq \frac{1}{n}\mathcal{I}^{-1}(\{\Pi_{y}\})
\geq \frac{1}{n}\mathcal{F}^{-1},
\end{equation}
where $\mathcal{I}(\{\Pi_{y}\})$ is the CFIM, $\mathcal{F}$ is the QFIM and $n$ is
the repetition of the experiment.
\end{theorem}
The second inequality is called the quantum multiparameter Cram\'{e}r-Rao bound.
In the derivation, we assume the QFIM can be inverted, which is reasonable since
a singular QFIM usually means not all the
unknown parameters are independent and the parameters cannot be estimated simultaneously.
In such cases one should first identify the set of parameters that
are independent, then calculate the  corresponding QFIM for those parameters.

For cases where the number of unknown parameters is large, it may be difficult or
even meaningless to know the error of every parameter, and the total variance or
the average variance is a more appropriate macroscopic quantity to study. Recall
that the $a$th diagonal entry of the covariance matrix is actually the variance of the
parameter $x_{a}$. Thus, the bound for the total variance can be immediately obtained as following.
\begin{corollary} \label{CR_lowbound1}
Denote $\mathrm{var}(\hat{x}_{a},\{\Pi_{y}\})$ as the variance of $x_{a}$, then the
total variance $\sum_{a}\mathrm{var}(x_{a},\{\Pi_{y}\})$ is bounded by the trace
of $\mathcal{F}^{-1}$, i.e.,
\begin{equation}
\sum_{a}\mathrm{var}(\hat{x}_{a},\{\Pi_{y}\}) \geq \frac{1}{n}\mathrm{Tr}
\left(\mathcal{I}^{-1}(\{\Pi_{y}\})\right) \geq
\frac{1}{n}\mathrm{Tr}\left(\mathcal{F}^{-1}\right).
\end{equation}
\end{corollary}
The inverse of QFIM sometimes is difficult to obtain analytically and one may need
a lower bound of $\mathrm{Tr}(\mathcal{F}^{-1})$ to roughly evaluate the precision limit.
Being aware of the property of QFIM (given in section~\ref{sec:definition}) that
$[\mathcal{F}^{-1}]_{aa} \geq 1/\mathcal{F}_{aa}$, one can easily obtain the following corollary.
\begin{corollary} \label{CR_lowbound2}
The total variance is bounded as
\begin{equation}
\sum_{a}\mathrm{var}(\hat{x}_{a},\{\Pi_{y}\}) \geq \frac{1}{n}\mathrm{Tr}
\left(\mathcal{F}^{-1} \right) \geq \sum_{a} \frac{1}{n \mathcal{F}_{aa}}.
\end{equation}
The second inequality can only be attained when $\mathcal{F}$ is diagonal. Similarly,
\begin{equation}
\sum_{a}\mathrm{var}(\hat{x}_{a},\{\Pi_{y}\}) \geq \frac{1}{n}\mathrm{Tr}
\left(\mathcal{I}^{-1}(\{\Pi_{y}\})\right) \geq \sum_{a}
\frac{1}{n \mathcal{I}_{aa}(\{\Pi_{y}\})}.
\end{equation}
\end{corollary}

The simplest example for the multi-parameter estimation is the case with two parameters.
In this case, $\mathcal{F}^{-1}$ can be calculated analytically as
\begin{equation}
\mathcal{F}^{-1}=\frac{1}{\det(\mathcal{F})}\left(\begin{array}{cc}
\mathcal{F}_{bb} & -\mathcal{F}_{ab} \\
-\mathcal{F}_{ab} & \mathcal{F}_{aa} \\
\end{array}\right).
\end{equation}
Here $\det(\cdot)$ denotes the determinant. With this equation, the corollary above
can reduce to the following form.
\begin{corollary} \label{two_parameter_theorem}
For two-parameter quantum estimation, corollary~\ref{CR_lowbound1} reduces to
\begin{equation}
\sum_{a}\mathrm{var}(\hat{x}_{a},\{\Pi_{y}\})\geq \frac{1}{n I_{\mathrm{eff}}
(\{\Pi_{y}\})}\geq \frac{1}{n F_{\mathrm{eff}}},
\end{equation}
where $I_{\mathrm{eff}}(\{\Pi_{y}\})=\det(\mathcal{I})/\mathrm{Tr}(\mathcal{I})$
and $F_{\mathrm{eff}}=\det(\mathcal{F})/\mathrm{Tr}(\mathcal{F})$ can be treated
as effective classical and quantum Fisher information.
\end{corollary}

\subsubsection{Attainability} \label{sec:attainability}

Attainability is a crucial problem in parameter estimation. An unattainable bound
usually means the given precision limit is too optimistic to be realized
in physics. In classical statistical estimations, the classical Cram\'{e}r-Rao
bound can be attained by the maximum likelihood estimator in the asymptotic limit,
i.e., $\lim_{n\rightarrow \infty} n\mathrm{cov}(\hat{\vec{x}}_{\mathrm{m}}(n))
=\mathcal{I}^{-1}$, where $\hat{\vec{x}}_{\mathrm{m}}(n)$ is the local maximum
likelihood estimator and a function of repetition or sample number, which is
unbiased in the asymptotic limit. Because of this, the parameter estimation
based on Cram\'{e}r-Rao bound is an asymptotic theory and requires infinite samples
or repetition of the experiment. For a finite sample case, the maximum likelihood
estimator is not unbiased and the Cram\'{e}r-Rao bound may not be attainable as
well. Therefore, the true attainability in quantum parameter estimation should
also be considered in the sense of asymptotic limit since the attainability
of quantum Cram\'{e}r-Rao bound usually requires the QFIM equals the CFIM
first. The general study of quantum parameter estimation from the asymptotic
aspect is not easy, and the recent progress can be found in reference~\cite{Hayashi2005}
and references therein. Here in the following, the attainability majorly refers to
that if the QFIM coincides with the CFIM in theory. Besides, another thing that
needs to emphasize is that the maximum likelihood estimator is optimal in a local
sense~\cite{Wasserman2004}, i.e., the estimated value is very close to the true
value, and a locally unbiased estimator only attains the bound locally but not
globally in the parameter space, thus, the attainability and optimal measurement
discussed below are referred to the local ones.

For the single-parameter quantum estimations, the quantum Cram\'{e}r-Rao bound can
be attained with a theoretical optimal measurement. However, for multi-parameter
quantum estimation, different parameters may have different optimal measurements,
and these optimal measurements may not commute with each other. Thus there may
not be a common measurement that is optimal for the estimation for all the
unknown parameters. The quantum Cram\'{e}r-Rao bound for the estimation of multiple
parameters is then not necessary attainable, which is a major obstacle for the
utilization of this bound in many years. In 2002, Matsumoto~\cite{Matsumoto2002}
first provided the necessary and sufficient condition for pure states. After this,
its generalization to mixed states was discussed in several specific
scenarios~\cite{Vidrighin2014,Crowley2014,Vaneph2013,Suzuki2016} and rigorously
proved via the Holevo bound firstly with the theory of local asymptotic
normality~\cite{Gill2013} and then the direct minimization of one term in Holevo
bound~\cite{Ragy2016}. We first show this condition in the following theorem.
\begin{theorem}
The necessary and sufficient condition for the saturation of the quantum
multiparameter Cram\'{e}r-Rao bound is
\begin{equation}
\mathrm{Tr}\left(\rho [L_{a},L_{b}]\right)=0,~~\forall~a,b.
\end{equation}
For a pure parameterized state $|\psi\rangle:=|\psi(\vec{x})\rangle$, this
condition reduces to
\begin{equation}
\langle\psi|[L_{a},L_{b}]|\psi\rangle=0,~~\forall~a,b,
\end{equation}
which is equivalent to the form
\begin{equation}
\mathrm{Im}(\langle\partial_{a}\psi|\partial_{b}\psi\rangle)=0.
\end{equation}
\end{theorem}
When this condition is satisfied, the Holevo bound is also attained and equivalent
to the Cram\'{e}r-Rao bound~\cite{Gill2013,Ragy2016}. Recall that the Berry curvature
introduced in section~\ref{sec:geometric_tensor} is of the form
\begin{eqnarray}
\Upsilon_{ab}&=&i\partial_{a}(\langle\psi|\partial_{b}\psi\rangle)
-i\partial_{b}(\langle\psi|\partial_{a}\psi\rangle) \nonumber \\
&=& -2\mathrm{Im}(\langle \partial_{a}\psi|\partial_{b}\psi\rangle).
\end{eqnarray}
Hence, the above condition can also be expressed as following.
\begin{corollary}
The multi-parameter quantum Cram\'{e}r-Rao bound for a pure parameterized state
can be saturated if and only if
\begin{equation}
\Upsilon=0,
\end{equation}
i.e., the matrix of Berry curvature is a null matrix.
\end{corollary}

For a unitary process $U$ with a pure probe state $|\psi_{0}\rangle$, this
condition can be expressed with the operator $\mathcal{H}_{a}$ and $\mathcal{H}_{b}$,
as shown in the following corollary~\cite{LiuSR}.
\begin{corollary} \label{corollary:attain_U}
For a unitary process $U$ with a pure probe state $|\psi_{0}\rangle$,
the necessary and sufficient condition for the attainability of quantum multiparameter
Cram\'{e}r-Rao bound is
\begin{equation}
\langle\psi_{0}|\left[\mathcal{H}_{a},\mathcal{H}_{b}\right]|\psi_{0}\rangle=0,
~~\forall~a,b.
\end{equation}
\end{corollary}
Here $\mathcal{H}_{a}$ was introduced in section~\ref{sec:unitary_process}.

\subsubsection{Optimal measurements} \label{sec:opt_measurement}

\begin{figure}[tp]
\centering\includegraphics[width=10cm]{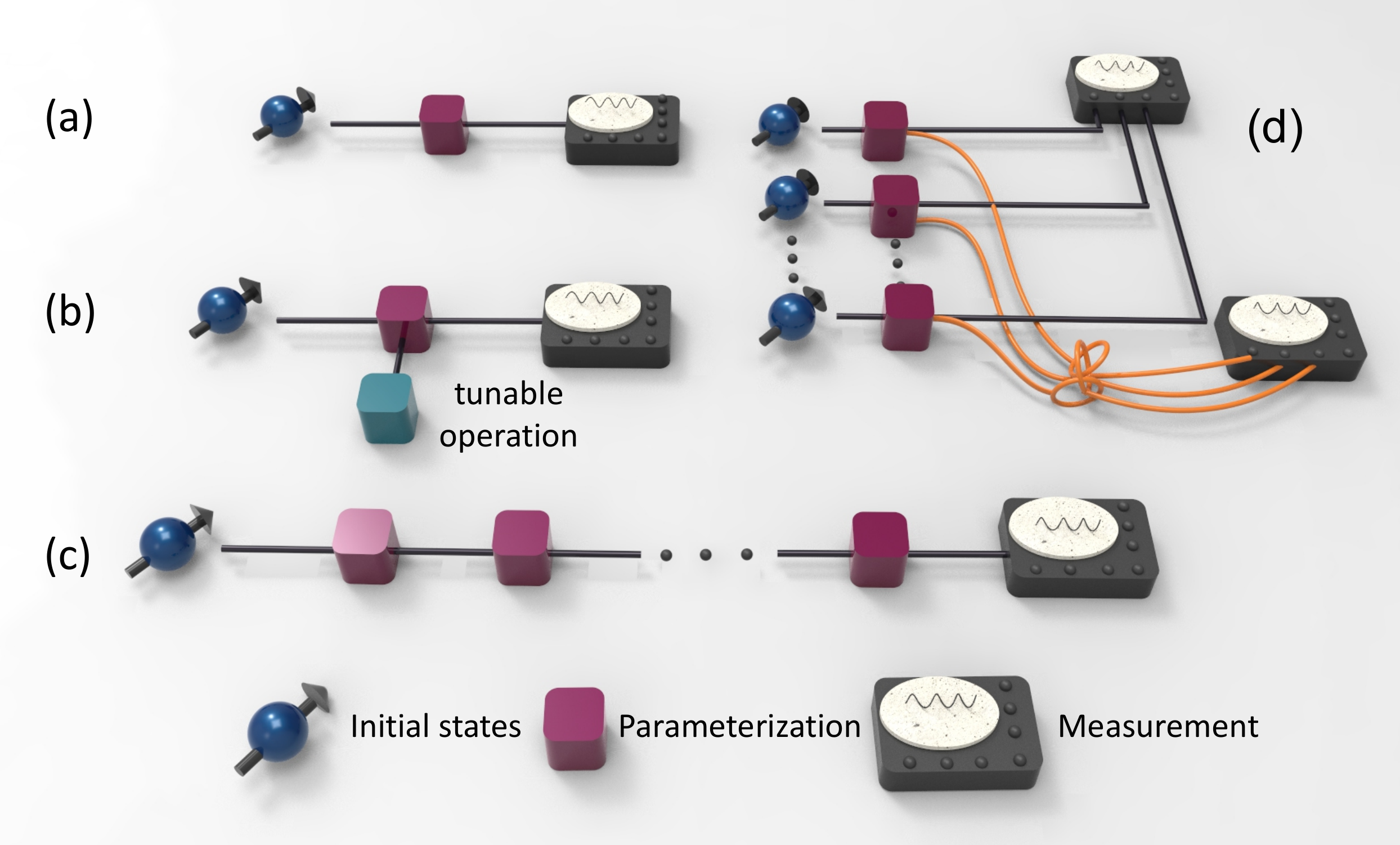}
\caption{Schematics for basic measurement schemes in quantum metrology, including
individual measurment, adaptive measurement and collective measurement.}
\label{fig:measurement}
\end{figure}

The satisfaction of attainability condition theoretically guarantees the existence
of some CFIM that can reach the QFIM. However, it still requires an optimal
measurement. The search of practical optimal measurements is always a core mission
in quantum metrology, and it is for the best that the optimal measurement is
independent of the parameter to be estimated. The most well studied measurement
strategies nowadays include the individual measurement, adaptive measurement
and collective measurement, as shown in figure~\ref{fig:measurement}. The individual
measurement refers to the measurement on a single copy (figure~\ref{fig:measurement}(a))
or local systems (black lines in figure~\ref{fig:measurement}(d)), and can be
easily extend the sequential scenario (figure~\ref{fig:measurement}(c)), which
is the most common scheme for controlled quantum metrology. The collective
measurement, or joint measurement, is the one performed simultaneously on
multi-copies or on the global system (orange lines in figure~\ref{fig:measurement}(d))
in parallel schemes. A typical example for collective measurement is the Bell
measurement. The adaptive measurement (figure~\ref{fig:measurement}(b)) usually
uses some known tunable operations to adjust the outcome. A well-studied case
is the optical Mach-Zehnder interferometer with a tunable path in one arm.
The Mach-Zehnder interferometer will be thoroughly introduced in the next
section.

For the single parameter case, a possible optimal measurement can be constructed
with the eigenstates of the SLD operator. Denote $\{|l_{i}\rangle\langle l_{i}|\}$
as the set of eigenstates of $L_{a}$, if we choose the set of POVM as the
projections onto these eigenstates, then the probability for the $i$th measurement
result is $\langle l_{i}|\rho |l_{i}\rangle$. In the case where $|l_{i}\rangle$
is independent of $x_{a}$, the CFI then reads
\begin{equation}
\mathcal{I}_{aa}= \sum_{i}\frac{\langle l_{i}|\partial_{a}\rho |l_{i}\rangle^{2}}
{\langle l_{i}|\rho |l_{i}\rangle}.
\end{equation}
Due to the equation $2\partial_{a}\rho =\rho L_{a}+L_{a}\rho $, the equation
above reduces to
\begin{equation}
\mathcal{I}_{aa}= \sum_{i}l^{2}_{i} \langle l_{i}|\rho |l_{i}\rangle
=\mathrm{Tr}(\rho L^{2}_{a})=\mathcal{F}_{aa},
\end{equation}
which means the POVM $\{|l_{i}\rangle\langle l_{i}|\}$ is the optimal measurement
to attain the QFI. However, if the eigenstates of the SLD are dependent on $x_{a}$,
it is no longer the optimal measurement. In the case with a high prior information,
the CFI with respect to $\{|l_{i}(\hat{x}_{a})\rangle\langle l_{i}(\hat{x}_{a})|\}$
($\hat{x}_{a}$ is the estimated value of $x_{a}$) may be very close to the QFI.
In practice, this measurement has to be used adaptively. Once we obtain a new
estimated value $\hat{x}_{a}$ via the measurement, we need to update the measurement
with the new estimated value and then perform the next round of measurement.
For a non-full rank parameterized density matrix, the SLD operator is not unique,
as discussed in section~\ref{sec:general_method}, which means the optimal measurement
constructed via the eigenbasis of SLD operator is not unique. Thus, finding a
realizable and simple optimal measurement is always the core mission in quantum
metrology. Update to date, only known states in single-parameter estimation that
own parameter-independent optimal measurement is the so-called
quantum exponential family~\cite{Petz2008,Hayashi2005}, which is of the form
\begin{equation}
\rho=e^{\frac{1}{2}\left(\int^{x}_{0}c(x^{\prime})\mathrm{d}x^{\prime} O
-\int^{x}_{0}x^{\prime}c(x^{\prime})\mathrm{d}x^{\prime}\right)}
\rho_0e^{\frac{1}{2}(\int^{x}_{0}c(x^{\prime})\mathrm{d}x^{\prime} O
-\int^{x}_{0}x^{\prime}c(x^{\prime})\mathrm{d}x^{\prime})},
\end{equation}
where $c(x)$ is a function of the unknown parameter $x$, $\rho_{0}$ is a
parameter-independent density matrix, and $O$ is an unbiased observable
of $x$, i.e., $\langle O\rangle=x$. For this family of states, the SLD
is $L_x = c(x)(O-x)$ and the optimal measurement is the eigenstates of $O$.

For multiparameter estimation, the SLD operators for different parameters may not
share the same eigenbasis, which means $\{|l_{i}(\hat{x}_{a})\rangle\langle
l_{i}(\hat{x}_{a})|\}$ is no longer an optimal choice for the estimation of all
unknown parameters, even with the adaptive strategy. Currently, most of the
studies in multiparameter estimation focus on the construction of the optimal
measurements for a pure parameterized state $|\psi\rangle$. In 2013, Humphreys
et al.~\cite{Humphreys2013} proposed a method to construct the optimal measurement,
a complete set of projectors containing the operator
$|\psi_{\vec{x}_{\mathrm{true}}}\rangle\langle\psi_{\vec{x}_{\mathrm{true}}}|$
($\vec{x}_{\mathrm{true}}$ is the true value of $\vec{x}$). Here
$|\psi_{\vec{x}_{\mathrm{true}}}\rangle$ equals the value of $|\psi\rangle$ by
taking $\vec{x}=\vec{x}_{\mathrm{true}}$ ($\vec{x}_{\mathrm{true}}$). All the
other projectors can be constructed via the Gram-Schmidt process. In practice,
since the true value is unknown, the measurement has to be performed adaptively
with the estimated values $\hat{\vec{x}}$, similar as the single parameter case.
Recently, Pezz\`{e} et al.~\cite{Pezze2017} provided the specific conditions
this set of projectors should satisfy to be optimal, which is organized in
the following three theorems.
\begin{theorem} \label{theorem:optimal_pure_1}
Consider a parameterized pure state $|\psi\rangle$.
$|\psi_{\vec{x}_{\mathrm{true}}}\rangle:=|\psi(\vec{x}=\vec{x}_{\mathrm{true}})\rangle$
with $\vec{x}_{\mathrm{true}}$ the true value of $\vec{x}$. The set of projectors
$\{|m_{k}\rangle\langle m_{k}|, |m_{0}\rangle=|\psi_{\vec{x}_{\mathrm{true}}}\rangle\}$
is an optimal measurement to let the CFIM reach QFIM if and only if~\cite{Pezze2017}
\begin{equation}
\lim_{\vec{x}\rightarrow\vec{x}_{\mathrm{true}}}
\frac{\mathrm{Im}(\langle \partial_{a}\psi |m_{k}\rangle\langle m_{k}|\psi \rangle)}
{|\langle m_{k}|\psi \rangle|}=0,~~\forall x_{a}~\mathrm{and}~k\neq 0,
\label{eq:optimal_measurement_0}
\end{equation}
which is equivalent to
\begin{equation}
\mathrm{Im}(\langle\partial_{a}\psi|m_{k}\rangle\langle m_{k}|\partial_{b}\psi\rangle)=0,
~\forall x_{a},x_{b}~\mathrm{and}~k\neq 0.
\end{equation}
\end{theorem}
The proof is given in~\ref{apx:Optimal_measurement}. This theorem shows that if
the quantum Cram\'{e}r-Rao bound can be saturated then it is always possible to
construct the optimal measurement with the projection onto the probe state itself
at the true value and a suitable choice of vectors on the orthogonal subspace~\cite{Pezze2017}.
\begin{theorem}
For a parameterized state $|\psi\rangle$, the set of projectors
$\{|m_{k}\rangle\langle m_{k}|,\langle\psi |m_{k}\rangle\neq 0~\forall k\}$
is an optimal measurement to let the CFIM reach QFIM if and only if~\cite{Pezze2017}
\begin{equation}
\mathrm{Im}(\langle\partial_{a}\psi |m_{k}\rangle\langle m_{k}|\psi\rangle)
=|\langle\psi |m_{k}\rangle|^{2}\mathrm{Im}(\langle\partial_{a}\psi|\psi\rangle),
~~\forall k, x_{a}. \label{eq:optimal_measurement_1}
\end{equation}
\end{theorem}
For the most general case that some projectors are vertical to $|\psi\rangle$
and some not, we have following theorem.
\begin{theorem}
For a parameterized pure state $|\psi\rangle$, assume a set of projectors
$\{|m_{k}\rangle\langle m_{k}|\}$ include two subsets $A=\{|m_{k}\rangle\langle m_{k}|,
\langle\psi|m_{k}\rangle=0~\forall k\}$ and $B=\{|m_{k}\rangle\langle m_{k}|,
\langle\psi|m_{k}\rangle\neq 0~\forall k\}$, i.e., $\{|m_{k}\rangle\langle m_{k}|\}=A\cup B$,
then it is an optimal measurement to let the CFIM to reach the QFIM if and
nly if~\cite{Pezze2017} equation~(\ref{eq:optimal_measurement_0}) is fulfilled
for all the projectors in set $A$ and~(\ref{eq:optimal_measurement_1}) is
fulfilled for all the projectors in set $B$.
\end{theorem}

Apart from the QFIM, the CFIM is also bounded by a measurement-dependent matrix with the $ab$th
entry $\sum_{k}\mathrm{Re}[\mathrm{Tr}(\rho L_{a}\Pi_{k}L_{b})]$~\cite{Yang2018}, where
$\{\Pi_{k}\}$ is a set of POVM. Recently, Yang et al.~\cite{Yang2018} provided the attainable
conditions for the CFIM to attain this bound by generalizing the approach in
reference~\cite{Braunstein1994}.

\subsection{Phase estimation in the Mach-Zehnder interferometer}

Mach-Zehnder interferometer (MZI) is one of the most important model in quantum technology.
It was first proposed in 1890th as an optical interferometer, and its quantum description
was given in 1986~\cite{Yurke1986}. With the development of quantum mechanics,
now it is not only a model for optical interferometer, but can also be realized
via other systems like spin systems and cold atoms. The recently developed gravitational
wave detector GEO 600 can also be mapped as a MZI in the absence of noise~\cite{Rafal2013}.
It is a little bit more complicated when the noise is involved, for which a valid bound has been
provided by Branford et al.~\cite{Branford2018} and is attainable by a frequency-dependent
homodyne detection. Phase estimation in MZI is the earliest case showing quantum advantages in metrology.
In 1981, Caves~\cite{Caves1981} showed that there exists an unused port in the MZI due
to quantum mechanics and the vacuum fluctuation in that port actually affects the
phase precision and limit it to the standard quantum limit (also known as shot-noise limit),
which is the ultimate limit for a classical apparatus. He continued to point out that
injecting a squeezed state in the unused port can lead to a high phase precision
beating the standard quantum limit. This pioneer work proved that quantum technologies
can be powerful in the field of precision measure, which was experimentally confirmed
in 1987~\cite{Xiao1987,Grangier1987}. Since then, quantum metrology has been seeing
a rapid development and grown into a major topic in quantum technology.

\begin{figure}[tp]
\centering\includegraphics[width=10cm]{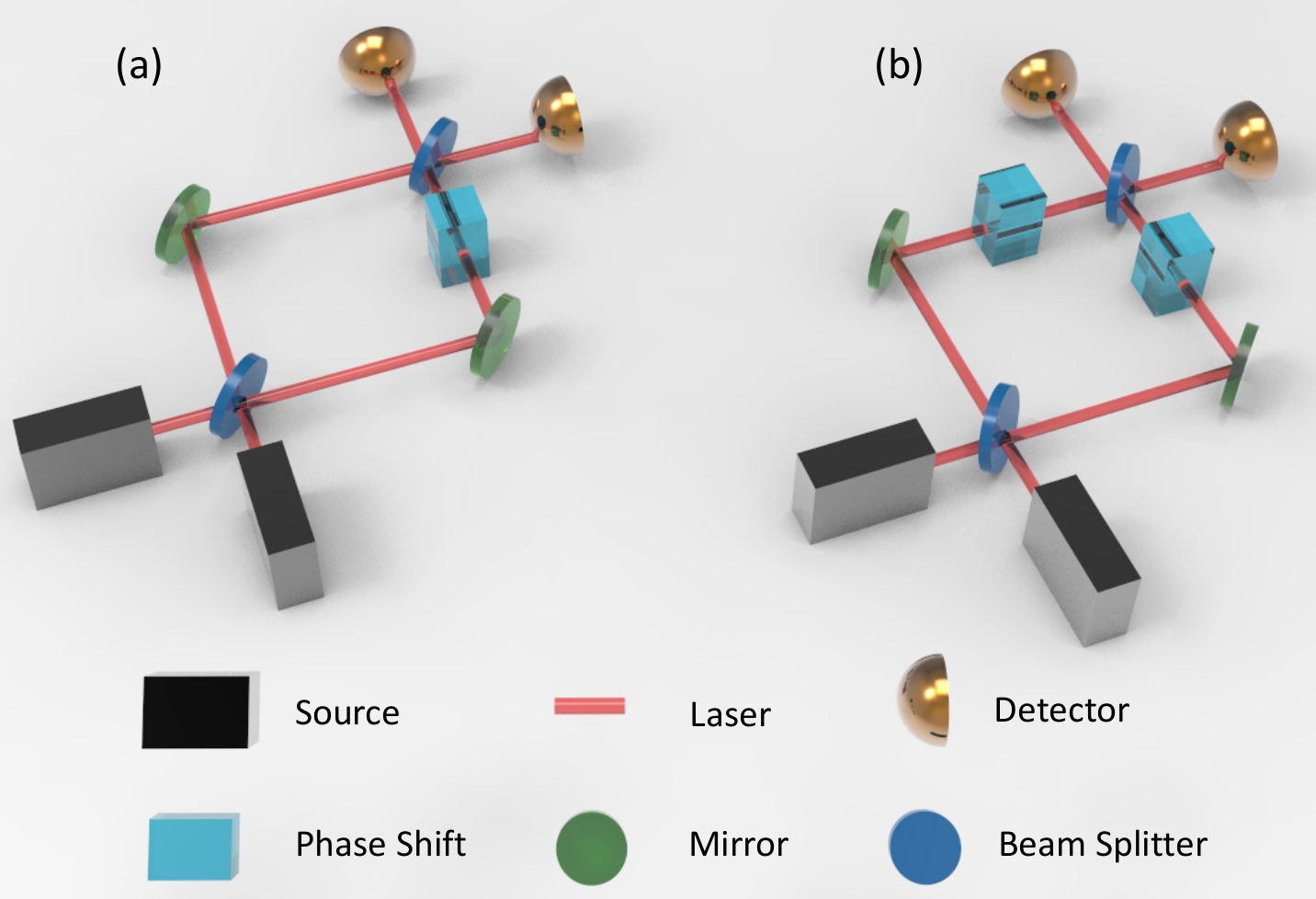}
\caption{(a) Schematic for a standard Mach-Zehnder interferometer, which generally consists of
two beam splitters and a phase shift; (b) Schematic for a double phase Mach-Zehnder interferometer
in which the phase shifts are put in all two paths. Here we emphasize that though the schematic is
given in an optical setup, the MZI model can also be realized via other systems like spin systems
and cold atoms.}
\label{fig:MZ_schematic}
\end{figure}

MZI is a two-path interferometer, which generally consists of two beam splitters and a
phase shift between them, as shown in Fig.~\ref{fig:MZ_schematic}(a). In theory, the beam
splitters and phase shift are unitary evolutions. A 50:50 beam splitter can
be theoretically expressed by $B=\exp(-i\frac{\pi}{2}J_{x})$ where$J_{x}=\frac{1}{2}
(\hat{a}^{\dagger}\hat{b}+\hat{b}^{\dagger}\hat{a})$ is a Schwinger operator with $\hat{a}$,
$\hat{b}$ ($\hat{a}^{\dagger}$, $\hat{b}^{\dagger}$) the annihilation (creation) operators
for two paths. Other Schwinger operators are $J_{y}=\frac{1}{2i}(\hat{a}^{\dagger}\hat{b}
-\hat{b}^{\dagger}\hat{a})$ and $J_{z}=\frac{1}{2}(\hat{a}^{\dagger}\hat{a}-\hat{b}^{\dagger}
\hat{b})$. The Schwinger operators satisfy the commutation $[J_{i}, J_{j}]=i\sum_{k}\epsilon_{ijk}J_{k}$
with $\epsilon_{ijk}$ the Levi-Civita symbol. The second beam splitter usually takes the form
as $B^{\dagger}$. For the standard MZI, the phase shift can be modeled as $P=\exp(i\theta J_{z})$,
which means the entire operation of MZI is $BPB^{\dagger}=\exp(-i\theta J_{y})$, which is a
SU(2) rotation, thus, this type of MZI is also referred as SU(2) interferometer. If the input state
is a pure state $|\psi_{0}\rangle$, the QFI for $\theta$ is just the variance
of $J_{y}$ with respect to $|\psi_{0}\rangle$, i.e., $\mathrm{var}_{|\psi_{0}\rangle}(J_{y})$.

\subsubsection{Double-phase estimation}
A double-phase MZI consists of two beam splitters and a phase shift in each path, as shown
in figure~\ref{fig:MZ_schematic}(b). In this setup, the operator of the two phase shifts is
$P(\phi_{a},\phi_{b})=\exp(i(\phi_{a}\hat{a}^{\dagger}\hat{a}+\phi_{b}\hat{b}^{\dagger}\hat{b}))$.
According to Proposition~\ref{prop:QFIM}, the QFIM for the phases is not affected by the second
beam splitter $B^{\dagger}$ since it is independent of the phases. Thus, the second
beam splitter can be neglected for the calculation of QFIM. Now take
$\phi_{\mathrm{tot}}=\phi_{a}+\phi_{b}$ and $\phi_{\mathrm{d}}=\phi_{a}-\phi_{b}$
as the parameters to be estimated. For a separable input state
$|\alpha\rangle\otimes|\chi\rangle$ with $|\alpha\rangle$ as
a coherent state, the entries of QFIM reads~\cite{Lang2013}
\begin{eqnarray}
\mathcal{F}_{\phi_{\mathrm{tot}},\phi_{\mathrm{tot}}}&=&
|\alpha|^{2}+\mathrm{var}_{|\chi\rangle}(b^{\dagger}b), \\
\mathcal{F}_{\phi_{\mathrm{d}},\phi_{\mathrm{d}}}&=&
2|\alpha|^{2}\mathrm{cov}_{|\chi\rangle}(b,b^{\dagger})
-2\mathrm{Re}\left(\alpha^{2}\mathrm{var}_{|\chi\rangle}(b^{\dagger})\right)+\langle
b^{\dagger}b\rangle, \\
\mathcal{F}_{\phi_{\mathrm{tot}},\phi_{\mathrm{d}}}&=& -i\alpha^{*}\langle
b\rangle-i\mathrm{Im}(\alpha^{*}(\langle
b^{\dagger}b^{2}\rangle
-\langle b^{\dagger}b\rangle\langle b\rangle)).
\end{eqnarray}
Focusing on the maximization of $\mathcal{F}_{\phi_{\mathrm{d}},\phi_{\mathrm{d}}}$
subject to a constraint of fixed mean photon number on $b$ mode, Lang and
Caves~\cite{Lang2013} proved that the squeezed vacuum state is the optimal choice
for $|\chi\rangle$ which leads to the maximum sensitivity of $\phi_{\mathrm{d}}$.

Since Caves already pointed out that the vacuum fluctuation will affect the phase
sensitivity~\cite{Caves1981}, it is then interesting to ask the question that how
bad the phase sensitivity could be if one input port keeps vacuum? Recently Takeoka
et al.~\cite{Takeoka2017} considered this question and gave the answer by proving
a no-go theorem stating that in the double phase MZI, if one input port is the vacuum
state, the sensitivity can never be better than the standard quantum limit regardless
of the choice of quantum state in the other port and the detection scheme. However,
this theorem does not hold for a single phase shift scenario in
figure~\ref{fig:MZ_schematic}(a). Experimentally, Polino et al.~\cite{Polino2019}
recently demonstrated quantum-enhanced double-phase estimation with a photonic chip.

Besides the two-phase estimation, another practical two-parameter scenario in
interometry is the joint measurement of phase and phase diffusion.
In 2014, Vidrighin et al.~\cite{Vidrighin2014} provide a trade-off bound on
the statistical variances for the joint estimation of phase and phase diffusion.
Later in 2015, Altorio et al.~\cite{Altorio2015} addressed the usefulness of
weak measurements in this case and in 2018 Hu et al.~\cite{Hu2018} discussed
the SU(1,1) interferometry in the presence of phase diffusion. In the same year,
Roccia et al.~\cite{Roccia2018} experimentally showed that some collective measurement,
like Bell measurement, can benefit the joint estimation of phase and phase diffusion.

\subsubsection{Multi-phase estimation}

\begin{figure}[tp]
\centering\includegraphics[width=10cm]{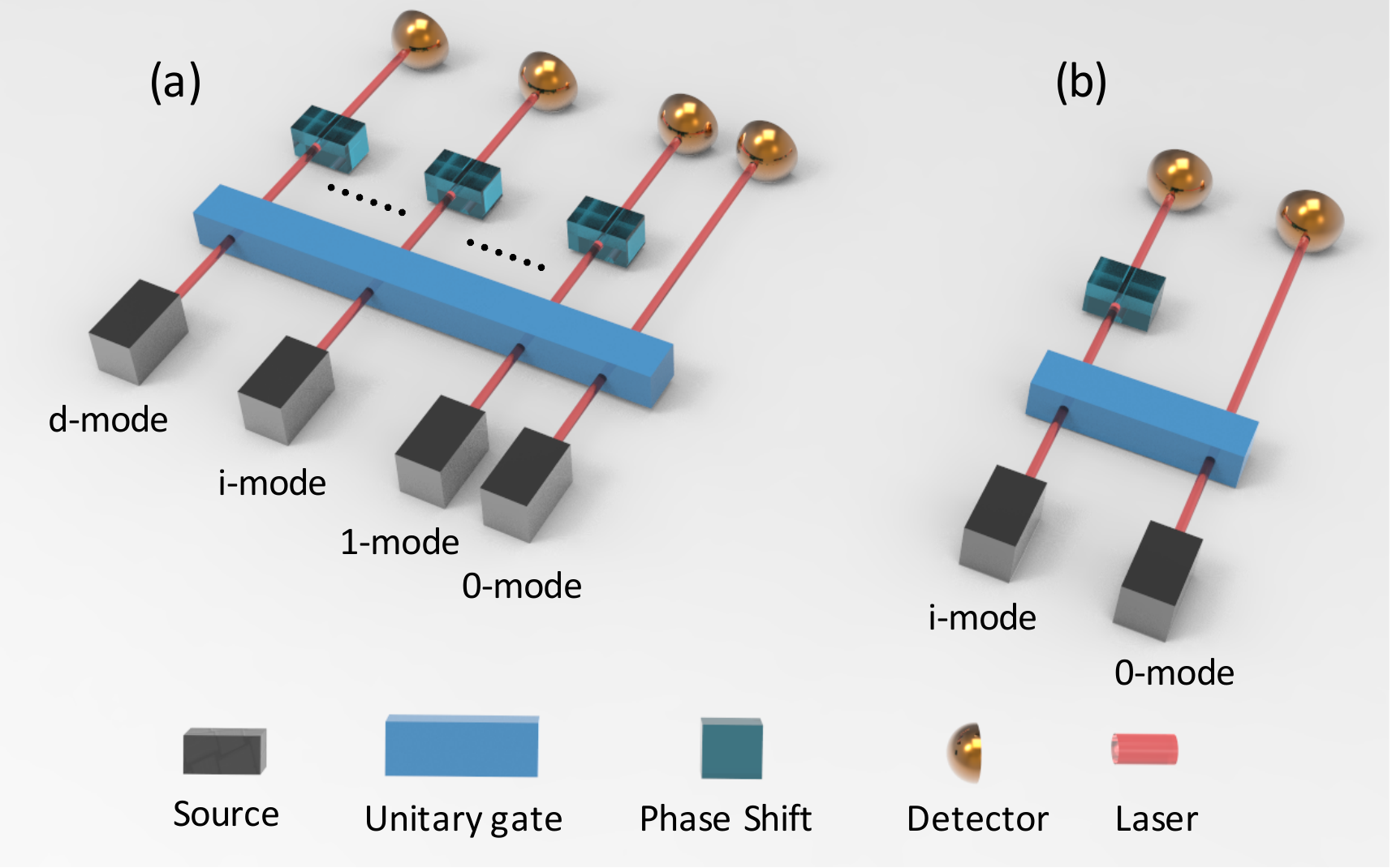}
\caption{Schematics of (a) simultaneous multi-phase estimation and (b) independent
estimation of multiple phases. $0$-mode light is the reference mode. In the independent
estimation, the phase in the $i$th mode is estimated via the MZI consisting of $0$-mode
and $i$-mode lights.}
\label{fig:multiphase_schematic}
\end{figure}

Apart from double phase estimation, multi-phase estimation is also an important
scenario in multiparameter estimation. The multi-phase estimation is usually
considered in the multi-phase interferometer shown in figure~\ref{fig:multiphase_schematic}(a).
Another recent review on this topic is reference~\cite{Szczykulska2016}. In this case,
the total variance of all phases is the major concern, which is bounded by
 $\frac{1}{n}\mathrm{Tr}(\mathcal{F}^{-1})$ according to Corollary~\ref{CR_lowbound1}.
For multiple phases, the probe state undergoes the parameterization, which can be
represented by $U=\exp(i \sum_{j}x_{j}H_{j})$ where $H_{j}$ is the generator of the $j$th mode.
In the optical scenario, this operation can be chosen as $\exp(i \sum_{j}x_{j}a^{\dagger}_{j}a_{j})$
with $a^{\dagger}_{j}$ ($a_{j}$) as the creation (annihilation) operator for the $j$th mode.
For a separable state $\sum_{k}p_{k}\bigotimes^{d}_{i=0}\rho^{(i)}_{k}$ where
$\rho^{(i)}_{k}$ is a state of the $i$th mode, and $p_{k}$ is the weight
with $p_{k}>0$ and $\sum_{k}p_{k}=1$, the QFI
satisfies $\mathcal{F}_{jj}\leq d (h_{j,\mathrm{max}}-h_{j,\mathrm{min}})^{2}$~\cite{Ciampini2016}
with $h_{j,\mathrm{max}}$ ($h_{j,\mathrm{min}}$) as the maximum (minimum) eigenvalue of $H_{j}$.
From Corollary~\ref{CR_lowbound2}, the total variance is then bounded by~\cite{Ciampini2016}
\begin{equation}
\sum_{j}\mathrm{var}(\hat{x}_{j}, \{\Pi_{y}\})\geq \frac{1}{d}\sum^{d}_{j=1}
\frac{1}{(h_{j,\mathrm{max}}-h_{j,\mathrm{min}})^{2}}.
\end{equation}
This bound indicates that the entanglement is crucial in the multi-phase estimation
to beat the standard quantum limit.

N00N state is a well known entangled state in quantum metrology which can saturate
the Heisenberg limit. In 2013, Humphreys et al.~\cite{Humphreys2013}
discussed a generalized $(d+1)$-mode N00N state in the form
$c_{0}|N_{0}\rangle+c_{1}\sum_{i}|N_{i}\rangle$ where $|N_{i}\rangle=|0...0N0...0\rangle$
is the state in which only the $i$th mode corresponds to a Fock state and all other modes
are left vacuum. $c_{0}$ and $c_{1}$ are real coefficients satisfying $c^{2}_{0}+dc^{2}_{1}=1$.
The 0-mode is the reference mode and the parametrization process is
$\exp(i \sum^{d}_{j=1}x_{j}a^{\dagger}_{j}a_{j})$.
The generation of this state has been proposed in reference~\cite{Zhang2018}.
Since this process is unitary, the QFIM can be calculated via Corollary~\ref{corollary_unitary}
with $\mathcal{H}_{j}=-a^{\dagger}_{j}a_{j}$. It
is then easy to see that the entry of QFIM reads~\cite{Humphreys2013} $\mathcal{F}_{ij}
=4N^{2}(\delta_{ij}c^{2}_{1}-c^{4}_{1})$, which further gives
\begin{equation}
\min_{c_{1}}\mathrm{Tr}(\mathcal{F}^{-1})=\frac{(1+\sqrt{d})^{2}d}{4N^{2}},
\end{equation}
where the optimal $c_{1}$ is $1/\sqrt{d+\sqrt{d}}$.

Apart from the scheme in figure~\ref{fig:multiphase_schematic}(a) with the simultaneous estimation,
the multi-phase estimation can also be performed by estimating the phases independently,
using the $i$th mode and the reference mode, as shown in figure~\ref{fig:multiphase_schematic}(b).
In this scheme, the phases are estimated one by one with the N00N state, which provides
the total precision limit $d^{3}/N^{2}$~\cite{Humphreys2013}. Thus, the simultaneous
estimation scheme in Fig.~\ref{fig:multiphase_schematic}(a) shows a $\mathcal{O}(d)$
advantage compared to the independent scheme in figure~\ref{fig:multiphase_schematic}(b).
However, the performance of the simultaneous estimation may be strongly affected by the
noise~\cite{Yao2014,Yao2014a} and the $\mathcal{O}(d)$ advantage may even disappear
under the photon loss noise~\cite{Yue2014}.

In the single phase estimation, it is known that the entangled coherent state
$\mathcal{N}(|0\alpha\rangle+|\alpha 0\rangle)$ ($|\alpha\rangle$ is a coherent
state and $\mathcal{N}$ is the normalization) can provide a better precision limit
than the N00N state~\cite{Joo2011}. Hence, it is reasonable to think the generalization
of entangled coherent state may also outperform the generalized N00N state. This result
was theoretically confirmed in reference~\cite{Liu2014a}. For a generalized entangled
coherent state written in the form $c_{0}|\alpha_{0}\rangle+c_{1}\sum^{d}_{j=1}|\alpha_{j}\rangle$
with $|\alpha_{j}\rangle=|0...0\alpha0...0\rangle$, the QFIM is $\mathcal{F}_{ij}=4|c_{1}|^{2}
|\alpha|^{2}[\delta_{ij}(1+|\alpha|^{2})-|c_{1}|^{2}|\alpha|^{2}]$~\cite{Liu2014a}.
For most values of $d$ and $|\alpha|$, the minimum $\mathrm{Tr}(\mathcal{F}^{-1})$
can be written as
\begin{equation}
\min_{c_{1}}\mathrm{Tr}(\mathcal{F}^{-1})=\frac{(1+\sqrt{d})^{2}d}{4(1+|\alpha|^{2})^{2}},
\end{equation}
which is smaller than the counterpart of the generalized N00N state, indicating
the performance of the generalized entangled coherent state is better than the
generalized N00N state. For a large $|\alpha|$, the total particle number $\sim |\alpha|^{2}$
and the performances of both states basically coincide. With respect to the independent scheme,
the generalized entangled coherent also shows a $\mathcal{O}(d)$ advantage in the absence
of noise.

In 2017, Zhang et al.~\cite{zhang2017} considered a general balanced state $c\sum_{j}|\psi_{j}\rangle$
with $|\psi_{j}\rangle=|0...0\psi0...0\rangle$. $c$ is the normalization coefficient. Four specific
balanced states: N00N state, entangled coherent state, entangled squeezed vacuum state and entangled
squeezed coherent state, were calculated and they found that the entangled squeezed vacuum state
shows the best performance, and the balanced type of this state outperforms the unbalanced one
in some cases.

A linear network is another common structure for multi-phase estimation~\cite{Knott2016,Ge2018}.
Different with the structure in figure~\ref{fig:multiphase_schematic}(a), a network requires the
phase in each arm can be detected independently, meaning that each arm needs a reference beam.
The calculation of Cram\'{e}r-Rao bound in this case showed~\cite{Ge2018} that the linear network
for miltiparameter metrology behaves classically even though endowed with well-distributed quantum
resources. It can only achieve the Heisenberg limit when the input photons are concentrated in a
small number of input modes. Moreover, the performance of a mode-separable state
$\mathcal{N}(|N\rangle+v|0\rangle)$ may also shows a high theoretical precision limit in this case
if $v\propto \sqrt{d}$.

Recently, Gessner~\cite{Gessner2018a} considered a general case for multi-phase estimation with
multimode interferometers. The general Hamiltonian is $H=\sum^{N}_{j=1}h^{(j)}_{k}$ with $h^{(j)}_{k}$
a local Hamiltonian for the $i$th particle in the $k$th mode. For the particle-separable states,
the maximum QFIM is a diagonal matrix with the $i$th diagonal entry $\langle n_{i}\rangle$
being the $i$th average particle number ($n_{i}$ is the $i$th particle number operator).
This bound could be treated as the shot-noise limit for the multi-phase estimation.
For the mode-separable states, the maximum QFIM is also diagonal, with the $i$th entry
$\langle n^{2}_{i}\rangle$, which means the maximum QFIM for mode-separable state is larger
than the particle-separate counterpart. Taking into account both the particle and mode
entanglement, the maximum QFIM is in the form $\mathcal{F}_{ij}=\mathrm{sgn}
(\langle n_{i}\rangle)\mathrm{sgn}(\langle n_{j}\rangle)\sqrt{\langle n^{2}_{i}\rangle
\langle n^{2}_{j}\rangle}$, which gives the Heisenberg limit in this case.

The comparison among the performances of different strategies usually requires the same
amount of resources, like the same sensing time, same particle number and so on.
Generally speaking, the particle number here usually refers to the average
particle number, which is also used to define the standard quantum limit and
Heisenberg limit in quantum optical metrology. Although the Heisenberg limit is
usually treated as a scaling behavior, people still attempt to redefine it as an
ultimate bound given via quantum mechanics by using the expectation of the square
of number operator to replace the square of average particle number~\cite{Hyllus2010},
i.e., the variance should be involved in the Heisenberg limit~\cite{Hyllus2010,Pasquale2015}
and be treated as a resource.

\subsection{Waveform estimation}

In many practical problems, like the detection of gravitational waves~\cite{Branford2018}
or the force detection~\cite{Tsang2012}, what needs to be estimated is not a parameter,
but a time-varing signal. The QFIM also plays an important role in the estimation of
such signals, also known as waveform estimation~\cite{Tsang2011}.
By discretizing time into small enough intervals, the estimation of a
time-continuous signal $x(t)$ becomes the estimation of multiparameters $x_j$'s.
The prior information of the waveform has to be taken into account in the estimation
problem, e.g., restricting the signal to a finite bandwidth, in order to make the
estimation error well-defined. The estimation-error covariance matrix $\Sigma$ is
then given by
\begin{equation}
\Sigma_{jk} = \int [\hat x_j(y)-x_j][\hat x_k(y)-x_k]\, \mathrm{d}y \mathrm{d}\vec x,
\end{equation}
where $\mathrm{d}\vec x =\prod_j \mathrm{d}x_j$ and $y$ denotes the measurement outcome.
Tsang proved the most general form of Bayesian quantum Cram\'er-Rao bound
\begin{equation}
\Sigma
\geq \left(\mathcal{F}^\mathrm{(C)}+\mathcal{F}^\mathrm{(Q)}\right)^{-1},
\label{eq:bayesian_qcrb}
\end{equation}
where the classical part
\begin{equation}
\mathcal{F}^\mathrm{(C)}_{ab} = \int\left[\partial_a \ln p(\vec{x})\right]
\left[\partial_b \ln p(\vec{x})\right] p(\vec{x})\,\mathrm{d}\vec{x}
\end{equation}
depends only on the prior information about the vector parameter $\vec x$ and
the quantum part~\cite{Fujiwara1995,Tsang2011}
\begin{equation}
\mathcal{F}^\mathrm{(Q)}_{ab}=\int\mathrm{Re}\left[\mathrm{Tr}(\tilde{L}_a^\dagger
\tilde{L}_b \rho)\right] p(\vec x)\,\mathrm{d}\vec x
\label{eq:FQ_waveform}
\end{equation}
depends on the parametric family $\rho$ of density operators with $\tilde{L}_a$
being determined via
\begin{equation}
\partial_a \rho=\frac{1}{2}\left(\tilde{L}_a\rho+\rho \tilde{L}_a^{\dagger}\right).
\end{equation}
Note that $\tilde{L}_a$ is an extended version of SLD and is not necessary to be Hermitian.
When all $\tilde{L}_a$'s are Hermitian, $\mathcal{F}^\mathrm{(Q)}$ is the average of the
QFIM over the prior distribution of $\vec x$. $\tilde{L}_a$ can also be anti-Hermiaition~\cite{Fujiwara1995}.
In this case, the corresponding bound is equivalent to the SLD one for pure states
but a potentially looser one for mixed states. For a unitary evolution, the entire
operator $U$ can be discretized into $U=U_{m}U_{m-1}\cdots U_{1}U_{0}$, where
$U_{a}=\exp(-iH(x_{a})\delta t)$. Denote $h_{a}=U^{\dagger}_{0}U^{\dagger}_{1}\cdots
U^{\dagger}_{a}(\partial_{a}H)U_{a}\cdots U_{1}U_{0}$ and
$\Delta h_{a}=h_{a}-\mathrm{Tr}(\rho_{0}h_{a})$ with $\rho_{0}$ the probe state,
the quantum part in equation~(\ref{eq:FQ_waveform}) reads~\cite{Tsang2011}
\begin{equation}
\mathcal{F}^{(Q)}_{ab}=2(\delta t)^{2}\int\mathrm{Tr}\left(\left\{\Delta h_{a},
\Delta h_{b}\right\}\rho_{0}\right)p(\vec{x})\mathrm{d}\vec{x}.
\end{equation}
Taking the continuous-time limit, it can be rewritten into~\cite{Tsang2011}
\begin{equation}
\mathcal{F}^{(Q)}(t,t^{\prime})=2\int\mathrm{Tr}\left(\left\{\Delta h(t),
\Delta h(t^{\prime})\right\}\rho_{0}\right)p(\vec{x})\mathrm{d}\vec{x}.
\end{equation}
Together with $\mathcal{F}^{(C)}(t,t^{\prime})$, the fundamental quantum limit to
waveform estimation based on QFIM is established~\cite{Tsang2011}. The waveform
estimation can also be solved with other tools, like the Bell-Ziv-Zakai lower
bounds~\cite{Berry2015}.

\subsection{Control-enhanced multiparameter estimation}


The dynamics of many artificial quantum systems, like the Nitrogen-vacancy center, trapped
ion and superconducting circuits, can be precisely altered by control. Hence, control
provides another freedom for the enhancement of the precision limit in these apparatuses.
It is already known that quantum control can help to improve the QFI to the Heisenberg
scaling with the absence of noise in some scenarios~\cite{Yuan2015,Pang2017,Yang2017}.
However, like other resources, this improvement could be sensitive to the noise, and
the performance of optimal control may strongly depend on the type of
noise~\cite{LiuGRAPE1,LiuGRAPE2}. In general, the master equation for a noisy
quantum system is described by
\begin{equation}
\partial_{t}\rho =\mathcal{E}_{\vec{x}}\rho.
\end{equation}
Here $\mathcal{E}_{\vec{x}}$ is a $\vec{x}$-dependent superoperator. For the Hamiltonian
estimation under control, the dynamics is
\begin{equation}
\mathcal{E}_{\vec{x}}\rho =-i[H_{0}(\vec{x})+H_{\mathrm{c}},\rho ]+\mathcal{L}\rho ,
\end{equation}
where $H_{\mathrm{c}}=\sum^{p}_{k=1}V_{k}(t)H_{k}$ is the control Hamiltonian with $H_{k}$
the $k$th control and $V_{k}(t)$ the corresponding time-dependent control amplitude.
For a general Hamiltonian, the optimal control can only be tackled via numerical methods.
One choice for this problem is the Gradient ascent pulse engineering algorithm.

\begin{figure}[tp]
\centering\includegraphics[width=10cm]{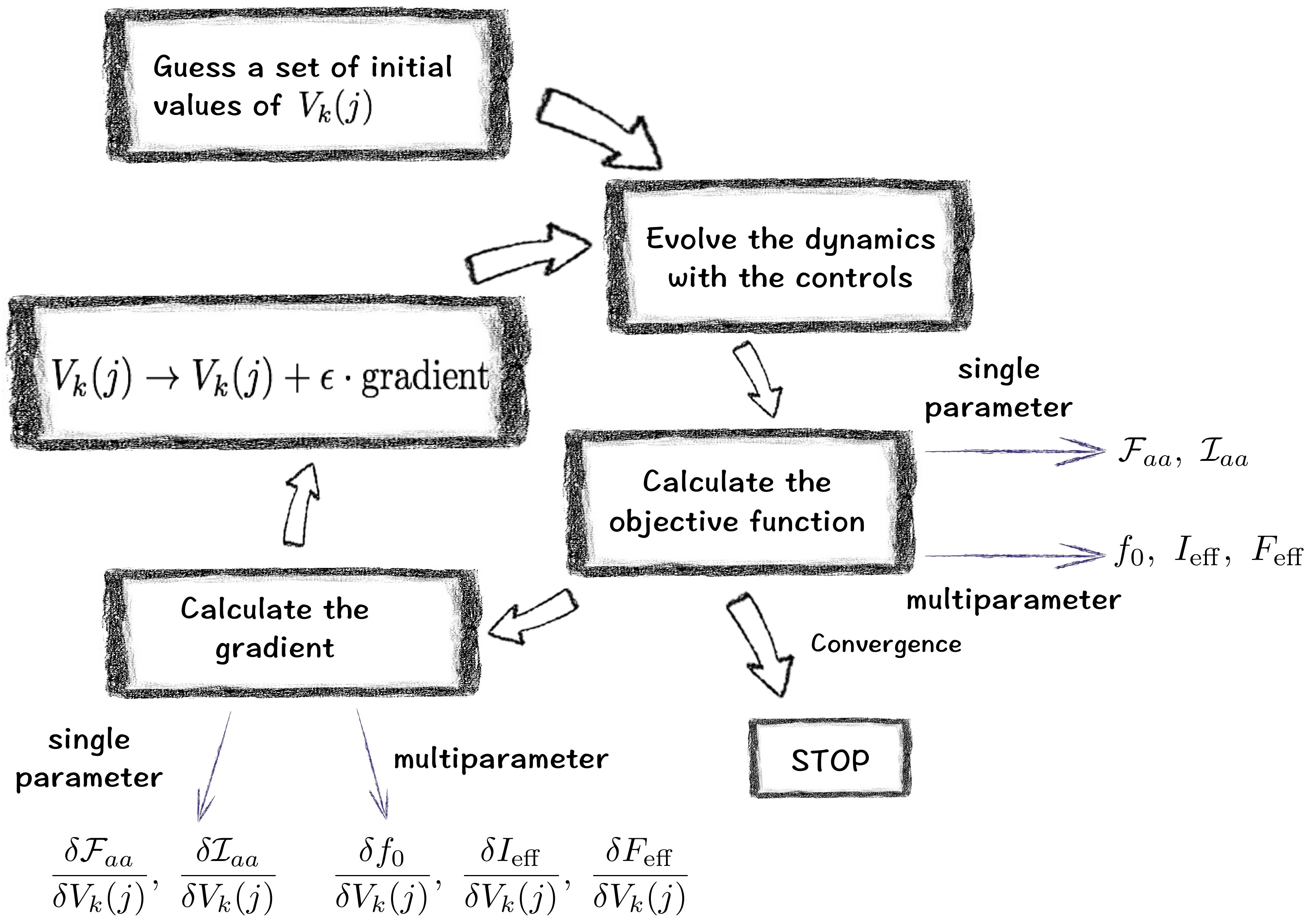}
\caption{Flow chart of GRAPE algorithm for controlled single-parameter
and multiparameter estimation. Reprinted figure with permission
from~\cite{LiuGRAPE1,LiuGRAPE2}. Copyright (2019) by the American Physical Society.}
\label{fig:GRAPE_schematic}
\end{figure}

Gradient ascent pulse engineering (GRAPE) algorithm is a gradient-based algorithm, which
was originally developed to search the optimal control for the design of nuclear magnetic
resonance pulse sequences~\cite{Khaneja2005}, and now is extended to the scenario of
quantum parameter estimation~\cite{LiuGRAPE1,LiuGRAPE2}. As a gradient-based method,
GRAPE requires an objective function and analytical expression of the corresponding
gradient. For quantum single-parameter estimation, the QFI is a natural choice for
the objective function. Of course, in some specific systems the measurement
methods might be very limited, which indicates the CFI is a proper objective function
in such cases. For quantum multiparameter estimation, especially those with a large
parameter number, it is difficult or even impossible to take into account the error of every
parameter, and thus the total variance $\sum_{a}\mathrm{var}(\hat{x}_{a},\{\Pi_{y}\})$
which represents the average error of all parameters, is then a good index to show
system precision. According to Corollary~\ref{CR_lowbound1},
$\mathrm{Tr}(\mathcal{F}^{-1})$ and $\mathrm{Tr}(\mathcal{I}^{-1})$ (for fixed measurement)
could be proper objective functions for GRAPE if we are only concern with the total variance.
For two-parameter estimation, $\mathrm{Tr}(\mathcal{F}^{-1})$ and $\mathrm{Tr}(\mathcal{I}^{-1})$
reduce to the effective QFI $F_{\mathrm{eff}}$ and CFI $I_{\mathrm{eff}}$ according to
Corollary~\ref{two_parameter_theorem}. However, as the parameter number increases,
the inverse matrix of QFIM and CFIM are difficult to obtain analytically,
and consequently superseded objective functions are required. $(\sum_{a}\mathcal{I}_{aa})^{-1}$
or $(\sum_{a}\mathcal{F}_{aa})^{-1}$, based on corollary~\ref{CR_lowbound2}, is then
a possible superseded objective function for fixed measurement. However, the use of
$(\sum_{a}\mathcal{F}_{aa})^{-1}$ should be very cautious since
$\mathrm{Tr}(\mathcal{F}^{-1})$ cannot always be achievable.
The specific expressions of gradients for these objective functions in Hamiltonian
estimation are given in~\ref{apx:GRAPE_gradient}.

The flow of the algorithm, shown in figure~\ref{fig:GRAPE_schematic}, is formulated
as follows~\cite{LiuGRAPE1,LiuGRAPE2}:
\begin{enumerate}
\item Guess a set of initial values for $V_{k}(j)$ [$V_{k}(j)$ is the $k$th control
at the $j$th time step].
\item Evolve the dynamics with the controls.
\item Calculate the objective function. For single-parameter estimation, the objective
function can be chosen as QFI $\mathcal{F}_{aa}$ or CFI $\mathcal{I}_{aa}$ (for mixed measurement).
For two-parameter estimation, it can be chosen as $I_{\mathrm{eff}}$ or $F_{\mathrm{eff}}$.
For large parameter number, it can be chosen as $f_{0}=\frac{1}{\sum_{a}\mathcal{I}^{-1}_{aa}}$
or $\frac{1}{\sum_{a}\mathcal{F}^{-1}_{aa}}$.
\item Calculate the gradient.
\item Update $V_{k}(j)$ to $V_{k}(j)+\epsilon\cdot\mathrm{gradient}$ ($\epsilon$ is a small
quantity) for all $j$ simultaneously.
\item Go back to step (ii) until the objective function converges.
\end{enumerate}

\begin{figure}[tp]
\centering\includegraphics[width=14cm]{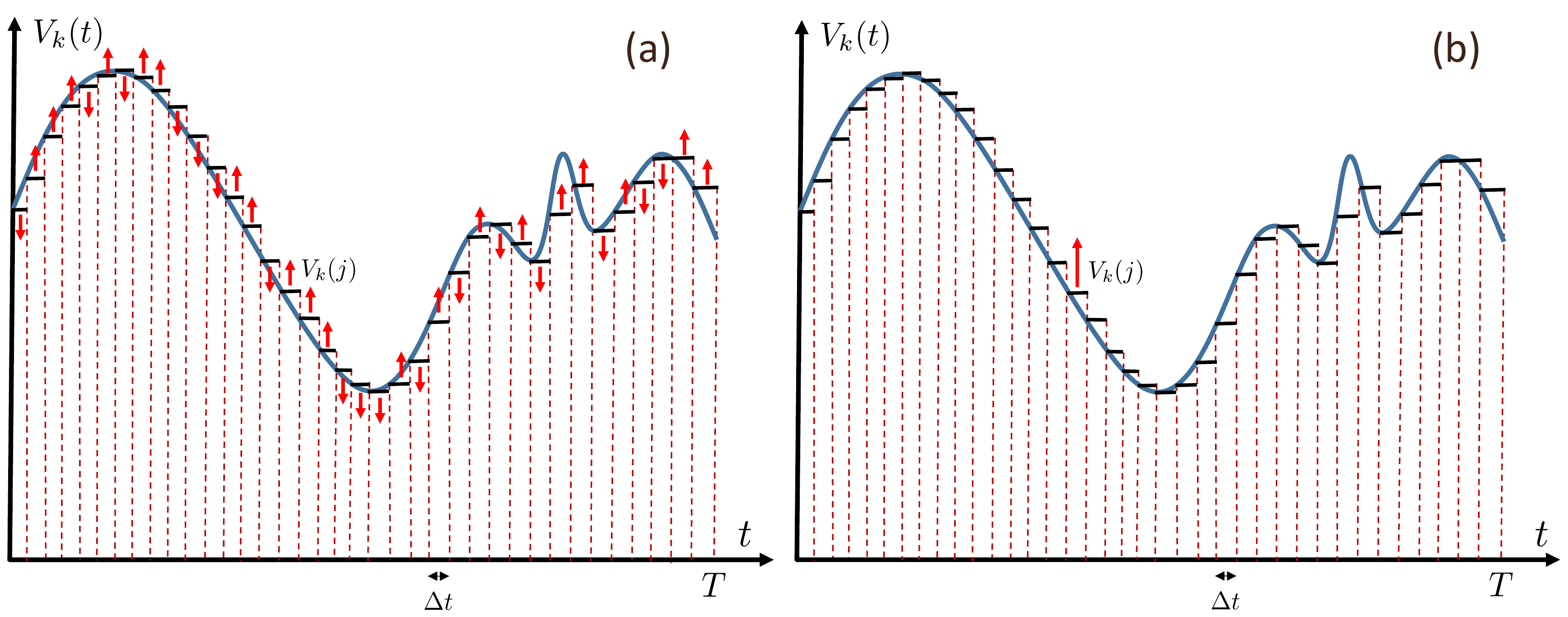}
\caption{(a) In GRAPE, the entire evolution time $T$ is cut into $m$ parts with time
interval $\Delta t$, i.e., $m \Delta t=T$. $V_{k}(t)$ within the $j$th time interval
is denoted as $V_{k}(j)$ and is assumed to be a constant. All the $V_{k}(j)$ are
simultaneously updated in each iteration. (b) Krotov's method updates one
$V_{k}(j)$ in each iteration.}
\label{fig:GRAPE_up}
\end{figure}

In step (v), all $V_{k}(j)$ are updated simultaneously in each time of
iteration~\cite{Khaneja2005}, as shown in figure~\ref{fig:GRAPE_up}(a),
which could improve the speed of convergence in some cases. However,
this parallel update method does not promise the monotonicity of convergence,
and the choice of initial guess of $V_{k}(j)$ is then important for a convergent
result. The specific codes for this algorithm can be found in reference~\cite{quanestimation}.

Krotov's method is another gradient-based method in quantum control, which promises
the monotonicity of convergence during the iteration~\cite{Reich2012}. Different from
GRAPE, only one $V_{k}(j)$ is updated in each iteration in Krotov's method, as shown
in figure~\ref{fig:GRAPE_up}(b). This method can also be extended to quantum
parameter estimation with the aim of searching the optimal control for a high precision
limit. It is known that the gradient-based methods can only harvest the local extremals.
Thus, it is also useful to involve gradient-free methods, including Monte Carlo method
and Particle Swarm Optimization, into the quantum parameter estimation, or apply a
hybrid method combining gradient-based and gradient-free methods as discussed
in reference~\cite{Goerz2015}.

\subsection{Estimation of a magnetic field}

Measurement of the magnetic fields is an important application of quantum metrology, as it
promises better performances than the classical counterparts. Various physical systems have
been used as quantum magnetometers, including but not limited to nitrogen-vacancy
centers~\cite{Rondin2014,Schloss2018}, optomechanical systems~\cite{Li2018},
and cold atoms~\cite{Sheng2013}.

A magnetic field can be represented as a vector,  $\vec{B}=B(\cos\theta\cos\phi,\cos\theta\sin\phi,
\sin\phi)$, in the spherical coordinates, where $B$ is the amplitude, and
$\theta,\phi$ define the direction of the field. Therefore, the estimation of a magentic field
is in general a three-parameter estimation problem. The estimation of the amplitude
with known angles is the most widely-studied case in quantum metrology, both in theory
and experiments. For many quantum systems, it is related to the estimation of the strength of
system-field coupling. In the case where one angle, for example $\phi$ is known,
the estimation of the field becomes a two-parameter estimation problem. The simplest quantum
detector for this case is a single-spin system~\cite{Pang2014,LiuSR}. An example of the
interaction Hamiltonian is $H=-B\vec{n}_{0}\cdot\vec{\sigma}$ with $n_{0}=(\cos\theta,0,\sin\theta)$
and $\vec{\sigma}=(\sigma_{x},\sigma_{y},\sigma_{z})$. The QFIM can
then be obtained by making use of the expressions of $\mathcal{H}$
operators~\cite{Pang2014,LiuSR}
\begin{equation}
\mathcal{H}_{B}=t\vec{n}_{0}\cdot\vec{\sigma},
~~\mathcal{H}_{\theta}=-\frac{1}{2}\sin (B t)\vec{n}_{1}\cdot\vec{\sigma},
\end{equation}
where $n_{1}=\left(\cos(B t )\sin\theta,\sin (B t),
-\cos (B t)\cos\theta \right)$. The entries of QFIM then read
\begin{eqnarray}
\mathcal{F}_{\theta\theta} &=& \sin^{2}(B t)[1-(\vec{n}_{1}\cdot\vec{r}_{\mathrm{in}})^{2}], \\
\mathcal{F}_{BB} &=& 4 t^{2}[1-(\vec{n}_{0}\cdot\vec{r}_{\mathrm{in}})^{2}], \\
\mathcal{F}_{B\theta} &=& 2t\sin(B t)(\vec{n}_{0}\cdot\vec{r}_{\mathrm{in}})
(\vec{n}_{1}\cdot\vec{r}_{\mathrm{in}}),
\end{eqnarray}
where $\vec{r}_{\mathrm{in}}$ is the Bloch vector of the probe state. The maximal
values of $\mathcal{F}_{\theta\theta}$ and $\mathcal{F}_{BB}$ can be attained when
$\vec{r}_{\mathrm{in}}$ is vertical to both $\vec{n}_{0}$ and $\vec{n}_{1}$.
However, as a two-parameter estimation problem, we have to check the value of
$\langle\psi_{0}|[\mathcal{H}_{B},\mathcal{H}_{\theta}]|\psi_{0}\rangle$ due to
Corollary~\ref{corollary:attain_U}. Specifically, for a pure probe state it reads
$\langle\psi_{0}|[\mathcal{H}_{B},\mathcal{H}_{\theta}]|\psi_{0}\rangle
=-\sin(Bt)(\vec{n}_{0}\times\vec{n}_{1})\cdot\vec{\sigma}$.
For the time $t=n\pi/B$ ($n=1,2,3...$), the expression above vanishes. However,
another consequence of this condition is that $\mathcal{F}_{\theta\theta}$ also vanishes.
Thus, single-qubit probe may not be an ideal magnetometer when at least one angle
is unknown. One possible candidate is a collective spin system, in which the
$\mathcal{H}$ operator and QFIM are provided in reference~\cite{Jing2015}.
Another simple and practical-friendly candidate is a two-qubit system.
One qubit is the probe and the other one is an ancilla, which does not interact
with the field. Consider the Hamiltonian $H=-\vec{B}\cdot\vec{\sigma}$,
the maximal QFIM in this case is (in the basis $\{B,\theta,\phi\}$)~\cite{Yuan2016}
\begin{equation}
\max\mathcal{F}=
4\left(\begin{array}{ccc}
t^{2} & 0 & 0 \\
0 & \sin^{2}(Bt) & 0 \\
0 & 0  & \sin^{2}(Bt)\cos^{2}\theta
\end{array}\right),
\end{equation}
which can be attained by any maximally entangled state and the Bell measurement
as the optimal measurement. With the assistance of an ancilla,
all three parameters $B,\theta,\phi$ of the field can be simultaneously
estimated. However, $\mathcal{F}_{\theta\theta}$ and $\mathcal{F}_{\phi\phi}$
are only proportional to $\sin^{2}(Bt)$, indicating that unlike
the estimation of the amplitude, a longer evolution time does not always lead to a better precision for
the estimation of $\theta$ or $\phi$.

In 2016, Baumgratz and Datta~\cite{Baumgratz2016} provided a framework
for the estimation of a multidimensional field with
noncommuting unitary generators. Analogous to the optical multi-phase estimation, simultaneous
estimation shows a better performance than separate and individual estimation.

To improve the performance of the probe system, additional control could be employed.
For unitary evolution, reference~\cite{Yuan2016} showed that the performance
can be significantly enhanced by inserting the anti-evolution operator as the
control. Specifically if after each evolution of a period of $\delta t$, we insert
a control which reverses the evolution as $U_c = U^\dagger(\delta t)=e^{i H(\vec{B})\delta t}$,
the QFIM can reach
\begin{equation}
\mathcal{F}=4N^{2}\left(\begin{array}{ccc}
(\delta t)^{2} & 0 & 0 \\
0 & \sin^{2}(B\delta t) & 0 \\
0 & 0  & \sin^{2}(B\delta t)\cos^{2}\theta
\end{array}\right),
\end{equation}
where $N$ is the number of the injected control pulses, and $N\delta t=t$. In practice the control
needs to be applied adaptively as $U_c = e^{iH(\hat{\vec{B}})\delta t}$ with $\hat{\vec{B}}$ as the
estimated value obtained from previous measurement results.
In the controlled scheme, the number of control pulses becomes a resource
for the estimation of all three parameters. For the amplitude, the precision
limit is the same as non-controlled scheme. But for the angles,
the precision limit is significantly improved, especially when $N$ is large.
When $\delta t\rightarrow 0$, the QFIM becomes
\begin{equation}
\mathcal{F}=4\left(\begin{array}{ccc}
t^{2} & 0 & 0 \\
0 & B^{2}t^{2} & 0 \\
0 & 0  & B^{2} t^{2}\cos^{2}\theta
\end{array}\right),
\end{equation}
which reaches the highest precision for the estimation of $B$, $\theta$ and $\phi$
simultaneously. Recently, Hou et al.~\cite{Hou2019} demonstrated this scheme up to
eight controls in an optical platform and demonstrate a precision near the
Heisenberg limit.

Taking into account the effect of noise, the optimal control for the estimation
of the magnetic field is then hard to obtain analytically. The aforementioned
numerical methods like GRAPE or Krotov's method could be used
in this case to find the optimal control. The performance of the controlled scheme
relies on the specific form of noise~\cite{LiuGRAPE1,LiuGRAPE2}, which could be
different for different magnetometers.

In practice, the magnetic field may not be a constant field in space.
In this case, the gradient of the field also needs to be estimated.
The corresponding theory for the estimation of gradients based on
Cram\'{e}r-Rao bound has been established in references~\cite{Apellaniz2018,Altenburg2017}.

\subsection{Other applications and alternative mathematical tools}

Apart from the aforementioned cases, quantum multiparameter estimation
is also studied in other scenarios, including the spinor systems~\cite{Zhuang2018},
unitary photonic systems~\cite{Liu2017}, networked quantum sensors~\cite{Proctor2018},
and the quantum thermometry~\cite{Correa2015}. In the case of noise estimation,
the simultaneous estimation of loss parameters has been studied in
reference~\cite{Nair2018}.

In 2016, Tsang et al.~\cite{Tsang2016} used quantum multiparameter estimation
for a quantum theory of superresolution for two incoherent optical point sources.
With weak-source approximation, the density operator for the optical field in the
imaging plane can be expressed as $\rho = (1-\epsilon)|\mathrm{vac}\rangle\langle\mathrm{vac}|
+\frac{\epsilon}{2}(|\psi_{1}\rangle \langle \psi_{1}|+|\psi_{2}\rangle\langle\psi_{2}|)$,
with $\epsilon$ the average photon number in a temporal mode, $|{\mathrm{vac}}\rangle$
the vacuum state, $|\psi_{j}\rangle = \int \mathrm{d}y\, \psi_{x_j}(y)|y\rangle$
the quantum state of an arrival photon from the point source located at $x_j$ of
the object plane, $\psi_{x_j}(y)$ the wave function in the image plane, and $|y\rangle$
the photon image-plane position eigenstate. Taking the position coordinates to be
one-dimensional, the parameters to be estimated are $x_1$ and $x_2$. The
performance of the underlying quantum measurement can be assessed by the CFIM
and its fundamental quantum limit can be revealed by the QFIM. In this case,
the QFIM is analytically calculated in the 4-dimensional Hilbert subspace spanned
by $|\psi_{1}\rangle$, $|\psi_{2}\rangle$, $|\partial_{1}\psi_{1}\rangle$, and
$|\partial_{2}\psi_{2}\rangle$. For convenience, the two parameters $x_1$ and
$x_2$ can be transformed to the centriod $\theta_1 = (x_1+x_2)/2$ and the
separation $\theta_2=x_2-x_1$. Assuming the imaging system is spatially invariant
such that the point-spread function of the form $\psi_{x_j}(y)=\psi(y-x_j)$ with
real-valued $\psi(y)$, the QFIM with respect to $\theta_1$ and $\theta_2$ is then
given by
\begin{equation}
\mathcal F =
\left(\begin{array}{cc}
4\epsilon (\Delta k^2-\gamma^2) & 0 \\
0 & \epsilon \Delta k^2
\end{array}\right),
\end{equation}
where $\Delta k^2 = \int_{-\infty}^\infty [\partial \psi(y) / \partial y]^2\,\mathrm{d}y$ and
$\gamma = \int_{-\infty}^\infty \psi(y-\theta_2)\partial \psi(y) / \partial y\,\mathrm{d}y$
~\cite{Tsang2016}. With this result, it has been shown that direct imaging performs poorly
for estimating small separations and other elaborate methods like spatial-mode demultiplexing
and image inversion interferometry can be used to achieve the quantum
limit~\cite{Tsang2016,Nair2016}.

The design of enhanced schemes for quantum parameter estimation would inevitably
face optimization processes, including the optimization of probe states,
the parameterization trajectories, the measurement and so on.
If control is involved, one would also need to harvest the optimal control.
Therefore, the stochastic optimization methods, including convex optimization,
Monte Carlo method and machine learning, could be used in quantum parameter estimation.

Machine learning is one of the most promising and cutting-edge methods nowadays.
In recent years, it has been applied to condensed matter physics for phase
transitions~\cite{Wang2016,Nieuwenburg2017}, design of a magneto-optical trap
~\cite{Tranter2018} and other aspects~\cite{Dunjko2018}. With respect to quantum
parameter estimation, in 2010, Hentschel and Sanders~\cite{Hentschel2010}
applied the particle swarm optimization algorithm~\cite{Eberhart1995} in the adaptive
phase estimation to determine the best estimation strategy, which was experimentally
realized by Lumino et al.~\cite{Lumino2018} in 2018. Meanwhile, in 2017, Greplova et
al.~\cite{Greplova2017} proposed to use neural networks to estimate the rates
of coherent and incoherent processes in quantum systems with continuous measurement
records. In the case of designing controlled schemes, similar to gradient-based
methods, machine learning could also help use to find the optimal control protocal
in order to achieve the best precision limit~\cite{Xu2019}. In particular, for quantum
multiparameter estimation, there still lacks of systematic research on the role of
machine learning. For instance, we are not assured if there exists different performance
compared to single-parameter estimation.

As a mathematical tool, quantum Cram\'{e}r-Rao bound is not the only one for quantum
multiparameter estimation. Bayesian approach~\cite{Macieszczak2014,Rzadkowski2017,
Kiilerich2016,Rubio2019,Rubio2019a}, including Ziv-Zakai family~\cite{Tsang2012a,Zhang2014,Berry2015}
and Weiss-Weinstein family~\cite{Tsang2011,Lu2016}, and some other
tools~\cite{Lu2016,LiuNJP,Yang2019} like Holevo bound~\cite{Holevo,Suzuki2016,Albarelli2019},
have also shown their validity in various multiparameter scenarios. These tools
beyond the Cram\'{e}r-Rao bound will be thoroughly reviewed by M. Gu\c{t}\u{a}
et al.~\cite{Guta2018} in the same special issue and hence we do not discuss them
in this paper. Please see reference~\cite{Guta2018} for further reading of this
topic. In some scenarios in quantum parameter estimations, there may exist some
parameters that not of interest but affect the precision of estimating other
parameters of interest, which is usually referred to as the nuisance parameters.
Suzuki et al.~\cite{Suzuki2019} has thoroughly reviewed the quantum state estimation
with nuisance parameters in the same special issue. Please see reference~\cite{Suzuki2019}
for further reading.

\section{Conclusion and outlook}

Quantum metrology has been recognized as one of the most promising quantum
technologies and has seen a rapid development in the past few decades. Quantum
Cram\'{e}r-Rao bound is the most studied mathematical tool for quantum parameter
estimation, and as the core quantity of quantum Cram\'{e}r-Rao bound, the QFIM
has drawn plenty of attentions. It is now well-known that the QFIM is not only
a quantity to quantify the precision limit, but also closely connected to many
different subjects in quantum physics. This makes it an important and fundamental
concept in quantum mechanics.

In recent years, many quantum multiparameter estimation schemes have been proposed
and discussed with regard to various quantum systems, and some of them have shown
theoretical advances compared to single-parameter schemes. However, there are
still many open problems, such as the design of optimal measurement, especially
simple and practical measurement which are independent of the unknown parameters,
as well as the robustness of precision limit against the noises and imperfect
controls. We believe that some of the issues will be solved in the near future.
Furthermore, quantum multiparameter metrology will then go deeply into the field
of applied science and become a basic technology for other aspects of sciences.

\ack
The authors thank Zibo Miao, Christiane Koch, Rafa\l{} Demkowicz-Dobrza\'{n}ski,
Yanming Che, Dominic Branford, Dominik \v{S}afr\'{a}nek, Jasminder Sidhu,
Wenchao Ge, Jesus Rubio, Wojciech Rzadkowski, Marco Genoni, Haggai Landa and
two anonymous referees for for helpful discussions and suggestions.
J.L. particularly thanks Yanyan Shao and Jinfeng Qin for the assistance on
drawing some of the schematics. J.L. acknowledges the support from National
Natural Science Foundation of China (Grant No. 11805073) and the startup grant
of HUST. H.Y. acknowledges the support from the Research Grants Council of Hong
Kong (GRF No. 14207717). X.M.L. acknowledges the support from the National Natural
Science Foundation of China (Grant Nos. 61871162 and 11805048), and Zhejiang
Provincial Natural Science Foundation of China (Grant No. LY18A050003).
X.W. acknowledges the support from National Key Research and Development
Program of China (Grant Nos. 2017YFA0304202 and 2017YFA0205700), the National
Natural Science Foundation of China (Grant Nos. 11935012 and 11875231),
and the Fundamental Research Funds for the Central Universities
(Grant No. 2017FZA3005).

\appendix

\section{Derivation of traditional form of QFIM} \label{apx:tradition_QFIM}
\setcounter{equation}{0}

The traditional calculation of QFIM usually assume a full rank density matrix,
of which the spectral decompostion is in the form
\begin{equation}
\rho =\sum^{\dim\rho -1}_{i=0}\lambda_{i}|\lambda_{i}\rangle\langle\lambda_{i}|,
\label{eq:app1_sd}
\end{equation}
where $\lambda_{i}$ and $|\lambda_{i}\rangle$ are $i$th eigenvalue and eigenstate
of $\rho $. $\dim\rho $ is the dimension of $\rho $. $|\lambda_{i}\rangle$ satisfies
$\sum^{\dim\rho -1}_{i=0}|\lambda_{i}\rangle\langle\lambda_{i}|=\openone$ with
$\openone$ the identity matrix. Substituting equation~(\ref{eq:app1_sd}) into the
equation of SLD ($\partial_{x_{a}}$ is the abbreviation of $\partial/\partial_{x_{a}}$)
\begin{equation}
\partial_{x_{a}}\rho =\frac{1}{2}\left(\rho L_{x_{a}}+L_{x_{a}}\rho \right),
\label{eq:app1_SLD}
\end{equation}
and taking $\langle\lambda_{i}|\cdot|\lambda_{j}\rangle$ on both sides of above equation,
one can obtain
\begin{equation}
\langle\lambda_{i}|L_{x_{a}}|\lambda_{j}\rangle=\frac{2\langle\lambda_{i}
|\partial_{x_{a}}\rho |\lambda_{j}\rangle}{\lambda_{i}+\lambda_{j}}.
\label{eq:app1_sldtp}
\end{equation}
Next, utlizing equation~(\ref{eq:app1_sd}), the QFIM
\begin{equation}
\mathcal{F}_{ab}=\frac{1}{2}\mathrm{Tr}\left(\rho\{L_{x_{a}},L_{x_{b}}\}\right)
\end{equation}
can be rewritten as
\begin{equation}
\mathcal{F}_{ab}=\frac{1}{2}\sum^{\mathrm{dim}\rho-1}_{i=0}\lambda_{i}\left(\langle\lambda_{i}
|L_{x_{a}}L_{x_{b}}|\lambda_{i}\rangle+\langle \lambda_{i}|L_{x_{b}}L_{x_{a}}
|\lambda_{i}\rangle\right).
\end{equation}
Inserting the equation $\openone=\sum^{\dim \rho -1}_{i=0}|\lambda_{i}\rangle
\langle\lambda_{i}|$ into above equation, one can obtain
\begin{equation}
\mathcal{F}_{ab}=\sum^{\dim\rho -1}_{i,j=0}\lambda_{i}
\mathrm{Re}(\langle\lambda_{i}|L_{x_{a}}|\lambda_{j}\rangle\langle\lambda_{j}
|L_{x_{b}}|\lambda_{i}\rangle).
\end{equation}
Substituting equation~(\ref{eq:app1_sldtp}) into this equation, one has
\begin{equation}
\mathcal{F}_{ab}=\sum^{\dim\rho -1}_{i,j=0}4\lambda_{i}
\frac{\mathrm{Re}(\langle\lambda_{i}|\partial_{x_{a}}\rho |\lambda_{j}\rangle
\langle\lambda_{j}|\partial_{x_{b}}\rho |\lambda_{i}\rangle)}{(\lambda_{i}+\lambda_{j})^{2}}.
\end{equation}
Exchange subscripts $i$ and $j$, the traditional form of QFIM is obtained as below
\begin{equation}
\mathcal{F}_{ab}=\sum^{\dim\rho -1}_{i,j=0}
\frac{2\mathrm{Re}(\langle\lambda_{i}|\partial_{x_{a}}\rho |\lambda_{j}\rangle
\langle\lambda_{j}|\partial_{x_{b}}\rho |\lambda_{i}\rangle)}{\lambda_{i}+\lambda_{j}}.
\end{equation}

\section{Derivation of QFIM for arbitrary-rank density matrices}
\label{apx:nonfull_QFIM}
\setcounter{equation}{0}

In this appendix we show the detailed derivation of QFIM for arbitrary-rank
density matrices. Here the spectral decompostion of $\rho$ is
\begin{equation}
\rho =\sum^{N-1}_{i=0}\lambda_{i}|\lambda_{i}\rangle\langle \lambda_{i}|,
\label{eq:app2_sd}
\end{equation}
where $\lambda_{i}$ and $|\lambda_{i}\rangle$ are $i$th eigenvalue and eigenstate of $\rho$.
Notice $N$ here is the dimension of $\rho $'s support. For a full-rank density matrix,
$N$ equals to $\dim\rho $, the dimension of $\rho $, and for a non-full rank density
matrix, $N < \dim\rho $, and $\sum^{N-1}_{i=0}|\lambda_{i}
\rangle\langle\lambda_{i}|\neq \openone$ ($\openone$ is the identity matrix and
$\openone=\sum^{\dim \rho -1}_{i=0}|\lambda_{i}\rangle\langle\lambda_{i}|$).
Furthermore, $\lambda_{i}\neq 0$ for $i\in[0,N-1]$.
With these notations, $\mathcal{F}_{ab}$ can be expressed by
\begin{eqnarray}
\mathcal{F}_{ab}&=&\sum^{N-1}_{i=0}\frac{(\partial_{x_{a}}\lambda_{i})
(\partial_{x_{b}}\lambda_{i})}{\lambda_{i}}+\sum^{N-1}_{i=0}4\lambda_{i}
\mathrm{Re}\left(\langle\partial_{x_{a}}\lambda_{i}|\partial_{x_{b}}
\lambda_{i}\rangle\right) \nonumber \\
& & -\sum^{N-1}_{i,j=0}\frac{8\lambda_{i}\lambda_{j}}{\lambda_{i}+\lambda_{j}}
\mathrm{Re}(\langle\partial_{x_{a}}\lambda_{i}|
\lambda_{j}\rangle\langle\lambda_{j}|\partial_{x_{b}}\lambda_{i}\rangle).
\label{eq:app2_QFIM}
\end{eqnarray}
Followings are the detailed proof of this equation.

Based on the equation of SLD ($\partial_{x_{a}}$ is the abbreviation
of $\partial/\partial x_{a}$)
\begin{equation}
\partial_{x_{a}}\rho =\frac{1}{2}\left(\rho L_{x_{a}}+L_{x_{a}}\rho \right),
\end{equation}
one can easily obtain
\begin{equation}
\langle\lambda_{i}|\partial_{x_{a}}\rho |\lambda_{j}\rangle=\frac{1}{2}
\langle\lambda_{i}|L_{x_{a}}|\lambda_{j}\rangle\left(\lambda_{i}+\lambda_{j}\right).
\end{equation}
The derivative of $\rho $ reads $\partial_{x_{a}}\rho =\sum^{N-1}_{i=0}
\partial_{x_{a}}\lambda_{i}|\lambda_{i}\rangle\langle \lambda_{i}|
+\lambda_{i}|\partial_{x_{a}}\lambda_{i}\rangle\langle\lambda_{i}|
+\lambda_{i}|\lambda_{i}\rangle\langle\partial_{x_{a}}\lambda_{i}|$, which gives
$\langle\lambda_{i}|\partial_{x_{a}}\rho |\lambda_{j}\rangle
=\partial_{x_{a}}\lambda_{i} \delta_{ij}
+\lambda_{j}\langle\lambda_{j}|\partial_{x_{a}}\lambda_{i}\rangle
+\lambda_{i}\langle\partial_{x_{a}}
\lambda_{j}|\lambda_{i}\rangle.$
$\delta_{ij}$ is the Kronecker delta function ($\delta_{ij}=1$ for $i=j$ and
zero otherwise). Substituting this expression into above equation,
$\langle\lambda_{i}|L_{x_{a}} |\lambda_{j}\rangle$ can be calculated as
\begin{equation}
\langle\lambda_{i}|L_{x_{a}}|\lambda_{j}\rangle\!=\!\!\Bigg\{\begin{array}{cc}
\!\!\!\delta_{ij}\frac{\partial_{x_{a}}\lambda_{i}}{\lambda_{i}}\!
+\!\frac{2(\lambda_{j}-\lambda_{i})}{\lambda_{i}+\lambda_{j}}\!\langle\lambda_{i}
|\partial_{x_{a}}\lambda_{j}\rangle, \!\!\!&\!\! i\,\mathrm{or}\,j\!\in\![0,N\!-\!1];  \\
\mathrm{arbitrary}~\mathrm{value}, & i,j\in[N,\dim\rho -1]. \\
\end{array}\!\! \label{eq:app2_SLD_proof}
\end{equation}
With the solution of SLD, we can now further calculate the QFIM.
Utlizing equation~(\ref{eq:app2_sd}), the QFIM can be rewritten into
\begin{equation}
\mathcal{F}_{ab}=\frac{1}{2}\sum^{N-1}_{i=0}\lambda_{i}\left(\langle\lambda_{i}
|L_{x_{a}}L_{x_{b}}|\lambda_{i}\rangle+\langle \lambda_{i}|L_{x_{b}}L_{x_{a}}
|\lambda_{i}\rangle\right).
\end{equation}
Inserting the equation $\openone=\sum^{\dim \rho -1}_{i=0}|\lambda_{i}
\rangle\langle\lambda_{i}|$ into above equation, one can obtain
\begin{equation}
\mathcal{F}_{ab}=\sum^{N-1}_{i=0}\sum^{\dim \rho -1}_{j=0}\lambda_{i}
\mathrm{Re}(\langle\lambda_{i}|L_{x_{a}}|\lambda_{j}\rangle\langle\lambda_{j}
|L_{x_{b}}|\lambda_{i}\rangle).
\label{eq:app1_Ftemp}
\end{equation}
In this equation, it can be seen that the arbitrary part of
$\langle\lambda_{i}|L_{x_{a}}|\lambda_{j}\rangle$
does not affect the value of QFIM since it is not involved in above equation, but it
provides a freedom for the optimal measurements, which will be further discussed later.
Substituting equation~(\ref{eq:app2_SLD_proof}) into above equation, we have
\begin{eqnarray}
\mathcal{F}_{ab}&=&\sum^{N-1}_{i,j=0}\delta_{ij}\frac{(\partial_{x_{a}}\lambda_{i})
(\partial_{x_{b}}\lambda_{i})}{\lambda_{i}}+\frac{4\lambda_{i}(\lambda_{i}-\lambda_{j})^{2}}
{(\lambda_{i}+\lambda_{j})^{2}}\mathrm{Re}\left(\langle\lambda_{i}|\partial_{x_{a}}
\lambda_{j}\rangle\langle\partial_{x_{b}}\lambda_{j}|\lambda_{i}\rangle\right) \nonumber  \\
& & +\sum^{N-1}_{i=0}\sum^{\dim \rho -1}_{j=N}4\lambda_{i}
\mathrm{Re}\left(\langle\partial_{x_{a}}\lambda_{i}|\lambda_{j}\rangle
\langle\lambda_{j}|\partial_{x_{b}}\lambda_{i}\rangle\right),
\label{eq:app2_middle_Fab}
\end{eqnarray}
where the fact $\langle \lambda_{i}|\partial_{x_{a}}\lambda_{j}\rangle
=-\langle\partial_{x_{a}}\lambda_{i}|\lambda_{j}\rangle$
has been applied. The third term in above equation can be further calculated as
\begin{eqnarray}
& & \sum^{N-1}_{i=0}\sum^{\dim \rho -1}_{j=N}4\lambda_{i}
\mathrm{Re}\left(\langle\partial_{x_{a}}\lambda_{i}|\lambda_{j}\rangle
\langle\lambda_{j}|\partial_{x_{b}}\lambda_{i}\rangle\right) \nonumber \\
&=& \sum^{N-1}_{i=0}4\lambda_{i}\mathrm{Re}\left(\langle\partial_{x_{a}}
\lambda_{i}|\left(\sum^{\dim \rho -1}_{j=N}|\lambda_{j}
\rangle\langle\lambda_{j}|\right)|\partial_{x_{b}}\lambda_{i}\rangle\right) \nonumber \\
&=& \sum^{N-1}_{i=0}4\lambda_{i}\mathrm{Re}\left(\langle\partial_{x_{a}}
\lambda_{i}|\left(\openone-\sum^{N-1}_{j=0}|\lambda_{j}
\rangle\langle\lambda_{j}|\right)|\partial_{x_{b}}\lambda_{i}\rangle\right) \nonumber \\
&=& \sum^{N-1}_{i=0}4\lambda_{i}\mathrm{Re}\left(\langle\partial_{x_{a}}\lambda_{i}|
\partial_{x_{b}}\lambda_{i}\rangle-\sum^{N-1}_{j=0}\langle\partial_{x_{a}}
\lambda_{i}|\lambda_{j}\rangle\langle\lambda_{j}|\partial_{x_{b}}\lambda_{i}\rangle\right).
\end{eqnarray}
Therefore equation~(\ref{eq:app2_middle_Fab}) can be rewritten into
\begin{eqnarray*}
\mathcal{F}_{ab} &=& \sum^{N-1}_{i,j=0}\delta_{ij}\frac{(\partial_{x_{a}}\lambda_{i})
(\partial_{x_{b}}\lambda_{i})}{\lambda_{i}}+\frac{4\lambda_{i}(\lambda_{i}-\lambda_{j})^{2}}
{(\lambda_{i}+\lambda_{j})^{2}}\mathrm{Re}\left(\langle\lambda_{i}|\partial_{x_{a}}
\lambda_{j}\rangle\langle\partial_{x_{b}}\lambda_{j}|\lambda_{i}\rangle\right) \\
& & +\sum^{N-1}_{i=0}4\lambda_{i}\mathrm{Re}\left(\langle\partial_{x_{a}}\lambda_{i}
|\partial_{x_{b}}\lambda_{i}\rangle\right)-\sum^{N-1}_{i,j=0}4\lambda_{i}\mathrm{Re}
\left(\langle\partial_{x_{a}}\lambda_{i}|\lambda_{j}\rangle\langle\lambda_{j}
|\partial_{x_{b}}\lambda_{i}\rangle\right).
\end{eqnarray*}
Exchange the subscripts $i$ and $j$ in the second and fourth terms of above equation,
it reduces into a symmetric form as shown in equation~(\ref{eq:app2_QFIM}).

\section{QFIM in Bloch representation} \label{apx:QFIM_Bloch}

Bloch representation is a well-used representation of quantum states in quantum
mechanics and quantum information theory. For a $d$-dimensional quantum state $\rho$,
it can be expressed by
\begin{equation}
\rho=\frac{1}{d}\left(\openone+\sqrt{\frac{d(d-1)}{2}}\vec{r}\cdot\vec{\kappa}\right),
\end{equation}
where $\vec{r}=(r_1,r_2...,r_k,...)^{\mathrm{T}}$ is the Bloch vector
($|\vec{r}|^2\leq 1$) and $\vec{\kappa}$ is a $(d^{2}-1)$-dimensional vector
of $\frak{su}(d)$ generator satisfying $\mathrm{Tr}(\kappa_i)=0$ and
\begin{eqnarray}
\{\kappa_i,\kappa_j\}&=&\frac{4}{d}\delta_{ij}\openone+\sum^{d^2-1}_{m=1}\mu_{ijm}\kappa_m,
\label{eq:apx_Blochanticommu} \\
\left[\kappa_i,\kappa_j\right]&=& i\sum^{d^2-1}_{m=1}\epsilon_{ijm}\kappa_m,
\end{eqnarray}
where $\delta_{ij}$ is the Kronecker delta function. $\mu_{ijm}$ and $\epsilon_{ijm}$
are the symmetric and antisymmetric structure constants, which can be calculated
via the following equations
\begin{eqnarray}
\mu_{ijl}&=&\frac{1}{2}\mathrm{Tr}\left(\{\kappa_i,\kappa_j\}\kappa_l\right), \\
\epsilon_{ijl}&=&-\frac{i}{2}\mathrm{Tr}\left([\kappa_i,\kappa_j]\kappa_l\right).
\end{eqnarray}
It is easy to find $\mathrm{Tr}(\kappa_{i}\kappa_{j})=2\delta_{ij}$ and $\mu_{ijl}$
keeps still for any order of $ijl$, which will be used in the following.

Now we express the SLD operator with the $\frak{su}(d)$ generators
as~\cite{Watanabe2010,Watanabe2011}
\begin{equation}
L_{x_a} = z_a\openone+\vec{y}_a\cdot\vec{\kappa},
\end{equation}
where $z_a$ is a number and $\vec{y}_a$ is a vector. In the following we only use
$z$ and $\vec{y}$ for convenience. From the equation
$\partial_{x_a}\rho=\frac{1}{2}(\rho L+L\rho)$, one can find
\begin{equation}
c\partial_{x_a}\vec{r}\cdot\vec{\kappa}=\frac{z}{d}\openone
+\frac{1}{d}\vec{y}\cdot\vec{\kappa}+c z\vec{r}\cdot\vec{\kappa}
+\frac{c}{2}\{\vec{r}\cdot\vec{\kappa},\vec{y}\cdot\vec{\kappa}\}, \label{eq:apx_Bloch1}
\end{equation}
where $c=\sqrt{(d-1)/(2d)}$. Furthermore,
\begin{equation}
\frac{c}{2}\{\vec{r}\cdot\vec{\kappa},\vec{y}\cdot\vec{\kappa}\}=\sum_{ij}r_i y_j
\frac{c}{2}\{\kappa_i, \kappa_j\}=\frac{2c}{d}\vec{y}\cdot\vec{r}\openone
+\frac{c}{2}\sum_{ijm}r_i y_j \mu_{ijm}\kappa_m.
\end{equation}
The coefficients of $\openone$ in equation~(\ref{eq:apx_Bloch1}) from both sides
have to be the same, which gives
\begin{equation}
z =-2c\vec{y}\cdot\vec{r}=-2c\vec{y}^{\,\mathrm{T}}\vec{r}, \label{eq:apx_Bloch_x}
\end{equation}
which can also be obtained from the equation $\mathrm{Tr}(\rho L_{x_a})=0$. Similarly,
the coefficients of $\kappa_i$ in equation~(\ref{eq:apx_Bloch1}) from both sides
have to be the same, which gives
\begin{equation}
c\partial_{x_a} r_i=\frac{y_i}{d}+czr_i+\frac{c}{2}\sum_{jm}\mu_{jmi}y_m r_j.
\label{eq:apx_Bloch_entry}
\end{equation}
This can also be obtained by multipling $\kappa_i$ on both sides of equation~(\ref{eq:apx_Bloch1})
and then taking the trace. Now we introduce the matrix $G$ with the entry
\begin{equation}
G_{ij}=\frac{1}{2}\mathrm{Tr}(\rho\{\kappa_i, \kappa_j\}).
\end{equation}
It is easy to see $G$ is real symmetric. Substituting equation~(\ref{eq:apx_Blochanticommu})
into the equation above, it can be further written into
\begin{equation}
G_{ij}= \frac{2}{d}\delta_{ij}+c\sum_{m}\mu_{ijm}r_m.
\end{equation}
Hence, the last term in equation~(\ref{eq:apx_Bloch_entry}) can be rewritten into
\begin{eqnarray}
\frac{c}{2}\sum_{jm}\mu_{jmi}y_m r_j
&=& \frac{c}{2}\sum_{m}y_m\sum_{j}\mu_{jmi} r_j \nonumber \\
&=& \frac{1}{2}\sum_{m}y_m\left(G_{im}-\frac{2}{d}\delta_{im}\right) \nonumber \\
&=& \frac{1}{2}\sum_m G_{im}y_m-\frac{y_i}{d}.
\end{eqnarray}
Substituting this equation into equation~(\ref{eq:apx_Bloch_entry}), one has
$c\partial_{x_a} r_i=cz r_i+\frac{1}{2}\sum_m G_{im}y_m$,
which immediately leads to
\begin{equation}
c\partial_{x_a} \vec{r}= c z\vec{r}+\frac{1}{2}G\vec{y}.
\label{eq:apx_Bloch_vector}
\end{equation}
Futhermore, one can check that
\begin{equation}
(\vec{r}\,\vec{r}^{\,\mathrm{T}})\vec{y}=\left(\vec{y}^{\,\mathrm{T}}\vec{r}\right)\vec{r}
=-\frac{1}{2c}z\vec{r}.
\end{equation}
Utilizing this equation, equation~(\ref{eq:apx_Bloch_vector}) reduces to
\begin{equation}
c\partial_{x_a} \vec{r}= -2c^2\left(\vec{r}\,\vec{r}^{\,\mathrm{T}}\right)\vec{y}
+\frac{1}{2}G\vec{y}.
\end{equation}
Therefore,
\begin{equation}
\vec{y}=\left(\frac{1}{2c}G-2c\,\vec{r}\,\vec{r}^{\,\mathrm{T}}\right)^{-1}
\partial_{x_a}\vec{r}. \label{eq:apx_Bloch_y}
\end{equation}

Next, we calculate the entry of QFIM and we will use the full notation $z_a$ and
$\vec{y}_a$ instead of $z$, $\vec{y}$. The entry of QFIM reads
\begin{eqnarray}
\mathcal{F}_{ab}&=&\frac{1}{2}\mathrm{Tr}(\rho\{L_a,L_b\}) \nonumber \\
&=& z_a z_b+2c(z_a \vec{y}_b\cdot\vec{r}+z_b \vec{y}_a\cdot\vec{r})+\frac{1}{2}
\sum_{ij}y_{a,i}y_{b,j}\mathrm{Tr}\left(\rho\{\kappa_i,\kappa_j\}\right) \nonumber \\
&=&-z_a z_b+\vec{y}^{\,\mathrm{T}}_a G\vec{y}_b.
\end{eqnarray}
Based on equations~(\ref{eq:apx_Bloch_x}) and~(\ref{eq:apx_Bloch_vector}), one
can obtain
\begin{equation}
c \vec{y}^{\,\mathrm{T}}_a\partial_{x_b} \vec{r}=cz_{b}\vec{y}^{\,\mathrm{T}}_a\vec{r}
+\frac{1}{2}\vec{y}^{\,\mathrm{T}}_a G\vec{y}_b
=-\frac{1}{2}z_a z_{b}+\frac{1}{2}\vec{y}^{\,\mathrm{T}}_a G\vec{y}_b.
\end{equation}
Hence, the QFIM can be written as
\begin{equation}
\mathcal{F}_{ab}=2c\vec{y}^{\,\mathrm{T}}_a\partial_{x_b} \vec{r}
=2c(\partial_{x_b} \vec{r})^{\mathrm{T}}\vec{y}_a.
\end{equation}
Utilizing equation~(\ref{eq:apx_Bloch_y}), the QFIM can be finally written into
\begin{eqnarray}
\mathcal{F}_{ab} &=& 2c(\partial_{x_b} \vec{r})^{\mathrm{T}}
\left(\frac{1}{2c}G-2c\,\vec{r}\,\vec{r}^{\,\mathrm{T}}\right)^{-1}\partial_{x_a}\vec{r}
\nonumber \\
&=& (\partial_{x_b} \vec{r})^{\mathrm{T}}
\left(\frac{1}{4c^2}G-\vec{r}\,\vec{r}^{\,\mathrm{T}}\right)^{-1}\partial_{x_a}\vec{r}
\nonumber \\
&=& (\partial_{x_b} \vec{r})^{\mathrm{T}}
\left(\frac{d}{2(d-1)}G-\vec{r}\,\vec{r}^{\,\mathrm{T}}\right)^{-1}\partial_{x_a}\vec{r}.
\label{eq:apx_bloch_final}
\end{eqnarray}
The theorem has been proved.

The simplest case here is single-qubit systems. The corresponding density matrix
is $\rho =\frac{1}{2}(\openone_2+\vec{r}\cdot\vec{\sigma})$ with
$\vec{\sigma}=(\sigma_{1},\sigma_{2},\sigma_{3})$ the vector of Pauli matrices
and $\openone_2$ is the 2-dimensional identity matrix. In this case,
$G=\openone_3$ ($\openone_3$ is the 3-dimensional identity matrix)
due to the fact $\{\sigma_i,\sigma_j\}=2\delta_{ij}\openone_2$.
Equation~(\ref{eq:apx_bloch_final}) then reduces to
\begin{equation}
\mathcal{F}_{ab} =(\partial_{x_b} \vec{r})^{\mathrm{T}}
\left(\openone_3-\vec{r}\,\vec{r}^{\,\mathrm{T}}\right)^{-1}\partial_{x_a}\vec{r}.
\end{equation}
It can be checked that
\begin{equation}
\left(\openone_3-\vec{r}\,\vec{r}^{\,\mathrm{T}}\right)^{-1}
=\openone_3+\frac{1}{1-|\vec{r}|^2}\vec{r}\,\vec{r}^{\,\mathrm{T}}.
\end{equation}
The QFIM then reads
\begin{equation}
\mathcal{F}_{ab}=(\partial_{x_b} \vec{r})^{\mathrm{T}}(\partial_{x_a}\vec{r})
+\frac{\left[(\partial_{x_a}\vec{r})^{\mathrm{T}}\vec{r}\right]
\left[(\partial_{x_b}\vec{r})^{\mathrm{T}}\vec{r}\right]}{1-|\vec{r}|^2},
\end{equation}
or equivalently,
\begin{equation}
\mathcal{F}_{ab}=(\partial_{x_{a}}\vec{r})\cdot(\partial_{x_{b}}\vec{r})
+\frac{(\vec{r}\cdot\partial_{x_{a}}\vec{r})(\vec{r}\cdot\partial_{x_{b}}\vec{r})}
{1-|\vec{r}|^{2}}.
\end{equation}

\section{One-qubit basis-independent expression of QFIM} \label{apx:one_qubit}
\setcounter{equation}{0}

This appendix shows the proof of Theorem~\ref{theorem:one_qubit_QFIM}. We first
prove that the QFIM for one-qubit mixed state can be written as~\cite{Dittmann1999}
\begin{equation}
\mathcal{F}_{ab}=\mathrm{Tr}\left[(\partial_{x_{a}}\rho )(\partial_{x_{b}}\rho )\right]
+\frac{1}{\det \rho }\mathrm{Tr}\left[(\partial_{x_{a}}\rho -\rho
\partial_{x_{a}}\rho )(\partial_{x_{b}}\rho -\rho \partial_{x_{b}}\rho )\right].
\label{apx:one_qubit0}
\end{equation}
Utilizing the spectral decomposition $\rho=\lambda_{0}|\lambda_{0}\rangle\langle
\lambda_{0}|+\lambda_{1}|\lambda_{1}\rangle\langle \lambda_{1}|$, the first term reads
\begin{eqnarray}
\mathrm{Tr}[(\partial_{x_{a}}\rho)(\partial_{x_{b}}\rho)]
&=& \langle \lambda_{0}|\partial_{x_{a}}\rho |\lambda_{0}\rangle\langle\lambda_{0}|
\partial_{x_{b}}\rho|\lambda_{0}\rangle+\langle \lambda_{1}|\partial_{x_{a}}\rho
|\lambda_{1}\rangle\langle\lambda_{1}|\partial_{x_{b}}\rho|\lambda_{1}\rangle \nonumber \\
& & +2 \mathrm{Re}(\langle\lambda_{0}|\partial_{x_{a}}\rho|\lambda_{1}\rangle
\langle\lambda_{1}|\partial_{x_{b}}\rho|\lambda_{0}\rangle).
\end{eqnarray}
The second term
\begin{eqnarray}
& & \frac{1}{\det \rho }\mathrm{Tr}\left[(\partial_{x_{a}}\rho -\rho
\partial_{x_{a}}\rho )(\partial_{x_{b}}\rho -\rho\partial_{x_{b}}\rho )\right] \nonumber \\
&=& \frac{1}{\det \rho }\mathrm{Tr}\left[(\partial_{x_{a}}\rho)(\partial_{x_{b}}\rho)
+\rho(\partial_{x_{a}}\rho)\rho(\partial_{x_{b}}\rho)-\rho\{\partial_{x_{a}}\rho,
\partial_{x_{b}}\rho\}\right].
\end{eqnarray}
Since
\begin{eqnarray}
& & \frac{1}{\det \rho }\mathrm{Tr}\left[\rho(\partial_{x_{a}}\rho)\rho(\partial_{x_{b}}\rho)\right]
\nonumber \\
&=& \frac{\lambda_{0}}{\lambda_{1}}\langle\lambda_{0}|\partial_{x_{a}}\rho|\lambda_{0}\rangle
\langle\lambda_{0}|\partial_{x_{b}}\rho|\lambda_{0}\rangle
+\frac{\lambda_{1}}{\lambda_{0}}\langle\lambda_{1}|\partial_{x_{a}}\rho
|\lambda_{1}\rangle\langle\lambda_{1}|\partial_{x_{b}}\rho|\lambda_{1}\rangle \nonumber \\
& & +2\mathrm{Re}(\langle\lambda_{0}|\partial_{x_{a}}\rho|\lambda_{1}\rangle\langle\lambda_{1}
|\partial_{x_{b}}\rho|\lambda_{0}\rangle),
\end{eqnarray}
and
\begin{eqnarray}
& & -\frac{1}{\det \rho }\mathrm{Tr}\left[\rho\{\partial_{x_{a}}\rho, \partial_{x_{b}}\rho\}\right]
\nonumber \\
&=& -\frac{2}{\lambda_{1}}\langle\lambda_{0}|\partial_{x_{a}}\rho|\lambda_{0}\rangle
\langle\lambda_{0}|\partial_{x_{b}}\rho|\lambda_{0}\rangle-\frac{2}{\lambda_{0}}
\langle\lambda_{1}|\partial_{x_{a}}\rho|\lambda_{1}\rangle\langle\lambda_{1}
|\partial_{x_{b}}\rho|\lambda_{1}\rangle \nonumber \\
& &-\frac{2}{\lambda_{0}\lambda_{1}}\mathrm{Re}(\langle\lambda_{0}|\partial_{x_{a}}\rho
|\lambda_{1}\rangle\langle\lambda_{1}|\partial_{x_{b}}\rho|\lambda_{0}\rangle ),
\end{eqnarray}
one can finally obtain
\begin{eqnarray}
& & \mathrm{Tr}\left[(\partial_{x_{a}}\rho )(\partial_{x_{b}}\rho )\right]
+\frac{1}{\det \rho }\mathrm{Tr}\left[(\partial_{x_{a}}\rho -\rho
\partial_{x_{a}}\rho )(\partial_{x_{b}}\rho -\rho \partial_{x_{b}}\rho )\right] \nonumber \\
&=& \frac{1}{\lambda_{0}}\langle\lambda_{0}|\partial_{x_{a}}\rho |\lambda_{0}\rangle\langle\lambda_{0}|
\partial_{x_{b}}\rho|\lambda_{0}\rangle+\frac{1}{\lambda_{1}}\langle\lambda_{1}|\partial_{x_{a}}\rho
|\lambda_{1}\rangle\langle\lambda_{1}|\partial_{x_{b}}\rho|\lambda_{1}\rangle \nonumber \\
& & +4\mathrm{Re}(\langle\lambda_{0}|\partial_{x_{a}}\rho|\lambda_{1}\rangle
\langle\lambda_{1}|\partial_{x_{b}}\rho|\lambda_{0}\rangle),
\end{eqnarray}
which coincides with the traditional formula of QFIM in Theorem~\ref{theorem:traditional_form}.
Equation~(\ref{apx:one_qubit0}) is then proved. Furthermore, one may notice that
$\langle\lambda_{0}|\partial_{x_{a}}\rho|\lambda_{0}\rangle=\partial_{x_{a}}\lambda_{0}$,
$\langle\lambda_{1}|\partial_{x_{a}}\rho|\lambda_{1}\rangle=\partial_{x_{a}}\lambda_{1}$, and
$\langle\lambda_{0}|\partial_{x_{a}}\rho|\lambda_{1}\rangle=\langle\partial_{x_{a}}\lambda_{0}|
\lambda_{1}\rangle+\langle\lambda_{0}|\partial_{x_{a}}\lambda_{1}\rangle=0$, which gives
\begin{eqnarray}
& & \mathrm{Tr}\left[(\partial_{x_{a}}\rho)(\partial_{x_{b}}\rho)-\rho\{\partial_{x_{a}}
\rho, \partial_{x_{b}}\rho)\}\right] \nonumber \\
&=& (\partial_{x_{a}}\lambda_{0})(\partial_{x_{b}}\lambda_{0})-\lambda_{0}
(\partial_{x_{a}}\lambda_{0})(\partial_{x_{b}}\lambda_{0})-\lambda_{1}
(\partial_{x_{a}}\lambda_{1})(\partial_{x_{b}}\lambda_{1}) \nonumber \\
&=& 0,
\end{eqnarray}
namely, the equality $\mathrm{Tr}\left[(\partial_{x_{a}}\rho)(\partial_{x_{b}}\rho)\right]=\mathrm{Tr}
\left(\rho\{\partial_{x_{a}}\rho, \partial_{x_{b}}\rho)\}\right)$ holds for single-qubit mixed states.
Thus, equation~(\ref{apx:one_qubit0}) can further reduce to
\begin{equation}
\mathcal{F}_{ab}=\mathrm{Tr}\left[(\partial_{x_{a}}\rho )(\partial_{x_{b}}\rho )\right]
+\frac{1}{\det \rho }\mathrm{Tr}\left[(\rho(\partial_{x_{a}}\rho)\rho(\partial_{x_{b}}\rho)\right].
\label{apx:qubit_reduce}
\end{equation}
The theorem has been proved.

\section{Derivation of SLD operator for Gaussian states}
\setcounter{equation}{0}

\subsection{SLD operator for multimode Gaussian states}  \label{apx:SLD_Gaussian}

The derivation in this appendix is majorly based on the calculation in
references~\cite{Monras2013,Serafini2017}. Let us first recall the notations before
the derivation. The vector of quadrature operators are defined as
$\vec{R}=(\hat{q}_{1},\hat{p}_{1},...,\hat{q}_{m},\hat{p}_{m})^{\mathrm{T}}$.
$\Omega$ is a symplectic matrix $\Omega=i\sigma^{\oplus m}_{y}$. $\chi(\vec{s})$ is
the characteristic function. Furthermore, denote $d=\langle \vec{R}\rangle$. To
keep the calculation neat, we use $\chi$, $\rho$, $L$ instead of $\chi(\vec{s})$,
$\rho $, $L_{x_{a}}$ in this appendix. Some other notations are
$\langle\cdot\rangle=\mathrm{Tr}(\rho\cdot)$ and $\dot{A}=\partial_{x_{a}}A$.

The characteristic function $\chi=\langle D\rangle$, where $D=e^{i\vec{R}^{\mathrm{T}}\Omega \vec{s}}$
with $\vec{s}\in \mathbb{R}^{2m}$. Substituting $D$ into the equation
$\partial_{x_{a}}\rho=\frac{1}{2}(\rho L+L\rho)$ and taking the trace, we obtain
\begin{equation}
\partial_{x_{a}}\chi=\frac{1}{2}\langle\{L, D\}\rangle.
\label{eq:apx_Gaussian_dchi}
\end{equation}
Next, assume the SLD operator is in the following form
\begin{equation}
L=L^{(0)}\openone+\vec{L}^{(1),\mathrm{T}}\vec{R}+\vec{R}^{\,\mathrm{T}}G\vec{R},
\end{equation}
where $L^{(0)}$ is a real number, $\vec{L}^{(1)}$ is a $2m$-dimensional real vector and $G$
is a $2m$-dimensional real symmetric matrix. These conditions promise the Hermiticity of SLD.
With this ansatz, equation~(\ref{eq:apx_Gaussian_dchi}) can then be rewritten into
\begin{equation}
\partial_{x_{a}}\chi=  L^{(0)}\chi+\frac{1}{2}\sum_{i}L^{(1)}_{i}\langle\{R_{i}, D\}\rangle
+\frac{1}{4}\sum_{ij}G_{ij}\langle \{ \{R_{i}, R_{j}\},D\}\rangle.
\label{eq:apx_Gaussian_expand}
\end{equation}

Now we calculate $\langle\{R_{i}, D\}\rangle$ in equation~(\ref{eq:apx_Gaussian_expand}).
Denote $[A,\cdot]=A^{\times}(\cdot)$, one can have
\begin{eqnarray}
\partial_{s_{k}}D &=& \sum_{i} i \Omega_{ik} \int^{1}_{0}
e^{i y \vec{R}^{\mathrm{T}}\Omega \vec{s}}R_{i}
e^{-i y \vec{R}^{\mathrm{T}}\Omega \vec{s}} \mathrm{d}y D \nonumber \\
&=& \sum_{i} i \Omega_{ik} \int^{1}_{0} \sum_{n}\frac{(iy\vec{R}^{\mathrm{T}}
\Omega\vec{s})^{\times,n}}{n!}R_{i} \mathrm{d}y D \nonumber \\
&=& \sum_{i} i \Omega_{ik} \int^{1}_{0} R_{i}+y\left[i\vec{R}^{\mathrm{T}}
\Omega\vec{s}, R_{i}\right] \mathrm{d}y D \nonumber \\
&=& \sum_{i} i \Omega_{ik} \left( R_{i}+\frac{1}{2}\sum_{jl}\Omega_{ij}
\Omega_{jl}s_{l}\right) D \nonumber \\
&=& i \sum_{i}\Omega_{ki}\left(\frac{1}{2}s_{i}-R_{i}\right) D.
\label{eq:apx_Gaussian_t0}
\end{eqnarray}
where we have used the fact $\Omega_{ij}=-\Omega_{ji}$ and $\sum_{j}\Omega_{ij}\Omega_{jk}
=-\delta_{ik}$ which comes from the equation $\Omega^{2}=-\openone_{2m}$. Substituting
the equation
\begin{equation}
\int^{1}_{0} e^{i y \vec{R}^{\mathrm{T}}\Omega \vec{s}}R_{i}e^{-i y
\vec{R}^{\mathrm{T}}\Omega \vec{s}}\mathrm{d}y D
= D\int^{1}_{0} e^{-i y \vec{R}^{\mathrm{T}}\Omega \vec{s}}R_{i}e^{i y
\vec{R}^{\mathrm{T}}\Omega \vec{s}}\mathrm{d}y,
\end{equation}
into the first line of equation~(\ref{eq:apx_Gaussian_t0}) and repeat the calculation,
one can obtain $\partial_{s_{k}}D=-i\sum_{i}\Omega_{ki}D\left(\frac{1}{2}s_{i}+R_{i}\right)$.
Combining this equation with equation~(\ref{eq:apx_Gaussian_t0}), $\partial_{s_{k}}D$ can be
finally written as
\begin{equation}
\partial_{s_{k}}D=-\frac{i}{2}\sum_{i}\Omega_{ki}\{R_{i}, D\},
\end{equation}
which further gives $\langle\partial_{s_{k}}D\rangle=-\frac{i}{2}\sum_{i}\Omega_{ki}
\langle\{R_{i},D\}\rangle$. Based on this equation, it can be found
$\sum_{k}\Omega_{jk}\langle\partial_{s_{k}}D\rangle =
-\frac{i}{2}\sum_{ik}\Omega_{jk}\Omega_{ki}\langle\{R_{i},D\}\rangle$. Again since
$\sum_{k}\Omega_{jk}\Omega_{ki}=-\delta_{ij}$, it reduces to
\begin{equation}
\langle\{R_{j},D\}\rangle=-i2\sum_{k}\Omega_{jk}\langle\partial_{s_{k}}D\rangle
=-i2\sum_{k}\Omega_{jk}\partial_{s_{k}}\chi.
\label{eq:apx_Gaussian_t2}
\end{equation}

Next, continue to take the derivative on $\partial_{s_{k}}D$, we have
\begin{equation}
\partial_{s_{k^{\prime}}}\partial_{s_{k}}D = -\frac{i}{2}
\sum_{i}\Omega_{ki}\{R_{i}, \partial_{s_{k^{\prime}}}D\}
-\frac{1}{4}\sum_{ij}\Omega_{k^{\prime}j}\Omega_{ki}
\{R_{i}, \{R_{j}, D\}\}. \label{eq:apx_Gaussian_t1}
\end{equation}
Due to the Baker-Campbell-Hausdorff formula, $D^{\dagger}R_{i}D=R_{i}
+[-i \vec{R}^{\mathrm{T}}\Omega\vec{s},R_{i}]=R_{i}+s_{i}$,
the commutation between $R_{i}$ and $D$ can be obtained as $[R_{i},D]=s_{i}D$,
which further gives $\{R_{i}, \{R_{j},D\}\}=\{ \{R_{i}, R_{j}\},D\}-s_{i}s_{j}D$.
Then equation~(\ref{eq:apx_Gaussian_t1}) can be rewritten into
\begin{equation}
\partial_{s_{k^{\prime}}}\partial_{s_{k}}D=-\frac{1}{4}\sum_{ij}\Omega_{k^{\prime}j}
\Omega_{ki}\{ \{R_{i}, R_{j}\},D\}
+\frac{1}{4}\sum_{ij}\Omega_{k^{\prime}j}\Omega_{ki}s_{i}s_{j}D.
\end{equation}
And one can finally obtain
\begin{eqnarray}
\langle\{R_{i}, \{R_{j}, D\}\}\rangle &=& -4\sum_{kk^{\prime}}\Omega_{ik}\Omega_{jk^{\prime}}
\langle\partial_{s_{k^{\prime}}}\partial_{s_{k}}D\rangle+s_{i}s_{j}\chi \nonumber \\
&=& -4\sum_{kk^{\prime}}\Omega_{ik}\Omega_{jk^{\prime}}
\left(\partial_{s_{k^{\prime}}}\partial_{s_{k}}\chi\right)+s_{i}s_{j}\chi.
\label{eq:apx_Gaussian_t3}
\end{eqnarray}
With equations~(\ref{eq:apx_Gaussian_t2}) and~(\ref{eq:apx_Gaussian_t3}),
equation~(\ref{eq:apx_Gaussian_expand}) can be expressed by
\begin{eqnarray}
\partial_{x_{a}}\chi &=&  \left(L^{(0)}+\frac{1}{4}\sum_{ij}G_{ij}s_{i}s_{j}\right)\chi
-i\sum_{ik}L^{(1)}_{i}\Omega_{ik}(\partial_{s_{k}}\chi)
\nonumber \\
& & -\sum_{ijkk^{\prime}}G_{ij}\Omega_{ik}\Omega_{jk^{\prime}}(\partial_{s_{k^{\prime}}}
\partial_{s_{k}}\chi). \label{eq:apx_Gaussian_dx0}
\end{eqnarray}

On the other hand, from the expression of characteristic function
\begin{equation}
\chi=e^{-\frac{1}{2}\vec{s}^{\mathrm{T}}\Omega C\Omega^{\mathrm{T}}\vec{s}
-i(\Omega d)^{\mathrm{T}}\vec{s}}.
\end{equation}
it can be found that
\begin{equation}
\partial_{x_a}\chi=\left[-\frac{1}{2}\vec{s}^{\,\mathrm{T}}\Omega
\dot{C}\Omega^{\mathrm{T}}\vec{s}-i(\Omega\dot{d})^{\mathrm{T}}\vec{s}\right]\chi,
\label{eq:apx_Gaussian_dx1}
\end{equation}
and
\begin{eqnarray*}
\partial_{s_k}\chi &=& -\sum_{ijl}\Omega_{kl}\Omega_{ij} C_{jl}s_{i}
\chi-i\sum_{i}\Omega_{ki} d_{i}\chi, \\
\partial_{s_{k^{\prime}}}\partial_{s_k}\chi &=&-\sum_{i_{1}j_{1}}\Omega_{ki_{1}}
\Omega_{k^{\prime}j_{1}} C_{i_{1}j_{1}}\chi-\sum_{i_{1}j_{1}}\Omega_{ki_{1}}
\Omega_{k^{\prime}j_{1}} d_{i_{1}}d_{j_{1}}\chi \\
& & +\!\!\sum_{i_{1}j_{1}l_{1}i_{2}j_{2}l_{2}}\!\!\!\!\Omega_{kl_{1}}\Omega_{k^{\prime}l_{2}}
\Omega_{i_{1}j_{1}}\Omega_{i_{2}j_{2}}
C_{j_{1}l_{1}} C_{j_{2}l_{2}}s_{i_{1}}s_{i_{2}}\chi \\
& &+i\!\!\sum_{i_{1}j_{1}l_{1}i_{2}}\!\!\!\!\left(\Omega_{ki_{2}}\Omega_{k^{\prime}l_{1}}
+\Omega_{k^{\prime}i_{2}}\Omega_{kl_{1}}\right)\Omega_{i_{1}j_{1}}
C_{j_{1}l_{1}}s_{i_{1}} d_{i_{2}}\chi.
\end{eqnarray*}
Then we have $-i\sum_{ik}L^{(1)}_{i}\Omega_{ik}(\partial_{s_{k}}\chi)=\sum_{i}L^{(1)}_{i}
\left(-i\sum_{jk}\Omega_{kj} C_{ij}s_{k}+d_{i}\right)\chi$,
and
\begin{eqnarray*}
& & -\sum_{ijkk^{\prime}}G_{ij}\Omega_{ik}\Omega_{jk^{\prime}}(\partial_{s_{k^{\prime}}}
\partial_{s_{k}}\chi) \\
&=& \sum_{ij} G_{ij}\left(C_{ij}+d_{i}d_{j}\right)\chi
-\!\!\sum_{iji_{1}j_{1}i_{2}j_{2}}\!\!G_{ij}\Omega_{i_{1}j_{1}}\Omega_{i_{2}j_{2}}
C_{ij_{1}}C_{jj_{2}}s_{i_{1}}s_{i_{2}}\chi \\
& &-i\sum_{iji_{1}j_{1}}G_{ij}\Omega_{i_{1}j_{1}}s_{i_{1}}\left(C_{jj_{1}}d_{i}
+C_{ij_{1}}d_{j}\right)\chi.
\end{eqnarray*}
Substituting these equations into equation~(\ref{eq:apx_Gaussian_dx0}), $\partial_{x_{a}}\chi$
can be expressed by
\begin{eqnarray}
\partial_{x_{a}}\chi &=& \left(L^{(0)}+\frac{1}{4}\sum_{ij}G_{ij}s_{i}s_{j}\right)\chi
+\sum_{i}L^{(1)}_{i}\left(-i\sum_{jk}\Omega_{kj} C_{ij}s_{k}+d_{i}\right)\chi \nonumber \\
& & +\sum_{ij} G_{ij}\left(C_{ij}+d_{i}d_{j}\right)\chi
-\!\!\sum_{iji_{1}j_{1}i_{2}j_{2}}\!\!G_{ij}\Omega_{i_{1}j_{1}}
\Omega_{i_{2}j_{2}} C_{ij_{1}} C_{jj_{2}}s_{i_{1}}s_{i_{2}}\chi \nonumber \\
& &-i\sum_{iji_{1}j_{1}}G_{ij}\Omega_{i_{1}j_{1}}s_{i_{1}}
\left(C_{jj_{1}}d_{i}+C_{ij_{1}}d_{j}\right)\chi.
\end{eqnarray}
This equation can be written into a more compact way as below
\begin{eqnarray}
\partial_{x_{a}}\chi &=& L^{(0)}\chi+\vec{L}^{(1),\mathrm{T}}d\chi
+\mathrm{Tr}(GC)\chi+d^{\mathrm{T}} G d \chi \nonumber \\
& & +i\vec{L}^{(1),\mathrm{T}}C\Omega\vec{s}\chi+i2d^{\mathrm{T}}G C
\Omega\vec{s}\chi+\vec{s}^{\mathrm{T}}\Omega C G
C\Omega\vec{s}\chi+\frac{1}{4}\vec{s}^{\mathrm{T}}G\vec{s}\chi.
\end{eqnarray}
Compare this equation with equation~(\ref{eq:apx_Gaussian_dx1}), it can be found that
\begin{eqnarray}
L^{(0)}+L^{(1),\mathrm{T}}d+\mathrm{Tr}(GC)+d^{\mathrm{T}}G d=0, & &
\label{eq:apx_Gaussian_L0} \\
\vec{L}^{(1),\mathrm{T}}C\Omega+2d^{\mathrm{T}}GC\Omega =-\dot{d}^{\mathrm{T}}\Omega^{\mathrm{T}},
& & \\
\frac{1}{4}G+\Omega C G C\Omega =-\frac{1}{2}\Omega \dot{C}\Omega^{\mathrm{T}}.
& & \label{eq:apx_Gaussian_G}
\end{eqnarray}
From the second equation above, one can obtain
\begin{equation}
\vec{L}^{(1)}=C^{-1}\dot{d}-2Gd.
\end{equation}
Using this equation and the equation~(\ref{eq:apx_Gaussian_L0}), it can be found that
\begin{equation}
L^{(0)}=d^{\mathrm{T}}Gd-\dot{d}^{\mathrm{T}}C^{-1}d-\mathrm{Tr}(GC).
\end{equation}

Once we obtain the expression of $G$, $L^{(0)}$ and $\vec{L}^{(1)}$ can be obtained correspondingly,
which means we need to solve equation~(\ref{eq:apx_Gaussian_G}), which can be rewritten into
following form
\begin{equation}
\Omega G\Omega +4C G C =2\dot{C}.
\end{equation}
Since $C=S C_{\mathrm{d}} S^{\mathrm{T}}$ and $S \Omega S^{\mathrm{T}}=\Omega$, above equation
can be rewritten into
\begin{equation}
\Omega G_{\mathrm{s}} \Omega+4C_{\mathrm{d}}G_{\mathrm{s}}C_{\mathrm{d}}=2S^{-1}
\dot{C}\left(S^{\mathrm{T}}\right)^{-1},
\label{eq:apx_Gaussian_gs}
\end{equation}
where $G_{\mathrm{s}}=S^{\mathrm{T}}G S$. Denote the map $\mathcal{E}(G_{\mathrm{s}}):=
\Omega G_{\mathrm{s}}\Omega+4C_{\mathrm{d}}G_{\mathrm{s}}C_{\mathrm{d}}$, then
$G_{\mathrm{s}}=\mathcal{E}^{-1}(\mathcal{E}(G_{\mathrm{s}}))$. The map
$\mathcal{E}(G_{\mathrm{s}})$ can be decomposed via the generators
$\{A^{(jk)}_{l}\}$ ($j,k=1,...,m$), where
\begin{equation}
A^{(jk)}_{l}=\frac{1}{\sqrt{2}} i\sigma^{(jk)}_{y},~\frac{1}{\sqrt{2}}
\sigma^{(jk)}_{z},~\frac{1}{\sqrt{2}} \openone^{(jk)}_{2},~\frac{1}{\sqrt{2}}
\sigma^{(jk)}_{x}
\end{equation}
for $l=0,1,2,3.$ $\sigma^{(jk)}_{i}$ is a $2m$-dimensional matrix with all
the entries zero expect the $2\times 2$ block shown as below
\begin{equation}
\sigma^{(jk)}_{i} = \left(\begin{array}{ccccc}
 & 1\mathrm{st} & \cdots & k\mathrm{th} & \cdots\\
1\mathrm{st} & 0_{2\times2} & 0_{2\times2} & 0_{2\times2} & 0_{2\times2}\\
\vdots & 0_{2\times2} & \vdots & \vdots & \vdots\\
j\mathrm{th} & 0_{2\times2} & \cdots & \sigma_{i} & \cdots\\
\vdots & \vdots & \vdots & \vdots & \vdots
\end{array}\right),
\end{equation}
where $0_{2\times 2}$ represents a 2 by 2 block with zero entries. $\openone^{(jk)}_{2}$
is similar to $\sigma^{(jk)}_{i}$ but replace the block $\sigma_{i}$ with $\openone_{2}$.
$A^{(jk)}_{l}$ satisfies the orthogonal relation
$\mathrm{Tr}(A^{(jk)}_{l}A^{(j^{\prime}k^{\prime})}_{l^{\prime}})=\delta_{jj^{\prime}}
\delta_{kk^{\prime}}\delta_{ll^{\prime}}$. Recall that
$C_{\mathrm{d}}=\bigoplus^{m}_{k=1}c_{k}\openone_{2}$, it is easy to check that
\begin{eqnarray}
\Omega A^{(jk)}_{l} \Omega &=&  (-1)^{l+1}A^{(jk)}_{l}, \label{eq:apx_Gaussian_gomega} \\
C_{\mathrm{d}} A^{(jk)}_{l} C_{\mathrm{d}} &=& c_{j}c_{k}A^{(jk)}_{l}. \label{eq:apx_Gaussian_cd}
\end{eqnarray}
Next decompose $S^{-1}\dot{C}(S^{\mathrm{T}})^{-1}$ with $\{ A^{(jk)}_{l} \}$ as
\begin{equation}
S^{-1}\dot{C}(S^{\mathrm{T}})^{-1}=\sum_{jkl}g^{(jk)}_{l}A^{(jk)}_{l},
\end{equation}
where $g^{(jk)}_{l}=\mathrm{Tr}[S^{-1}\dot{C}(S^{\mathrm{T}})^{-1}A^{(jk)}_{l}]$.
Decomposing $G_{\mathrm{s}}$ as $G_{\mathrm{s}}=\sum_{jkl}\tilde{g}^{(jk)}_{l}A^{(jk)}_{l}$,
and substituting it into equation~(\ref{eq:apx_Gaussian_gs}), we have
\begin{equation}
\sum_{jkl}g^{(jk)}_{l}\left(\Omega A^{(jk)}_{l}\Omega
+4 C_{\mathrm{d}}A^{(jk)}_{l}C_{\mathrm{d}}\right)=\sum_{jkl}g^{(jk)}_{l}A^{(jk)}_{l}.
\end{equation}
Utilizing equations~(\ref{eq:apx_Gaussian_gomega}) and~(\ref{eq:apx_Gaussian_cd}),
above equation reduces to
\begin{equation}
\sum_{jkl}\tilde{g}^{(jk)}_{l}\left[4 c_{j}c_{k}+(-1)^{l+1}\right]A^{(jk)}_{l}
=\sum_{jkl}g^{(jk)}_{l}A^{(jk)}_{l},
\end{equation}
which indicates
\begin{equation}
\tilde{g}^{(jk)}_{l}=\frac{g^{(jk)}_{l}}{4 c_{j}c_{k}+(-1)^{l+1}}.
\end{equation}
Thus, $G=(S^{\mathrm{T}})^{-1}G_{\mathrm{s}}S^{-1}$ can be solved as below
\begin{equation}
G=\sum_{jkl}\frac{g^{(jk)}_{l}}{4 c_{j}c_{k}+(-1)^{l+1}}
\left(S^{\mathrm{T}}\right)^{-1}A^{(jk)}_{l}S^{-1}.
\end{equation}

In summary, the SLD can be expressed by
\begin{equation}
L=L^{(0)}\openone+\vec{L}^{(1),\mathrm{T}}\vec{R}+\vec{R}^{\mathrm{T}}G\vec{R},
\end{equation}
where
\begin{equation}
G=\sum^{m}_{j,k=1}\sum^{3}_{l=0}\frac{g^{(jk)}_{l}}{4 c_{j}c_{k}+(-1)^{l+1}}
\left(S^{\mathrm{T}}\right)^{-1}A^{(jk)}_{l}S^{-1}
\end{equation}
with
\begin{equation}
A^{(jk)}_{l}=\frac{1}{\sqrt{2}} i\sigma^{(jk)}_{y},~\frac{1}{\sqrt{2}}
\sigma^{(jk)}_{z},~\frac{1}{\sqrt{2}}\openone^{(jk)}_{2},
~\frac{1}{\sqrt{2}} \sigma^{(jk)}_{x}
\end{equation}
for $l=0,1,2,3$ and $g^{(jk)}_{l}=\mathrm{Tr}[S^{-1}\dot{C}(S^{\mathrm{T}})^{-1}A^{(jk)}_{l}]$.
And
\begin{eqnarray}
\vec{L}^{(1)} &=& C^{-1}\dot{d}-2Gd, \\
L^{(0)} &=& d^{\mathrm{T}}Gd-\dot{d}^{\mathrm{T}}C^{-1}d-\mathrm{Tr}(GC).
\end{eqnarray}

\subsection{SLD operator for single-mode Gaussian state}  \label{apx:SLD_smGuassian}

For a single-mode case, a 2-dimensional symplectic $S$ matrix satisfies $\det S=1$. Based on
the equation $C=SC_{\mathrm{d}}S^{\mathrm{T}}=c SS^{\mathrm{T}}$ ($c$ is the symplectic value),
the following equations can be obtained
\begin{eqnarray}
C_{00} &=& c(S^{2}_{00}+S^{2}_{01}), \nonumber \\
C_{11} &=& c(S^{2}_{10}+S^{2}_{11}), \label{eq:Gauss_2by2}\\
C_{01} &=& c(S_{00}S_{10}+S_{01}S_{11}).
\end{eqnarray}
Together with the equation $\det S=S_{00}S_{11}-S_{01}S_{10}=1$, the symplectic value $c$
can be solved as $c=\sqrt{\det C}$. Assume the form
\begin{eqnarray}
S_{00} &=& \sqrt{\frac{C_{00}}{c}}\cos\theta,\quad S_{01}=\sqrt{\frac{C_{00}}{c}}\sin\theta, \\
S_{10} &=& \sqrt{\frac{C_{11}}{c}}\cos\phi, \quad S_{11}=\sqrt{\frac{C_{11}}{c}}\sin\phi.
\end{eqnarray}
Substituting above expressions into equation~(\ref{eq:Gauss_2by2}), it can be found
$\theta,\phi$ need to satisfy
\begin{equation}
\cos(\phi-\theta)=\frac{C_{01}}{\sqrt{C_{00}C_{11}}},
~~\sin(\phi-\theta)=\frac{c}{\sqrt{C_{00}C_{11}}}.
\end{equation}
Since there is only four constrains for these five variables, one of them is free.
We take $\theta=\pi/2-\phi$, and the symplectic matrix reduces to
\begin{equation}
S=\frac{1}{\sqrt{c}}\left(\begin{array}{cc}
\sqrt{C_{00}}\sin\phi & \sqrt{C_{00}}\cos\phi \\
\sqrt{C_{11}}\cos\phi & \sqrt{C_{11}}\cos\phi
\end{array}\right),
\end{equation}
where $\phi$ satisfies $\sin(2\phi)=\frac{C_{01}}{\sqrt{C_{00}C_{11}}}$ and
$\cos(2\phi)=-\frac{c}{\sqrt{C_{00}C_{11}}}$. Based on Theorem~\ref{SLD_Gaussian},
we obtain
\begin{eqnarray}
g_{0} &=& 0, \\
g_{1} &=& \frac{1}{\sqrt{2C_{00}C_{11}}}C_{11}\dot{C}_{00}-C_{00}\dot{C}_{11}, \\
g_{2} &=& \sqrt{2} \dot{c}, \\
g_{3} &=& \frac{\sqrt{2}(c\dot{C}_{01}-\dot{c}C_{01})}{\sqrt{C_{00}C_{11}}},
\end{eqnarray}
where $\dot{C}$, $\dot{c}$ are short for $\partial_{x_{a}}C$ and $\partial_{x_a}c$.
Meanwhile, we have
\begin{eqnarray}
(S^{\mathrm{T}})^{-1}A_{1}S^{-1} &=& \frac{\sqrt{2C_{00}C_{11}}}{2}
\left(\begin{array}{cc}
\frac{1}{C_{00}} & 0 \\
0 & -\frac{1}{C_{11}}
\end{array}\right),  \\
(S^{\mathrm{T}})^{-1}A_{2}S^{-1} &=& \frac{1}{\sqrt{2}c}
\left(\begin{array}{cc}
C_{11} & -C_{01} \\
-C_{01} & C_{00}
\end{array}\right),  \\
(S^{\mathrm{T}})^{-1}A_{3}S^{-1} &=& \frac{\sqrt{C_{00}C_{11}}}{\sqrt{2}c}
\left(\begin{array}{cc}
-\frac{C_{01}}{C_{00}} & 1 \\
1 & -\frac{C_{01}}{C_{11}}
\end{array}\right).
\end{eqnarray}
These expressions immediately give $G_{x_a}$ as below
\begin{eqnarray}
[G_{x_a}]_{00} &=& \frac{1}{4c^{2}+1}\left[\frac{C_{11}\dot{C}_{00}-C_{00}\dot{C}_{11}}{2C_{00}}
-\frac{C_{01}}{C_{00}}(\dot{C}_{01}-\frac{\dot{c}}{c}C_{01})\right]
+\frac{1}{4c^{2}-1}\frac{\dot{c}}{c}C_{11} \nonumber \\
&=& \frac{1}{4c^{2}+1}\left(\frac{\dot{c}}{c}C_{11}-\dot{C}_{11}\right)+
\frac{1}{4c^{2}-1}\frac{\dot{c}}{c}C_{11} \nonumber \\
&=& -\frac{1}{16c^{4}-1}\left[(4c^{2}-1)C_{11}-8c\dot{c}C_{11}\right].
\end{eqnarray}
Define a matrix
\begin{equation}
J:=\frac{1}{4c^{2}-1}C,
\end{equation}
$[G_{x_a}]_{00}$ can be rewritten into
\begin{equation}
[G_{x_a}]_{00}= -\frac{4c^{2}-1}{4c^{2}+1}\partial_{x_a}J_{11}, \\
\end{equation}
Similarly, one can obtain
\begin{eqnarray}
\left[G_{x_a}\right]_{11} &=& -\frac{4c^{2}-1}{4c^{2}+1}\partial_{x_a}J_{00},
\end{eqnarray}
and
\begin{equation}
\left[G_{x_a}\right]_{01}=\left[G_{x_a}\right]_{10}
=\frac{4c^{2}-1}{4c^{2}+1}\partial_{x_a}J_{01}.
\end{equation}
These expressions indicate
\begin{equation}
G_{x_a}=\frac{4c^{2}-1}{4c^{2}+1}\Omega(\partial_{x_a}J)\Omega,
\end{equation}
with
\begin{equation}
\Omega= \left(\begin{array}{cc}
0 & 1 \\
-1 & 0
\end{array}\right).
\end{equation}

\section{QFIM and Bures metric} \label{apx:fidelity_QFIM}
\setcounter{equation}{0}

The Bures distance between two quantum states $\rho_{1}$ and $\rho_{2}$ is defined as
\begin{equation}
D^{2}_{\mathrm{B}}(\rho_{1},\rho_{2})= 2-2f(\rho_{1},\rho_{2}),
\end{equation}
where $f(\rho_{1},\rho_{2})=\mathrm{Tr}\sqrt{\sqrt{\rho_{1}}\rho_{2}\sqrt{\rho_{1}}}$
is the quantum fidelity. Now we calcuate the fidelity for two close quantum states
$\rho(\vec{x})$ and $\rho(\vec{x}+\mathrm{d}\vec{x})$. The Taylor series of
$\rho(\vec{x}+\mathrm{d}\vec{x})$ (up to the second order) reads
\begin{equation}
\rho(\vec{x}+\mathrm{d}\vec{x})=\rho(\vec{x})+\sum_{a}\partial_{x_{a}}
\rho(\vec{x})\mathrm{d}x_{a}+\frac{1}{2}\sum_{ab}\frac{\partial^{2}\rho(\vec{x})}
{\partial x_{a}\partial x_{b}}\mathrm{d}x_{a}\mathrm{d}x_{b},
\end{equation}
In the following $\rho$ will be used as the abbreviation of $\rho(\vec{x})$.
Utilizing the equation above, one can obtain
\begin{eqnarray}
& & \sqrt{\rho}\rho(\vec{x}+\mathrm{d}\vec{x})\sqrt{\rho} \nonumber \\
&=& \rho^{2}+\sum_{a}\sqrt{\rho}\partial_{x_{a}}\rho
\sqrt{\rho}\mathrm{d}x_{a}+\frac{1}{2}\sum_{ab}\left(\sqrt{\rho}\frac{\partial^{2}\rho}
{\partial x_{a}\partial x_{b}}\sqrt{\rho}\right)\mathrm{d}x_{a}\mathrm{d}x_{b}.
\label{eq:apx_Bures0}
\end{eqnarray}
Now we assume
\begin{equation}
\sqrt{\sqrt{\rho}\rho(\vec{x}+\mathrm{d}\vec{x})\sqrt{\rho}}
=\rho+\sum_{a}W_{a}\mathrm{d}x_{a}+\sum_{ab}Y_{ab}\mathrm{d}x_{a}\mathrm{d}x_{b}.
\label{eq:apx_Bures1}
\end{equation}
Taking the square of above equation and compare it to equation~(\ref{eq:apx_Bures0}),
one can obtain
\begin{eqnarray}
\sqrt{\rho}\partial_{x_{a}}\rho\sqrt{\rho}&=& \rho W_{a}+W_{a}\rho, \label{eq:apx_Bures2} \\
\frac{1}{2}\sqrt{\rho}\partial^{2}\rho\sqrt{\rho} &=& \rho Y_{ab}+Y_{ab}\rho
+\frac{1}{2}\{W_{a}, W_{b}\},
\label{eq:apx_Bures3}
\end{eqnarray}
where $\partial^{2}\rho$ is short for $\frac{\partial^{2}\rho}{\partial x_{a}\partial x_{b}}$.
$\frac{1}{2}\{W_{a},W_{b}\}$ (not $W_{a}W_{b}$) is used to make sure the equation is
unchanged when the subscripts $a$ and $b$ exchange. Meawhile, from equation~(\ref{eq:apx_Bures1}),
the Bures metric $D_{\mathrm{B}}(\rho(\vec{x}),\rho(\vec{x}+\mathrm{d}\vec{x}))$
(the abbreviation $D_{\mathrm{B}}$ will bu used below) can be calculated as
\begin{equation}
D^{2}_{\mathrm{B}}=-2\sum_{a}(\mathrm{Tr}W_{a})\mathrm{d}x_{a}
-2\sum_{ab}(\mathrm{Tr}Y_{ab})\mathrm{d}x_{a}\mathrm{d}x_{b}.
\end{equation}
As long as we obtain the specific expressions of $W_{a}$ and $Y_{ab}$ from
equations~(\ref{eq:apx_Bures2}) and~(\ref{eq:apx_Bures3}), $D_{\mathrm{B}}$
can be obtained immediately. To do that, we utilize the spectral decomposition
$\rho=\sum_{i}\lambda_{i}|\lambda_{i}\rangle\langle\lambda_{i}|$. In the basis
$\{|\lambda_{i}\rangle\}$, the matrix entries ($[\cdot]_{ij}=\langle\lambda_{i}
|\cdot|\lambda_{j}\rangle$) read
\begin{eqnarray}
\left[\sqrt{\rho}\partial_{x_{a}}\rho\sqrt{\rho}\right]_{ij} &=&
\sqrt{\lambda_{i}\lambda_{j}}[\partial_{x_{a}}\rho]_{ij}, \\
\left[\sqrt{\rho}\partial^{2}\rho\sqrt{\rho}\right]_{ij} &=&
\sqrt{\lambda_{i}\lambda_{j}}\left[\partial^{2}\rho\right]_{ij}.
\end{eqnarray}
Here $[\partial_{x_{a}}\rho]_{ij}=\delta_{ij}\partial_{x_{a}}
\lambda_{i}-(\lambda_{i}-\lambda_{j})\langle\lambda_{i}|\partial_{x_{a}}\lambda_{j}\rangle$,
and
\begin{eqnarray}
\left[\partial^{2}\rho \right]_{ij} &=& \partial^{2}\lambda_{i}\delta_{ij}
+\lambda_{i}\langle\partial^{2}\lambda_{i}|\lambda_{j}\rangle
+\lambda_{j}\langle\lambda_{i}|\partial^{2}\lambda_{j}\rangle \nonumber \\
& &+(\partial_{x_{a}}\lambda_{j}-\partial_{x_{a}}\lambda_{i})\langle\lambda_{i}
|\partial_{x_{b}}\lambda_{j}\rangle +(\partial_{x_{b}}\lambda_{j}
-\partial_{x_{b}}\lambda_{i})\langle\lambda_{i}|\partial_{x_{a}}\lambda_{j}\rangle \nonumber \\
& &+\sum_{k}\lambda_{k}(\langle\lambda_{i}|\partial_{x_{a}}\lambda_{k}\rangle
\langle\partial_{x_{b}}\lambda_{k}|\lambda_{j}\rangle+\langle\lambda_{i}
|\partial_{x_{b}}\lambda_{k}\rangle\langle\partial_{x_{a}}\lambda_{k}|\lambda_{j}\rangle).
\end{eqnarray}
where $\partial^{2}\lambda_{i}$ and $|\partial^{2}\lambda_{i}\rangle$ are short for
$\frac{\partial^{2}\lambda_{i}}{\partial x_{a}\partial x_{b}}$ and
$\frac{\partial^{2}}{\partial x_{a}\partial x_{b}}|\lambda_{i}\rangle$.
Now we denote $\rho$'s dimension as $N$ and $\lambda_{i}\in\mathcal{S}$ for
$i=0,1,2...,M-1$. Under the assumption that the support $\mathcal{S}$ is not affected
by the values of $\vec{x}$, i.e., the rank of $\rho(\vec{x})$ equals to that of
$\rho(\vec{x}+\mathrm{d}\vec{x})$, $\sqrt{\rho}\partial_{x_{a}}\rho\sqrt{\rho}$
and $\sqrt{\rho}\partial^{2}\rho\sqrt{\rho}$ are both block diagonal.
Based on equation~(\ref{eq:apx_Bures2}), the $ij$th matrix entry of $W_{a}$ can
be calculated as
\begin{eqnarray}
[W_{a}]_{ij}=\frac{\left[\sqrt{\rho}\partial_{x_{a}}\rho\sqrt{\rho}\right]_{ij}}
{\lambda_{i}+\lambda_{j}}=\frac{1}{2}\partial_{x_{a}}\lambda_{i}\delta_{ij}-
\frac{\sqrt{\lambda_{i}\lambda_{j}}(\lambda_{i}-\lambda_{j})}{\lambda_{i}+\lambda_{j}}
\langle\lambda_{i}|\partial_{x_{a}}\lambda_{j}\rangle,
\label{eq:apx_Bures2.5}
\end{eqnarray}
for $i,j\in [0,M-1]$ and $[W_{a}]_{ij}=0$ for others. One can observe that
$W_{a}$ is a Hermitian matrix, and
\begin{equation}
\mathrm{Tr}W_{a}=\sum_{ii}[W_{a}]_{ii}=\frac{1}{2}\sum_{i}\partial_{x_{a}}\lambda_{i}=0,
\end{equation}
which means there is no first order term in Bures metric. With respect to the
second order term, we need to know the value of $[Y_{ab}]_{ii}$. From
equation~(\ref{eq:apx_Bures3}), one can obtain
\begin{equation}
\mathrm{Tr}Y_{ab}=\frac{1}{4}\sum_{i}[\partial^{2}\rho]_{ii}
-\sum_{ik}\frac{1}{2\lambda_{i}}\mathrm{Re}\left([W_{a}]_{ik}[W_{b}]_{ki}\right).
\end{equation}
Due to the fact $\langle\partial^{2}\lambda_{i}|\lambda_{i}\rangle
+\langle\lambda_{i}|\partial^{2}\lambda_{i}\rangle=-2\mathrm{Re}
(\langle\partial_{x_{a}}\lambda_{i}|\partial_{x_{b}}\lambda_{i}\rangle)$,
one can have
\begin{equation}
[\partial^{2}\rho]_{ii} = \partial^{2}\lambda_{i}-2\lambda_{i}\mathrm{Re}
(\langle\partial_{x_{a}}\lambda_{i}|\partial_{x_{b}}\lambda_{i}\rangle)
+\sum_{k}2\lambda_{k}\mathrm{Re}(\langle\lambda_{i}|\partial_{x_{a}}\lambda_{k}\rangle
\langle\partial_{x_{b}}\lambda_{k}|\lambda_{i}\rangle),
\end{equation}
which further gives
\begin{eqnarray}
\frac{1}{4}\sum_{i}[\partial^{2}\rho]_{ii} &=& -\frac{1}{2}\sum_{i}\lambda_{i}\mathrm{Re}
(\langle\partial_{x_{a}}\lambda_{i}|\partial_{x_{b}}\lambda_{i}\rangle) \nonumber \\
& & +\sum_{ik}\frac{1}{4}(\lambda_{i}+\lambda_{k})\mathrm{Re}(\langle\lambda_{i}
|\partial_{x_{a}}\lambda_{k}\rangle\langle\partial_{x_{b}}\lambda_{k}|\lambda_{i}\rangle),
\end{eqnarray}
where the fact $\sum_{i}\partial^{2}\lambda_{i}=0$ has been applied. Next, from
equation~(\ref{eq:apx_Bures2.5}) one can obtain
\begin{equation*}
[W_{a}]_{ik}[W_{b}]_{ki} = \frac{1}{4}(\partial_{x_{a}}\lambda_{i})
(\partial_{x_{b}}\lambda_{i})\!+\!\sum_{k}\frac{\lambda_{i}\lambda_{k}
(\lambda_{i}-\lambda_{k})^{2}}{(\lambda_{i}+\lambda_{k})^{2}}
\langle\lambda_{i}|\partial_{x_{a}}\lambda_{k}\rangle
\langle\partial_{x_{b}}\lambda_{k}|\lambda_{i}\rangle,
\end{equation*}
which means
\begin{eqnarray*}
& & \sum_{ik}\frac{1}{2\lambda_{i}}\mathrm{Re}([W_{a}]_{ik}[W_{b}]_{ki}) \\
&=& \sum_{i}\frac{1}{8\lambda_{i}}(\partial_{x_{a}}\lambda_{i})
(\partial_{x_{b}}\lambda_{i})+\sum_{ik}\frac{\lambda_{k}(\lambda_{i}-\lambda_{k})^{2}}
{2(\lambda_{i}+\lambda_{k})^{2}}\mathrm{Re}(\langle\lambda_{i}|\partial_{x_{a}}
\lambda_{k}\rangle\langle\partial_{x_{b}}\lambda_{k}|\lambda_{i}\rangle) \\
&=& \sum_{i}\frac{1}{8\lambda_{i}}(\partial_{x_{a}}\lambda_{i})
(\partial_{x_{b}}\lambda_{i})+\sum_{ik}\frac{(\lambda_{i}-\lambda_{k})^{2}}
{4(\lambda_{i}+\lambda_{k})}\mathrm{Re}(\langle\lambda_{i}|\partial_{x_{a}}
\lambda_{k}\rangle\langle\partial_{x_{b}}\lambda_{k}|\lambda_{i}\rangle)
\end{eqnarray*}
Thus, $\mathrm{Tr}Y_{ab}$ can then be expressed by
\begin{eqnarray}
\mathrm{Tr}Y_{ab} &=& -\frac{1}{8}\Bigg[\sum_{i}\frac{(\partial_{x_{a}}\lambda_{i})
(\partial_{x_{b}}\lambda_{i})}{\lambda_{i}}+\sum_{i}4\lambda_{i}\mathrm{Re}
(\langle\partial_{x_{a}}\lambda_{i}|\partial_{x_{b}}\lambda_{i}\rangle) \nonumber \\
& & -\sum_{ik}\frac{8\lambda_{i}\lambda_{k}}{\lambda_{i}+\lambda_{k}}\mathrm{Re}
\langle\lambda_{i}|\partial_{x_{a}}\lambda_{k}\rangle\langle\partial_{x_{b}}
\lambda_{k}|\lambda_{i}\rangle\Bigg] \nonumber \\
&=& -\frac{1}{8}\mathcal{F}_{ab}.
\end{eqnarray}
With this equation, we finally obtain
\begin{equation}
D_{\mathrm{B}}(\rho(\vec{x}),\rho(\vec{x}+\mathrm{d}\vec{x}))=
\sum_{ab}\frac{1}{4}\mathcal{F}_{ab}\mathrm{d}x_{a}x_{b}.
\end{equation}
One should notice that this proof shows that this relation is established for
density matrices with any rank as long as the rank of $\rho(\vec{x})$ is
unchanged with the varying of $\vec{x}$. In the case the rank can change, a
thorough discussion can be found in reference~\cite{Safranek2017}.

\section{Relation between QFIM and cross-correlation functions} \label{apx:dynamic_sus}
\setcounter{equation}{0}

This appendix gives the thorough calculation of the relation between QFIM
and dynamic susceptibility in reference~\cite{Hauke2016}. Consider the unitary
parameterization $U=\exp(i\sum_{a}x_{a}O_{a})$ with a thermal state
$\rho=\frac{1}{Z}e^{-\beta H}$. Here $Z=\mathrm{Tr}(e^{-\beta H})$ is
the partition function. $O_{a}$ is a Hermitian generator for $x_{a}$. In the following
we set $k_{B}=1$ and assume all $O_{a}$ are commutative, i.e., $[O_{a},O_{b}]=0$
for any $a$ and $b$. Denote $O_{a}(t)=e^{iHt}O_{a}e^{-iHt}$, and
$\langle\cdot\rangle=\mathrm{Tr}(\rho\cdot)$, the symmetric cross-correlation
spectrum in this case reads
\begin{equation}
S_{ab}(\omega)=\frac{1}{2}\int^{\infty}_{-\infty}
\langle \{ Q_{a}(t), O_{b}\}\rangle e^{i\omega t}\mathrm{d}t.
\end{equation}
Utilizing the energy basis $\{|E_{i}\rangle \}$ (with $E_{i}$ the $i$th energy),
it can be rewritten into
\begin{equation*}
S_{ab}(\omega) = \frac{1}{2Z}\sum_{ij}(e^{-\beta E_{i}}+e^{-\beta E_{j}})
\langle E_{i}|O_{a}|E_{j}\rangle\langle E_{j}|O_{b}|E_{i}\rangle\!\!
\int^{\infty}_{-\infty}\!\!e^{i(\omega+E_{i}-E_{j})}\mathrm{d}t.
\end{equation*}
Further use the equation $\int^{\infty}_{-\infty}e^{i(\omega+E_{i}-E_{j})}
\mathrm{d}t=2\pi\delta(\omega+E_{i}-E_{j})$, $S_{ab}(\omega)$ can reduce to
\begin{equation}
S_{ab}(\omega)=\frac{\pi}{Z}\sum_{ij}
(e^{-\beta E_{i}}+e^{-\beta E_{j}})\delta(\omega+E_{i}-E_{j})
\langle E_{i}|O_{a}|E_{j}\rangle\langle E_{j}|O_{b}|E_{i}\rangle.
\end{equation}
With this expression, one can find
\begin{eqnarray*}
& & \int^{\infty}_{-\infty}\tanh^{2}\left(\frac{\omega}{2T}\right)
\mathrm{Re}(S_{ab}(\omega))\mathrm{d}\omega \\
&=& \sum_{ij}\int^{\infty}_{-\infty}\tanh^{2}\left(\frac{\omega}{2T}\right)
\delta(\omega+E_{i}-E_{j})\mathrm{d}\omega  \\
& & \times \frac{\pi}{Z}(e^{-\beta E_{i}}+e^{-\beta E_{j}})
\mathrm{Re}(\langle E_{i}|O_{a}|E_{j}\rangle\langle E_{j}|O_{b}|E_{i}\rangle)  \\
&=&  \sum_{ij}\tanh^{2}\left(\frac{E_{j}-E_{i}}{2T}\right)
\frac{\pi}{Z}(e^{-\beta E_{i}}+e^{-\beta E_{j}})
\mathrm{Re}(\langle E_{i}|O_{a}|E_{j}\rangle\langle E_{j}|O_{b}|E_{i}\rangle).
\end{eqnarray*}
Since
\begin{eqnarray}
\tanh\left(\frac{E_{i}-E_{j}}{2T}\right)&=&\frac{[e^{\frac{1}{2}\beta (E_{i}-E_{j})}-
e^{-\frac{1}{2}\beta (E_{i}-E_{j})}]e^{-\frac{1}{2}\beta (E_{i}+E_{j})}}
{[e^{\frac{1}{2}\beta (E_{i}-E_{j})}+e^{-\frac{1}{2}\beta (E_{i}-E_{j})}]
e^{-\frac{1}{2}\beta (E_{i}+E_{j})}} \nonumber \\
&=& \frac{e^{-\beta E_{j}}-e^{-\beta E_{i}}}{e^{-\beta E_{i}}+e^{-\beta E_{j}}},
\end{eqnarray}
one can obtain
\begin{eqnarray}
& & \int^{\infty}_{-\infty}\tanh^{2}\left(\frac{\omega}{2T}\right)
\mathrm{Re}(S_{ab}(\omega))\mathrm{d}\omega \\
&=& \pi\sum_{ij}\frac{\left(\frac{1}{Z}e^{\beta E_{i}}-\frac{1}{Z}e^{-\beta E_{j}}\right)^{2}}
{\frac{1}{Z}e^{-\beta E_{i}}+\frac{1}{Z}e^{-\beta E_{j}}}
\mathrm{Re}(\langle E_{i}|O_{a}|E_{j}\rangle\langle E_{j}|O_{b}|E_{i}\rangle).
\end{eqnarray}
From the expression of QFIM, it can be found that
\begin{equation}
\mathcal{F}_{ab}=\frac{4}{\pi}\int^{\infty}_{-\infty}
\tanh^{2}\left(\frac{\omega}{2T}\right)\mathrm{Re}(S_{ab}(\omega))\mathrm{d}\omega.
\end{equation}
It can also be checked that
\begin{equation}
\mathrm{Re}(S_{ab}(\omega))=\frac{1}{2}\int^{\infty}_{-\infty}
\langle Q_{a}(t)O_{b}+O_{b}(t)O_{a}\rangle e^{i\omega t}\mathrm{d}t.
\end{equation}

In the mean time, the asymmetric cross-correlation spectrum is in the form
\begin{eqnarray}
\chi_{ab}(\omega) &=& \frac{i}{2}\int^{\infty}_{-\infty} e^{i\omega t}
\langle [O_{a}(t), O_{b}]\rangle\mathrm{d}t \nonumber \\
&=& i\frac{\pi}{Z}\sum_{ij}\delta(\omega+E_{i}-E_{j})(e^{-\beta E_{i}}-e^{-\beta E_{j}})
\langle E_{i}|O_{a}|E_{j}\rangle\langle E_{j}|O_{b}|E_{i}\rangle, \nonumber
\end{eqnarray}
which directly gives the relation between $\chi_{ab}$ and $S_{ab}$ as below
\begin{eqnarray}
\mathrm{Im}(\chi_{ab}(\omega)) &=& \frac{\pi}{Z}\sum_{ij}\tanh\left(\frac{E_{j}-E_{i}}{2T}\right)
(e^{-\beta E_{i}}+e^{-\beta E_{j}})\delta(\omega+E_{i}-E_{j}) \nonumber \\
& & \times \mathrm{Re}(\langle E_{i}|O_{a}|E_{j}\rangle\langle E_{j}|O_{b}|E_{i}\rangle)
\nonumber \\
&=& \frac{\pi}{Z}\tanh\left(\frac{\omega}{2T}\right)\sum_{ij}
(e^{-\beta E_{i}}+e^{-\beta E_{j}})\delta(\omega+E_{i}-E_{j})
\nonumber \\
& & \times \mathrm{Re}(\langle E_{i}|O_{a}|E_{j}\rangle
\langle E_{j}|O_{b}|E_{i}\rangle) \nonumber \\
&=& \tanh\left(\frac{\omega}{2T}\right)\mathrm{Re}(S_{ab}(\omega)),
\end{eqnarray}
namely,
\begin{equation}
\mathrm{Im}(\chi_{ab}(\omega))=\tanh\left(\frac{\omega}{2T}\right)\mathrm{Re}(S_{ab}(\omega)),
\end{equation}
which is just the fluctuation-dissipation theorem. Using this relation, one can
further obtain the result in reference~\cite{Hauke2016} as below
\begin{equation}
\mathcal{F}_{ab}=\frac{4}{\pi}\int^{\infty}_{-\infty}
\tanh\left(\frac{\omega}{2T}\right)\mathrm{Im}(\chi_{ab}(\omega))\mathrm{d}\omega.
\end{equation}

\section{Derivation of quantum multiparameter Cram\'{e}r-Rao bound} \label{apx:multiCR}
\setcounter{equation}{0}

The derivation of quantum multiparameter Cram\'{e}r-Rao bound is based on
the Cauchy-Schwarz inequality below
\begin{equation}
\mathrm{Tr}(X^{\dagger}X)\mathrm{Tr}(Y^{\dagger}Y)\geq
\frac{1}{4}|\mathrm{Tr}(X^{\dagger}Y+XY^{\dagger})|^{2},
\label{eq:appCR_CS}
\end{equation}
which comes from the complete form
\begin{equation}
\mathrm{Tr}(X^{\dagger}X)\mathrm{Tr}(Y^{\dagger}Y)\geq
\frac{1}{4}|\mathrm{Tr}(X^{\dagger}Y+XY^{\dagger})|^{2}+
\frac{1}{4}|\mathrm{Tr}(X^{\dagger}Y-XY^{\dagger})|^{2}.
\end{equation}
Define $X$ and $Y$ as
\begin{eqnarray}
X &:=& \sum_{m}f_{m}L_{m}\sqrt{\rho }, \label{eq:appCR_X} \\
Y &:=& \sum_{m}g_{m}(O_{m}-\langle O_{m}\rangle)\sqrt{\rho },
\end{eqnarray}
where $f_{m},g_{m}$ are real numbers for any $m$, and
$\langle \cdot\rangle=\mathrm{Tr}(\cdot\rho)$. Here $O_{m}$ is an observable defined as
\begin{equation}
O_{m}:=\sum_{k}\hat{x}_{m}(k)\Pi_{k}.
\end{equation}
$\hat{x}_{m}(k)$ is the estimator of $x_{m}$ and is the function of $k$th result.
Based on equation~(\ref{eq:appCR_X}), it can be calculated that
$\mathrm{Tr}(X^{\dagger}X)=\sum_{ml}f_{m}f_{l}\mathrm{Tr}(L_{m}L_{l}\rho )$,
which can be symmetrized into
$\mathrm{Tr}(X^{\dagger}X)=\frac{1}{2}\sum_{ml}f_{m}f_{l}\mathrm{Tr}(\{L_{m}, L_{l}\}\rho )
=\sum_{ml}f_{m}f_{l}\mathcal{F}_{ab}$. Define
$\vec{f}=(f_{0},f_{1},...,f_{m},...)^{\mathrm{T}}$, $\mathrm{Tr}(X^{\dagger}X)$ can
be further rewritten into
\begin{equation}
\mathrm{Tr}(X^{\dagger}X)=\vec{f}^{\,\mathrm{T}}\mathcal{F}\vec{f}. \label{eq:appCR_xvec}
\end{equation}
Similarly, define $\vec{g}=(g_{0},g_{1},...,g_{m},...)^{\mathrm{T}}$ and through some algebra,
$\mathrm{Tr}(Y^{\dagger}Y)$ can be calculated as
\begin{equation}
\mathrm{Tr}(Y^{\dagger}Y)=\vec{g}^{\, \mathrm{T}}C\vec{g},  \label{eq:appCR_yvec}
\end{equation}
where $C$ is the covariance matrix for $\{O_{m}\}$, and is defined as
$C_{ml}=\frac{1}{2}\langle \{O_{m}, O_{l}\}\rangle-\langle O_{m}\rangle\langle O_{l}\rangle$.
Furthermore, one can also obtain
\begin{equation}
\frac{1}{2}\mathrm{Tr}(X^{\dagger}Y+XY^{\dagger})=\vec{f}^{\,\mathrm{T}}B\vec{g},
\label{eq:appCR_xy}
\end{equation}
where the entry of $B$ is defined as $B_{ml}=\frac{1}{2}\mathrm{Tr}(\rho \{L_{m}, \Pi_{l}\})
=\frac{1}{2}\mathrm{Tr}(\{\rho ,L_{m}\}O_{l})$.
Utilizing $\partial_{x_{a}}\rho =\frac{1}{2}\{\rho ,L_{m}\}$, $B_{ml}$ reduces to
$B_{ml}=\mathrm{Tr}(\Pi_{l}\partial_{x_{m}}\rho )$. Since we consider unbiased
estimators, i.e.,
$\langle O_{m}\rangle=\sum_{m}\hat{x}_{m}\mathrm{Tr}(\rho \Pi_{k})=x_{m}$, $B_{ml}$ further
reduces to $B_{ml}=\partial_{x_{m}}\langle O_{l}\rangle=\delta_{ml}$, with $\delta_{ml}$
the Kronecker delta function. Hence, for unbiased estimators $B$ is actually the
identity matrix $\openone$.

Now substituting equations~(\ref{eq:appCR_xvec}), (\ref{eq:appCR_yvec})
and~(\ref{eq:appCR_xy}) into the Cauchy-Schwarz inequality~(\ref{eq:appCR_CS}),
one can obtain
\begin{equation}
\vec{f}^{\, \mathrm{T}}\mathcal{F}\vec{f}\vec{g}^{\, \mathrm{T}}C\vec{g}\geq
\left(\vec{f}^{~\mathrm{T}}\vec{g}\right)^{2}.
\end{equation}
Assuming $\mathcal{F}$ is invertable, i.e., it is positive-definite, and taking
$\vec{f}=\mathcal{F}^{-1}\vec{g}$, above inequality reduces to
$\vec{g}^{\, \mathrm{T}}\mathcal{F}^{-1}\vec{g}\vec{g}^{\, \mathrm{T}}C\vec{g}\geq
\left(\vec{g}^{\, \mathrm{T}}\mathcal{F}^{-1}\vec{g}\right)^{2}$.
Since $\mathcal{F}$ is positive-definite, $\mathcal{F}^{-1}$ is also positive-definite,
which means $\vec{g}^{\, \mathrm{T}}\mathcal{F}^{-1}\vec{g}$
is a positive number, thus, the above equation can further reduce to
$\vec{g}^{\, \mathrm{T}}C\vec{g}\geq \vec{g}^{\, \mathrm{T}}\mathcal{F}^{-1}\vec{g}$,
namely,
\begin{equation}
C \geq \mathcal{F}^{-1}. \label{eq:appCR_CF}
\end{equation}

Next we discuss the relation between $C$ and $\mathrm{cov}(\hat{\vec{x}},\{\Pi_{k}\})$.
Utilizing the defintion of $O_{m}$,
$C_{ml}$ can be written as
\begin{eqnarray}
C_{ml} &=& \sum_{k k^{\prime}}\hat{x}_{m}(k)\hat{x}_{l}(k^{\prime})\frac{1}{2}
\mathrm{Tr}(\rho \{\Pi_{k},\Pi_{k^{\prime}}\}) \nonumber \\
& & -\left[\sum_{k}\hat{x}_{m}\mathrm{Tr}(\rho \Pi_{k})\right]
\left[\sum_{k}\hat{x}_{l}\mathrm{Tr}(\rho \Pi_{k})\right],
\end{eqnarray}
and $\mathrm{cov}(\hat{\vec{x}},\{\Pi_{k}\})$ for unbiased estimators reads
\begin{equation}
\mathrm{cov}(\hat{\vec{x}},\{\Pi_{k}\})=\sum_{k}\hat{x}_{m}\hat{x}_{l}\mathrm{Tr}
(\rho \Pi_{k})-x_{m}x_{l}.
\end{equation}
If $\{\Pi_{k}\}$ is a set of projection operators, it satisfies
$\Pi_{k}\Pi_{k^{\prime}}=\Pi_{k}\delta_{kk^{\prime}}$. For unbiased estimators,
$\sum_{k}\hat{x}_{m(l)}\mathrm{Tr}(\rho \Pi_{k})=x_{m(l)}$, which gives
$C_{ml}=\sum_{k}\hat{x}_{m}\hat{x}_{l}\mathrm{Tr}(\rho \Pi_{k})-x_{m}x_{l}$, i.e.,
\begin{equation}
C=\mathrm{cov}(\hat{\vec{x}},\{\Pi_{k}\}).
\end{equation}
If $\{\Pi_{k}\}$ is a set of POVM, one can see
\begin{eqnarray}
\vec{g}^{\, \mathrm{T}}\mathrm{cov}(\hat{\vec{x}},\{\Pi_{k}\})\vec{g}
&=& \sum_{k}\sum_{ml}g_{m}g_{l}\hat{x}_{m}\hat{x}_{l}\mathrm{Tr}
(\rho \Pi_{k})-\sum_{ml}g_{m}g_{l}x_{m}x_{l} \nonumber \\
&=& \sum_{k}\left(\sum_{m}g_{m}\hat{x}_{m}\right)^{\!\!2}
\mathrm{Tr}(\rho \Pi_{k})-\sum_{ml}g_{m}g_{l}x_{m}x_{l}.
\label{eq:appCR_temp1}
\end{eqnarray}
In the mean time,
\begin{equation}
\vec{g}^{\, \mathrm{T}}C\vec{g}=\frac{1}{2}\sum_{ml}g_{m}g_{l}
\mathrm{Tr}\left(\rho O_{m}O_{l}+\rho O_{l}O_{m}\right)
-\sum_{ml}g_{m}g_{l}x_{m}x_{l}.
\label{eq:appCR_temp2}
\end{equation}
Now define $\mathcal{B}_{k}:=\sqrt{\Pi_{k}}\left(\sum_{m}g_{m}\hat{x}_{m}
-\sum_{m}g_{m}O_{m}\right)$. Based on the Cauchy-Schwarz inequality
$\mathrm{Tr}(A^{\dagger}A)\geq 0$, which is valid for any operator $A$,
one can immediately obtain $\mathrm{Tr}(\rho\mathcal{B}^{\dagger}_{k}\mathcal{B}_{k})
=\mathrm{Tr}(\sqrt{\rho }\mathcal{B}^{\dagger}_{k}\mathcal{B}_{k}\sqrt{\rho })\geq 0$,
which further gives $\sum_{k}\mathrm{Tr}(\rho \mathcal{B}^{\dagger}_{k}\mathcal{B}_{k})\geq 0$.
Through some calculations, the expression of $\sum_{k}\mathrm{Tr}(\rho
\mathcal{B}^{\dagger}_{k}\mathcal{B}_{k})$ is in the form
\begin{equation}
\sum_{k}\mathrm{Tr}(\rho \mathcal{B}^{\dagger}_{k}\mathcal{B}_{k})=\sum_{k}
\!\left(\!\sum_{m}g_{m}\hat{x}_{m}\!\right)^{\!2}\!\!
\mathrm{Tr}(\rho \Pi_{k})-\sum_{ml}g_{m}g_{l}\mathrm{Tr}(\rho O_{m}O_{l}).
\end{equation}
Now taking the difference of equations~(\ref{eq:appCR_temp1}) and (\ref{eq:appCR_temp2}),
it can be found
\begin{equation}
\sum_{k}\mathrm{Tr}(\rho \mathcal{B}^{\dagger}_{k}\mathcal{B}_{k})=\vec{g}^{\,\mathrm{T}}
\left(\mathrm{cov}(\hat{\vec{x}},\{\Pi_{k}\})-C\right)\vec{g}.
\end{equation}
Finally, we obtain the following inequality
$\vec{g}^{\, \mathrm{T}}(\mathrm{cov}(\hat{\vec{x}},\{\Pi_{k}\})-C)\vec{g}\geq 0$,
namely, for any POVM measurement,
\begin{equation}
\mathrm{cov}(\hat{\vec{x}},\{\Pi_{k}\})\geq C.
\end{equation}
Based on inequality~(\ref{eq:appCR_CF}) and the property of quadratic form,
we finally obtain $\mathrm{cov}(\hat{\vec{x}},\{\Pi_{k}\}) \geq \mathcal{F}^{-1}$.
Consider the repetition of experiments (denoted as $n$), above bound needs to add a
factor of $1/n$. Hence the quantum multiparameter Cram\'{e}r-Rao bound can be finally
expressed by
\begin{equation}
\mathrm{cov}(\hat{\vec{x}},\{\Pi_{k}\}) \geq \frac{1}{n}\mathcal{F}^{-1}.
\end{equation}

\section{Construction of optimal measurement for pure states}
\label{apx:Optimal_measurement}

Assume the true values of the vector of unknown parameters $\vec{x}$ is $\vec{x}_{\mathrm{true}}$,
we now provide the proof that for a pure parameterized state $|\psi\rangle$,  a set of projectors
containing the state $|\psi_{\vec{x}_{\mathrm{true}}}\rangle:=|\psi(\vec{x}
=\vec{x}_{\mathrm{true}})\rangle$, i.e., $\{|m_{k}\rangle\langle m_{k},|m_{0}\rangle
=|\psi_{\vec{x}_{\mathrm{true}}}\rangle\}$ is possible to be an optimal measurement to
attain the quantum Cram\'{e}r-Rao bound, as shown in reference~\cite{Humphreys2013,Pezze2017}.
The calculation in this appendix basically coincides with the appendix in
reference~\cite{Pezze2017}.

Since $\{|m_{k}\rangle\langle m_{k}|\}$ contains the information of the true value,
in practice one need to use the estimated value of $\vec{x}$ (denoted as $\hat{\vec{x}}$)
to construct $|m_{0}\rangle=|\psi(\hat{\vec{x}})\rangle$ to perform this measurement, then
improve the accuracy of $\hat{\vec{x}}$ adaptively. Thus, it is reasonable to assume
$|\psi_{\vec{x}_{\mathrm{true}}}\rangle=|m_{0}\rangle
+\sum_{x_{a}}\delta x_{a}|\partial_{x_{a}}\psi\rangle|_{\vec{x}=\hat{\vec{x}}}$.
The probability for $|m_{k}\rangle\langle m_{k}|$ is $p_{k}=|\langle\psi|m_{k}\rangle|^{2}$,
which gives the CFIM as below
\begin{equation}
\mathcal{I}_{ab}(\hat{\vec{x}})=\sum_{k}\frac{4\mathrm{Re}(\langle m_{k}|\partial_{x_{a}}
\psi\rangle\langle\psi|m_{k}\rangle) \mathrm{Re}(\langle m_{k}|\partial_{x_{b}}\psi\rangle
\langle\psi|m_{k}\rangle)}{|\langle\psi|m_{k}\rangle|^{2}}.
\end{equation}
At the limit $\hat{\vec{x}}\rightarrow \vec{x}_{\mathrm{true}}$, it is
\begin{equation}
\mathcal{I}_{ab}(\vec{x}_{\mathrm{true}})=\lim_{\hat{\vec{x}}\rightarrow \vec{x}_{\mathrm{true}}}
\!\sum_{k}\frac{4\mathrm{Re}(\langle m_{k}|\partial_{x_{a}}\psi\rangle\langle\psi|m_{k}\rangle)
\mathrm{Re}(\langle m_{k}|\partial_{x_{b}}\psi\rangle\langle\psi|m_{k}\rangle)}
{|\langle\psi|m_{k}\rangle|^{2}}.
\end{equation}
For the $k=0$ term,
\begin{eqnarray}
\lim_{\hat{\vec{x}}\rightarrow \vec{x}_{\mathrm{true}}}
\mathrm{Re}(\langle m_{0}|\partial_{x_{a}}\psi\rangle\langle\psi|m_{0}\rangle)
&=& \mathrm{Re}(\langle \psi_{\vec{x}_{\mathrm{true}}}|\partial_{x_{a}}\psi\rangle) \nonumber \\
&=& \mathrm{Re}(\langle \psi|\partial_{x_{a}}\psi\rangle)|_{\vec{x}=\vec{x}_{\mathrm{true}}}=0,
\end{eqnarray}
and $\lim_{\vec{x}\rightarrow\vec{x}_{\mathrm{true}}}\langle\psi|m_{0}\rangle=1$.
Thus, the CFIM is
\begin{equation}
\mathcal{I}_{ab}(\vec{x}_{\mathrm{true}})=\lim_{\hat{\vec{x}}\rightarrow \vec{x}_{\mathrm{true}}}
\!\sum_{k\neq 0}\frac{4\mathrm{Re}(\langle m_{k}|\partial_{x_{a}}\psi\rangle\langle\psi|m_{k}\rangle)
\mathrm{Re}(\langle m_{k}|\partial_{x_{b}}\psi\rangle\langle\psi|m_{k}\rangle)}
{|\langle\psi|m_{k}\rangle|^{2}}\!.
\end{equation}
Due to the fact
\begin{eqnarray}
& & 4\mathrm{Re}(\langle m_{k}|\partial_{x_{a}}\psi\rangle\langle\psi|m_{k}\rangle)
\mathrm{Re}(\langle m_{k}|\partial_{x_{b}}\psi\rangle\langle\psi|m_{k}\rangle) \nonumber \\
&=& 2\mathrm{Re}(\langle\partial_{x_{a}}\psi|m_{k}\rangle\langle m_{k}|\partial_{x_{b}}\psi\rangle)
|\langle\psi|m_{k}\rangle|^{2} \nonumber \\
& & +2\mathrm{Re}(\langle m_{k}|\partial_{x_{a}}\psi\rangle\langle m_{k}
|\partial_{x_{b}}\psi\rangle\langle\psi|m_{k}\rangle^{2}),
\end{eqnarray}
one can have
\begin{eqnarray}
\mathcal{I}_{ab}(\vec{x}_{\mathrm{true}})=\mathcal{F}_{ab}(\vec{x}_{\mathrm{true}})-Q_{ab},
\end{eqnarray}
where
\begin{eqnarray}
Q_{ab} &:=& \lim_{\hat{\vec{x}}\rightarrow \vec{x}_{\mathrm{true}}} \sum_{k\neq 0}\frac{2\mathrm{Re}
(\langle m_{k}|\partial_{x_{a}}\psi\rangle\langle\partial_{x_{b}}\psi|m_{k}\rangle)
|\langle\psi|m_{k}\rangle|^{2}}{|\langle\psi|m_{k}\rangle|^{2}}
\nonumber \\
& & -\frac{2\mathrm{Re}(\langle m_{k}|\partial_{x_{a}}\psi\rangle\langle m_{k}|\partial_{x_{b}}\psi\rangle
\langle\psi|m_{k}\rangle^{2})}{|\langle\psi|m_{k}\rangle|^{2}}
\nonumber \\
&=& \lim_{\hat{\vec{x}}\rightarrow \vec{x}_{\mathrm{true}}}
\sum_{k\neq 0}\frac{4\mathrm{Im}(\langle\partial_{x_{a}}\psi|m_{k}\rangle\langle m_{k}|\psi\rangle)
\mathrm{Im}(\langle\partial_{x_{b}}\psi|m_{k}\rangle\langle m_{k}|\psi\rangle)}
{|\langle\psi|m_{k}\rangle|^{2}}.
\end{eqnarray}
To let $\mathcal{I}(\vec{x}_{\mathrm{true}})=\mathcal{F}(\vec{x}_{\mathrm{true}})$,
$Q$ has to be a zero matrix. Since the diagonal entry
\begin{equation}
Q_{aa}=\lim_{\hat{\vec{x}}\rightarrow \vec{x}_{\mathrm{true}}}
\sum_{k\neq 0}\frac{4\mathrm{Im}^{2}(\langle\partial_{x_{a}}\psi|m_{k}\rangle\langle m_{k}|\psi\rangle)}
{|\langle\psi|m_{k}\rangle|^{2}},
\end{equation}
in which all the terms within the summation are non-negative. Therefore, its value is zero if and only if
\begin{equation}
\lim_{\hat{\vec{x}}\rightarrow \vec{x}_{\mathrm{true}}}
\frac{\mathrm{Im}(\langle\partial_{x_{a}}\psi|m_{k}\rangle\langle m_{k}|\psi\rangle)}
{|\langle\psi|m_{k}\rangle|}=0,~\forall x_{a}, k\neq 0.  \label{apx:optimal_measdurement_lim}
\end{equation}
Furthermore, this condition simultaneously makes the non-diagonal entries of $Q$ vanish. Thus, it is
the necessary and sufficient condition for $Q=0$, which means it is also the necessary and sufficient
condition for $\mathcal{I}=\mathcal{F}$ at the point of true value. One may notice that the
limitation in equation~(\ref{apx:optimal_measdurement_lim}) is a $0/0$ type. Thus, we use the
formula $|\psi_{\vec{x}_{\mathrm{true}}}\rangle=|m_{0}\rangle+\sum_{x_{j}}
\delta x_{j}|\partial_{x_{j}}\psi\rangle|_{\vec{x}=\hat{\vec{x}}}$ to further calculate
above equation. With this formula, one has
\begin{equation}
\frac{\mathrm{Im}(\langle\partial_{x_{a}}\psi|m_{k}\rangle\langle m_{k}|\psi\rangle)}
{|\langle\psi|m_{k}\rangle|}=\frac{\sum_{j}\delta x_{j}\mathrm{Im}(\langle\partial_{x_{a}}\psi|
m_{k}\rangle\langle m_{k}|\partial_{x_{j}}\psi\rangle)}{|\sum_{j}\delta x_{j}
\langle\partial_{x_{j}}\psi|m_{k}\rangle|}.
\end{equation}
$\langle\partial_{x_{j}}\psi|m_{k}\rangle$ cannot generally be zero for all $x_{j}$ since
all $\partial_{x_{j}}\psi\rangle$ are not orthogonal in general. Thus,
equation~(\ref{apx:optimal_measdurement_lim}) is equivalent to
\begin{equation}
\mathrm{Im}(\langle\partial_{x_{a}}\psi|m_{k}\rangle\langle m_{k}|\partial_{x_{b}}\psi\rangle)=0,
~\forall x_{a},x_{b}, k\neq 0.
\end{equation}

\section{Gradient in GRAPE for Hamiltonian estimation} \label{apx:GRAPE_gradient}
\setcounter{equation}{0}

\subsection{Gradient of CFIM}

The core of GRAPE algorithm is to obtain the expression of gradient. The dynamics
of the system is described by
\begin{equation}
\partial_{t}\rho =\mathcal{E}_{\vec{x}}\rho.
\end{equation}
For the Hamiltonian estimation under control, the dynamics is
\begin{equation}
\partial_{t}\rho =-i[H_{0}(\vec{x})+H_{\mathrm{c}},\rho ]+\mathcal{L}\rho,
\end{equation}
where $H_{\mathrm{c}}=\sum^{p}_{k=1}V_{k}(t)H_{k}$ is the control Hamiltonian with $H_{k}$ the $k$th
control and $V_{k}(t)$ the corresponding time-dependent control amplitude. To perform the algorithm,
the entire evolution time $T$ is cut into $m$ parts with time interval $\Delta t$, i.e.,
$m \Delta t=T$. $V_{k}(t)$ within the $j$th time interval is denoted as $V_{k}(j)$ and is assumed
to be a constant.

For a set of probability distribution $p(y|\vec{x})=\mathrm{Tr}(\rho \Pi_{y})$ with $\{\Pi_{y}\}$
a set of POVM. The gradient of $\mathcal{I}_{ab}$ at target time $T$ reads~\cite{LiuGRAPE2}
\begin{eqnarray}
\frac{\delta \mathcal{I}_{ab}(T)}{\delta V_{k}(j)}&=&\Delta t
\mathrm{Tr}\left(\tilde{L}_{2,ab}\mathcal{M}^{(1)}_{j}\right)-\Delta^{2}t
\mathrm{Tr}\left[\tilde{L}_{1,b}\left(\mathcal{M}^{(2)}_{j,a}
+\mathcal{M}^{(3)}_{j,a}\right)\right] \nonumber \\
& & -\Delta^{2}t\mathrm{Tr}\left[\tilde{L}_{1,a}\left(\mathcal{M}^{(2)}_{j,b}
+\mathcal{M}^{(3)}_{j,b}\right)\right],
\end{eqnarray}
where
\begin{eqnarray}
\tilde{L}_{1,a(b)} &=& \sum_{y}\left[\partial_{x_{a(b)}}\ln p(y|\vec{x})\right]\Pi_{y}, \\
\tilde{L}_{2,ab} &=& \sum_{y}\left[\partial_{x_{a}}\ln p(y|\vec{x})\right]
\left[\partial_{x_{b}}\ln p(y|\vec{x})\right]\Pi_{y},
\end{eqnarray}
and $\mathcal{M}^{(1)}_{j}$, $\mathcal{M}^{(2)}_{j,a(b)}$ and
$\mathcal{M}^{(3)}_{j,a(b)}$ are Hermitian operators and can
be expressed by
\begin{eqnarray}
\mathcal{M}^{(1)}_{j} &=& i\mathcal{D}^{m}_{j+1}H^{\times}_{k}(\rho_{j}), \\
\mathcal{M}^{(2)}_{j,a(b)} &=& \sum^{j}_{i=1}\mathcal{D}^{m}_{j+1}H^{\times}_{k}
\mathcal{D}^{j}_{i+1}(\partial_{x_{a(b)}}H_{0})^{\times}(\rho_{i}),  \\
\mathcal{M}^{(3)}_{j,a(b)} &=&(1-\delta_{jm})\sum^{m}_{i=j+1}\mathcal{D}^{m}_{i+1}
(\partial_{x_{a(b)}}H_{0})^{\times}\mathcal{D}^{i}_{j+1}H^{\times}_{k}(\rho_{j}).
\end{eqnarray}
The notation $A^{\times}(\cdot):=[A,\cdot]$ is a superoperator. $\delta_{jm}$ is
the Kronecker delta function. $D^{j^{\prime}}_{j}$ is the propagating superoperator
from the $j$th time point to the $j^{\prime}$th time with the definition
$D^{j^{\prime}}_{j}:=\prod^{j^{\prime}}_{i=j}\exp(\Delta t\mathcal{E}_{i})$ for
$j\leq j^{\prime}$. We define $D^{j^{\prime}}_{j}=\openone$ for $j>j^{\prime}$.
$\rho_{j}=D^{j}_{1}\rho(0)$ is the quantum state at $j$th time.

For a two-parameter case $\vec{x}=(x_{0},x_{1})$, the objective function can be
chosen as $F_{\mathrm{eff}}(T)$ according to corollary~\ref{two_parameter_theorem},
and the corresponding gradient is
\begin{equation}
\frac{\delta \mathcal{F}_{\mathrm{eff}}(T)}{\delta V_{k}(j)}=
\frac{\mathcal{I}^{2}_{11}+\mathcal{I}^{2}_{01}}{(\mathcal{I}_{00}+\mathcal{I}_{11})^{2}}
\frac{\delta \mathcal{I}_{00}}{\delta V_{k}(j)}
+\frac{\mathcal{I}^{2}_{00}+\mathcal{I}^{2}_{01}}{(\mathcal{I}_{00}+\mathcal{I}_{11})^{2}}
\frac{\delta \mathcal{I}_{11}}{\delta V_{k}(j)}-\frac{2\mathcal{I}_{01}}
{\mathcal{I}_{00}+\mathcal{I}_{11}}\frac{\delta \mathcal{I}_{01}}{\delta V_{k}(j)}.
\end{equation}
For the objective function
\begin{equation}
f_{0}(T)=\left(\sum_{a} \frac{1}{\mathcal{I}_{aa}(T)}\right)^{-1},
\end{equation}
the gradient reads
\begin{equation}
\frac{\delta f_{0}(T)}{\delta V_{k}(j)}=\sum_{a}\left(\frac{f_{0}}
{\mathcal{I}_{aa}}\right)^{2}\frac{\delta \mathcal{I}_{aa}}{\delta V_{k}(j)}.
\end{equation}

\subsection{Gradient of QFIM}

Now we calculate the gradient of the QFIM. Based on the equation
\begin{equation}
\partial_{x_{a}}\rho (T)=\frac{1}{2}(\rho (T)L_{x_{a}}(T)+L_{x_{a}}(T)\rho (T)),
\end{equation}
we can obtain
\begin{eqnarray}
\mathrm{Tr}\left(\partial_{x_{a}}\frac{\delta \rho (T)}{\delta V_{k}(j)}L_{x_{b}}\right)&=&
\frac{1}{2}\mathrm{Tr}\left(\frac{\delta \rho (T)}{\delta V_{k}(j)}
\left\{L_{x_{a}}(T),L_{x_{b}}(T)\right\}\right) \nonumber \\
& & +\frac{1}{2}\mathrm{Tr}\left(\frac{\delta L_{x_{a}}(T)}{\delta V_{k}(j)}
\left\{\rho (T),L_{x_{b}}(T)\right\}\right).
\label{apx:grad1}
\end{eqnarray}
Similarly, we have
\begin{eqnarray}
\mathrm{Tr}\left(\partial_{x_{b}}\frac{\delta \rho (T)}{\delta V_{k}(j)}L_{x_{a}}\right) &=&
\frac{1}{2}\mathrm{Tr}\left(\frac{\delta \rho (T)}{\delta V_{k}(j)}
\left\{L_{x_{a}}(T),L_{x_{b}}(T)\right\}\right) \nonumber \\
& & +\frac{1}{2}\mathrm{Tr}\left(\frac{\delta L_{x_{b}}(T)}{\delta V_{k}(j)}
\left\{\rho (T),L_{x_{a}}(T)\right\}\right).
\label{apx:grad2}
\end{eqnarray}
Next, from the definition of QFIM, the gradient for $\mathcal{F}_{ab}$ at target time
$T$ reads~\cite{LiuGRAPE2}
\begin{eqnarray}
\frac{\delta \mathcal{F}_{ab}(T)}{\delta V_{k}(j)} &=& \frac{1}{2}\mathrm{Tr}
\!\left(\!\frac{\delta \rho (T)}{\delta V_{k}(j)}\!\left\{L_{x_{a}}(T),L_{x_{b}}(T)\right\}
\!\!\right)\!+\!\frac{1}{2}\mathrm{Tr}\!\left(\!\frac{\delta L_{x_{a}}(T)}{\delta V_{k}(j)}\!
\left\{\rho (T), L_{x_{b}}(T) \right\}\!\!\right) \nonumber \\
& & +\frac{1}{2}\mathrm{Tr}\!\left(\frac{\delta L_{x_{b}}(T)}{\delta V_{k}(j)}
\left\{\rho (T), L_{x_{a}}(T) \right\} \right).
\label{apx:grad3}
\end{eqnarray}
Combing equations~(\ref{apx:grad1}), (\ref{apx:grad2}) and (\ref{apx:grad3}), one can obtain
\begin{eqnarray}
\frac{\delta \mathcal{F}_{ab}(T)}{\delta V_{k}(j)}&=&
\mathrm{Tr}\left(\partial_{x_{a}}\frac{\delta \rho (T)}{\delta V_{k}(j)}L_{x_{b}}(T)\right)
+\mathrm{Tr}\left(\partial_{x_{b}}\frac{\delta \rho (T)}{\delta V_{k}(j)}L_{x_{a}}(T)\right)
\nonumber \\
& & -\frac{1}{2}\mathrm{Tr}\left(\frac{\delta \rho (T)}
{\delta V_{k}(j)}\left\{L_{x_{a}}(T),L_{x_{b}}(T)\right\}\right).
\end{eqnarray}
Substituing the specific expressions of $\frac{\delta \rho (T)}{\delta V_{k}(j)}$ given in
reference~\cite{LiuGRAPE2}, one can obtain the gradient of $\mathcal{F}_{ab}$ as below
\begin{eqnarray}
\frac{\delta \mathcal{F}_{ab}(T)}{\delta V_{k}(j)} &=&  \frac{1}{2}\Delta t
\mathrm{Tr}\left(\left\{L_{x_{a}}(T),L_{x_{b}}(T)\right\}
\mathcal{M}^{(1)}_{j}\right) \nonumber \\
& & -\Delta^2 t \mathrm{Tr}\left[L_{x_{b}}(T)\left(\mathcal{M}^{(2)}_{j,a}
+\mathcal{M}^{(3)}_{j,a}\right)\right] \nonumber \\
& & -\Delta^2 t \mathrm{Tr}\left[L_{x_{a}}(T)\left(\mathcal{M}^{(2)}_{j,b}
+\mathcal{M}^{(3)}_{j,b}\right)\right].
\end{eqnarray}
The gradient for the diagonal entry $\mathcal{F}_{aa}$ reduces to the form in
reference~\cite{LiuGRAPE2}, i.e.,
\begin{eqnarray}
\frac{\delta \mathcal{F}_{aa}(T)}{\delta V_{k}(j)} &=& \Delta t
\mathrm{Tr}\!\left(L^{2}_{x_{a}}(T)\mathcal{M}^{(1)}_{j}\right)
-2\Delta^2 t \mathrm{Tr}\!\left[\!L_{x_{a}}(T)\!\left(\mathcal{M}^{(2)}_{j,a}
+\mathcal{M}^{(3)}_{j,a}\right)\!\right]\!.
\end{eqnarray}

\end{document}